\documentclass[twocolumn,showpacs,superscriptaddress,floatfix,prb]{revtex4-2}
\usepackage{graphicx,amsfonts,amssymb,amsmath,enumerate}
\usepackage{braket}
\usepackage{graphicx}
\usepackage{dcolumn}
\usepackage{bm}
\usepackage{color}
\usepackage[colorlinks,allcolors=blue]{hyperref}
\usepackage{physics}
\usepackage{subfigure}
\begin{document}

\preprint{APS/123-QED}

\title{Dissipationless Spin-Charge Conversion in Excitonic Pseudospin Superfluid}

\author{Yeyang Zhang}
\affiliation{International Center for Quantum Materials, School of Physics, Peking 
 University, Beijing 100871, China}
\affiliation{Collaborative Innovation Center of Quantum Matter, Beijing 100871, China}
\author{Ryuichi Shindou}
\email{rshindou@pku.edu.cn}
\affiliation{International Center for Quantum Materials, School of Physics, Peking University, Beijing 100871, China}
\affiliation{Collaborative Innovation Center of Quantum Matter, Beijing 100871, China}


\date{\today}

\begin{abstract}
Spin-charge conversion by inverse spin Hall effect or inverse Rashba-Edelstein effect is prevalent in spintronics but dissipative. We propose a dissipationless spin-charge conversion mechanism by an excitonic pseudospin superfluid in an electron-hole double layer system. Magnetic exchange fields lift singlet-triplet degeneracy of interlayer exciton levels in the double layer system. Condensation of the singlet-triplet hybridized excitons breaks both a U(1) gauge symmetry and a pseudospin rotational symmetry around the fields, leading to spin-charge coupled superflow in the system. We demonstrate the mechanism by 
deriving spin-charge coupled Josephson equations for the excitonic superflow from a coupled quantum-dot model.

\end{abstract}

\maketitle


\textit{Introduction.}---
Exploring novel approaches to information storage and transport is one of the major challenge in condensed matter physics and quantum information~\cite{baibich1988,binasch1989,loss1998,zutic2004}.  
Spintronics utilize spin degree of freedom of electrons~\cite{kim2016,kim2016b,tserkovnyak2018,tserkovnyak2019,zou2019,dasgupta2020}.
As spin voltage or spin current is hardly direct observable, 
efficient spin-charge conversion becomes a prerequisite for spintronics applications. Inverse spin Hall~\cite{saitoh2006,costache2006,kimura2007,takei2015,cornelissen2015,yuan2018} and Rashba-Edelstein~\cite{sanchez2013,shen2014,zhang2015,sanchez2016,lesne2016,isasa2016,song2017} effects are 
widely used to convert spin current and spin voltage into charge current respectively. These effects are accompanied 
by diffusive quasiparticle transport so that the spin-charge conversions by them are generally lossy.

A dissipationless spin-charge conversion can be realized in superfluids that have both 
charge~\cite{bardeen1957,bednorz1986} and spin~\cite{volovik2003,halperin1968,sonin1978_1,sonin1978_2,konig2001,hakioglu2007,can2009,sonin2010,sun2011,bender2012,sun2013,takei2014,takei2014b,chen2014,nakata2014,hoffman2015,duine2015,takei2016,liu2016,chung2018} superflow properties. Spin-triplet superconductors~\cite{sigrist1991} and ferromagnetic Josephson junctions~\cite{waintal2002,grein2009,halterman2016,eschrig2011,eschrig2015,linder2015} are 
among such systems, where spin-polarized Cooper pairs in superconductors are 
induced by spontaneous symmetry breakings or by magnetic proximity effects~\cite{tokuyasu1988,buzdin2005,bergeret2005,ohnishi2020,cai2021} from 
ferromagnetic interfaces. In ferromagnetic 
Josephson junctions, ferromagnetic moments in the interfaces control a 
relative superconducting phase between spin up and down Cooper pairs, leading to 
dissipationless Josephson charge and spin currents~\cite{grein2009,eschrig2015}. Nonetheless, 
the relative phase is a massive mode. Thereby, the finite mass hinders the low-energy conversion from spin voltage 
to charge current.


Exciton condensates in two-dimensional (2D) electron-hole double-layer (EHDL) systems are ideal platforms 
for dissipationless conversion between spin voltage and charge current. In the 2D EHDL, electron and hole layers are separated from each other by an insulating layer~\cite{zhu1995,laikhtman1996,wu2019_1,wu2019_2}.
Electrons and holes interact only through Coulomb attraction, which binds them into bound states (excitons). In the presence of a
spin-rotational symmetry in either one of the two layers, the bound states have an energy degeneracy between singlet and triplet
{\it excitonic pseudospin} (electrons and holes with opposite spins) levels.
Condensation of such excitons breaks not only 
a relative U(1) gauge symmetry between the two layers but also a pseudospin 
rotational symmetry, a combination of two spin rotational symmetries in the two layers. 
The broken gauge symmetry gives rise to electric supercurrents flowing in opposite directions in the two 
layers~\cite{lozovik1975,eisenstein2004,zhu1995}, while the broken pseudospin rotational symmetry 
leads to spin supercurrents. 
An experimental observation of the charge supercurrents without magnetic field remains illusive at this moment~\cite{wu2019_1,wu2019_2,burg2018}, while it has been observed in the quantum limit~\cite{spielman2000,tutuc2004,kellogg2004,wiersma2004,yoon2010,nandi2012,liu2017,li2017}.


In this paper, we propose a dissipationless 
spin-charge conversion in the 2D EHDL system under magnetic {\it exchange fields}. The exchange fields induce a polarization of an excitonic pseudospin.
A condensate of such excitons 
break the pseudospin rotational symmetry around the exchange fields
and the U(1) gauge symmetry, 
having two gapless Goldstone modes. 
We clarify relations among the pseudospin polarization, physical 
symmetries and the Goldstone modes in the condensate. 
We derive spin-charge coupled 
Josephson equations by a quantum-dot junction model~\cite{altland2010,eckern1984}. Based on the 
coupled Josephson equations, we show that a finite static spin voltage 
(a spatial gradient of the exchange field)
leads to an unconventional time-dependent charge supercurrent, giving a microscopic 
mechanism of the dissipationless spin-charge conversion. 
We also clarify that SOC~\cite{winkler2003,liu2008}
gives rise to spatial textures of the pseudospin polarization in the condensate~\cite{chen2019}, where 
the finite static spin voltage induces not only the charge supercurrent but also a 
dissipationless sliding of the textures.


\textit{Model.}---
The 2D EHDL system (in $xy$ plane) is described by a Hamiltonian ($\hat{H}$):

\begin{align}
\label{eqn1}
\hat{K}&\equiv \hat{H}-\mu \hat{N}\nonumber\\
&=\int\mathrm{d}^2\vec{r}\bm{a}^\dagger(\vec{r})[(-\frac{\hbar^2\nabla^2}{2m_a}-E_g)\bm{\sigma}_0+H_a\bm{\sigma}_x]\bm{a}(\vec{r})\nonumber\\
&+\int\mathrm{d}^2\vec{r}\bm{b}^\dagger(\vec{r})[(\frac{\hbar^2\nabla^2}{2m_b}+E_g)\bm{\sigma}_0+H_b\bm{\sigma}_x]\bm{b}(\vec{r})\nonumber\\
&+g\sum_{\sigma,\sigma'=\uparrow,\downarrow}\int\mathrm{d}^2\vec{r} \!\ 
a^\dagger_\sigma(\vec{r})b^\dagger_{\sigma'}(\vec{r})
b_{\sigma'}(\vec{r})a_\sigma(\vec{r}).
\end{align}
Here $\bm{a}\equiv(a_\uparrow,a_\downarrow)$ and
$\bm{b}\equiv(b_\uparrow,b_\downarrow)$ are annihilation operators of spin-1/2 electrons in
the electron and hole layer with a positive effective mass $m_a$ and a negative effective mass $-m_b$ respectively.
$2E_g$ is an energy difference between the bottom of the electron band and the top of the hole band.
Electrons in both layers have chemical potential $\mu$, and $\hat{N}$ is a total number of electrons in the EHDL system.
$H_a$ and $H_b$ are magnetic exchange fields in the two layers. The exchange 
fields can be experimentally induced by magnetic proximity effect from 
magnetic substrates. The interlayer interaction is modelled by a short-range 
interaction with a coupling constant $g$, while no tunneling 
between the two layers is allowed.
The interaction leads to
interlayer $s$-wave exciton pairing, that can be described by a four-component
exciton pairing field  
$\phi_\mu\equiv\frac{g}{2}\expval{\bm{b}^\dagger\bm{\sigma}_\mu\bm{a}}$ 
with pseudospin singlet ($\mu=0$) and triplet ($\mu=x,y,z$)
components. The exchange fields lift  
four-fold degeneracy of the exciton levels, which causes a 
singlet-triplet hybridization.

\begin{figure}[t]
\centering
\includegraphics[width=0.4\textwidth]{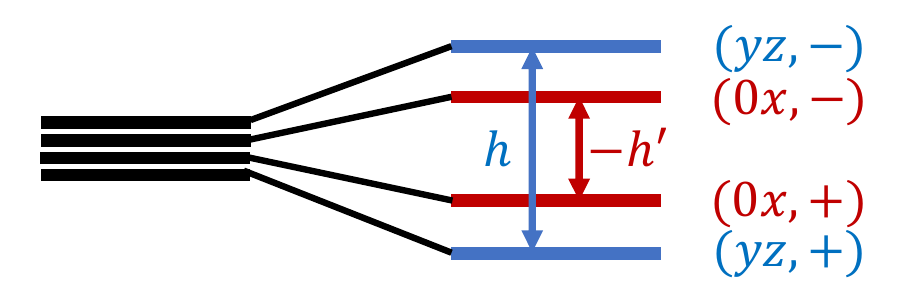}
\caption{The four-fold spin degeneracy is lifted 
by the exchange fields. When $|h|>|h'|$ ($|h|<|h'|$), the lowest 
band is transverse (longitudinal) hybrid mode, where the pseudospin 
polarization is in the $yz$ ($0x$) plane. $(yz/0x,\pm)$ are exciton 
levels whose pseudospin polarization field $\vec{\phi}$ 
are in the $yz/0x$ plane and $\pm$ specifies a relative position 
between the real and imaginary part of the four-components exciton field 
$\vec{\phi}$ within the $yz/0x$ plane. The figure is for $h>-h^{\prime}>0$.}
\label{fig.splitting}
\end{figure}


The hybdrization 
can be seen from a $\phi^4$-type effective Lagrangian 
for the four-component exciton pairing field $\vec{\Phi} 
\equiv (-\mathrm{i}\phi_0,\phi_x,\phi_y,\phi_z)$~\cite{chen2019,SM} derived from the Hamiltonian (Eq.~(\ref{eqn1})):
\begin{align}
\label{eqn2}
\mathcal{L}=&-\eta \vec{\Phi}^\dagger\cdot\partial_\tau \vec{\Phi}
-\Big(\alpha-\frac{2}{g}\Big)|\vec{\Phi}|^2-\gamma\Big[\big(\vec{\Phi}'^2\big)^2+\big(\vec{\Phi}''^2\big)^2\nonumber\\
& \hspace{-0.4cm} +6\vec{\Phi}'^2\vec{\Phi}''^2
-4\big(\vec{\Phi}'\cdot\vec{\Phi}''\big)^2\Big]+ 
\lambda|\nabla\vec{\Phi}|^2 \nonumber\\
&\hspace{-1.0cm} 
-2h\big(\Phi'_y\Phi''_z-\Phi'_z\Phi''_y\big)
+2h'\big(\Phi'_0\Phi''_x-\Phi'_x\Phi''_0\big)+\mathcal{O}(H_{a/b}^2),
\end{align}
where $|\vec{\Phi}|^2 \equiv \vec{\Phi}^{\dagger}\cdot \vec{\Phi}$ and $\tau$ is the imaginary time. 
Here $\vec{\Phi}'$ and $\vec{\Phi}''$ are real and imaginary parts of the complex-valued 
four-component exciton 
field, i.e. $\vec{\Phi} \equiv \vec{\Phi}' +{\rm i}\vec{\Phi}''$. 
$h$ and $h^{\prime}$ are weighted averages between the 
exchange fields in the electron and hole layers, 
while their coefficients as well as other 
parameters ($\eta,\alpha,\gamma,\lambda$)
in the Lagrangian depend on detailed material properties.
We assume that $\eta<0$, $\gamma<0$, and $\lambda>0$~\cite{chen2019}. 

The pseudospin degeneracy is lifted by the $h$ and $h^{\prime}$ 
terms (Fig.~\ref{fig.splitting}). 
Energy levels of the singlet-triplet hybridized modes depend on a 
competition between $h$ and $h'$, which favor $\vec{\Phi}$ 
polarized within the $yz$ and $0x$ planes respectively. 
When a mass of the lowest hybridized mode becomes negative,
the excitons undergo condensation. 
In the condensate phase with finite amplitude of $\vec{\Phi}$, the term $\gamma(\vec{\Phi}'\cdot\vec{\Phi}'')^2$
in the action competes with the 
exchange terms; the quartic term favors a parallel arrangement of 
$\vec{\Phi}'$ and $\vec{\Phi}''$, while the two exchange terms favor 
a perpendicular arrangement. The competition 
results in a finite angle between $\vec{\Phi}'$ and $\vec{\Phi}''$.

The nature of the excitonic condensate can be clarified 
by a minimization of an action $S=\int d\tau \mathrm{d}^2\vec{r}\mathcal{L}$ by  
a $\tau$-independent classical configuration~\cite{SM}. 
For $|h|>|h^{\prime}|$, the action is minimized by a transverse configuration,
\begin{align}
\label{eqn3}
\vec{\phi}_{\perp}(\theta,\varphi,\varphi_0)&=\rho\mathrm{cos}\theta(\mathrm{cos}\varphi_0\!\ 
{\vec e}_y+\mathrm{sin}\varphi_0 \!\ 
{\vec e}_z)\nonumber\\
&\hspace{-1cm} 
+\mathrm{i}\rho\mathrm{sin}\theta[\mathrm{cos}(\varphi+\varphi_0)
\!\ {\vec e}_y+\mathrm{sin}(\varphi+\varphi_0)\!\ 
{\vec e}_z], 
\end{align}
while for $|h|<|h^{\prime}|$, it is minimized 
by a longitudinal configuration, 
\begin{align}
\label{eqn4}
\vec{\phi}_{\parallel}(\theta,\varphi,\varphi_0)&=
\rho[-\mathrm{sin}\theta\mathrm{cos}(\varphi+\varphi_0)  
\!\ {\vec e}_0+\mathrm{cos}\theta\mathrm{sin}\varphi_0 \!\
{\vec e}_x]\nonumber\\
&\hspace{-1cm} 
+\mathrm{i}\rho[\mathrm{cos}\theta\mathrm{cos}\varphi_0 \!\ 
{\vec e}_0+\mathrm{sin}\theta\mathrm{sin}(\varphi+\varphi_0) \!\ 
{\vec e}_x], 
\end{align}
with $\rho\equiv \sqrt{h_c/(2|\gamma|)}$ and $h_c \equiv \alpha-2/g$.
Here ${\vec e}_\mu$ $(\mu=0,x,y,z)$ are unit vectors in the 
four-component vector space.
Note also the difference between two coordinate 
spaces, $\vec{\Phi}\equiv 
(-{\rm i}\phi_0,\phi_x,\phi_y,\phi_z)$ and $\vec{\phi}\equiv (\phi_0,\phi_x,\phi_y,\phi_z)$. 
Eqs.~(\ref{eqn3}, \ref{eqn4}) are given in the latter coordinate space.
$\varphi$ in the equations is 
the angle between $\vec{\Phi}^{\prime}$ and $\vec{\Phi}^{\prime\prime}$. 
$\theta$ defines a ratio between $|\vec{\Phi}^{\prime}|$ and 
$|\vec{\Phi}^{\prime\prime}|$ through 
$\tan\theta \equiv |\vec{\Phi}^{\prime\prime}|/|\vec{\Phi}^{\prime}|$. 
$\varphi$ and $\theta$ in Eqs.~(\ref{eqn3}, \ref{eqn4}) form a loop of
a minimum-energy degeneracy;
\begin{equation}
\label{eqn5}
\tilde{h}\equiv\mathrm{sin}\varphi\mathrm{sin}2\theta= 
{\sf{h}} \equiv 
\left\{\begin{array}{cc} 
h/h_c & {\rm for} \!\ \!\ \!\  \vec{\phi}_{\perp} \\
- h^{\prime}/h_c & {\rm for}  \!\ 
\!\ \!\ \vec{\phi}_{\parallel}. \\
\end{array}\right.
\end{equation}
We call the exciton condensate 
with one of these two configurations (``$\perp$" and ``$\parallel$") as 
transverse ($yz$) and longitudinal ($0x$) phases
respectively. Both configurations have two arbitrary 
phase variables. One is $\varphi_0$, an overall rotational phase
of the pseudospin vector within the $yz$ or $0x$ plane~\cite{SM}.
The other is a combination of $\theta$ and $\varphi$  
that satisfies the constraint Eq.~(\ref{eqn5}). These two 
are nothing but gapless Goldstone 
modes associated with broken continuous symmetries. 
A first-order transition
happens at $|h|=|h'|\neq 0$, where the general classical solution 
is given by a linear superposition of the two configurations~\cite{SM}.

\textit{Spontaneously Broken Symmetries.}---
Both $yz$ and $0x$ phases break  
the relative $U(1)$ gauge symmetry between the two 
layers. They also break the pseudospin rotational symmetry 
in which spins in the electron and hole layers are rotated around 
the field in the same and opposite direction(s) respectively.  
The two arbitrary phase variables in Eqs.~(\ref{eqn3}--\ref{eqn5}) correspond 
to the Goldstone modes associated with these symmetry breakings.
In fact, they can be absorbed into the relative gauge transformation and 
the pseduospin rotation by way of a mean-field coupling,  
$\vec{\phi}_{\omega}(\theta,\varphi,\varphi_0)
\cdot\bm{a}^\dagger\vec{\bm{\sigma}}\bm{b}$ 
($\omega=\perp,\parallel$). Namely, the coupling is 
invariant under spin rotations around the $x$ axis 
together with a change of $\varphi_0$ by $\delta \varphi_0$,  
\begin{align}
\label{eqn7}
&\vec{\phi}_{\omega}(\theta,\varphi,\varphi_0)\rightarrow
\vec{\phi}_{\omega}(\theta,\varphi,\varphi_0+\delta\varphi_0),\quad\nonumber\\
&\bm{a}\rightarrow\mathrm{e}^{\mathrm{i}\varphi_a\bm{\sigma}_x}\bm{a},\quad\bm{b}
\rightarrow
\mathrm{e}^{\mathrm{i}\varphi_b\bm{\sigma}_x}\bm{b}=
\mathrm{e}^{\mp\mathrm{i}(\varphi_a+\delta\varphi_0)\bm{\sigma}_x}\bm{b}. 
\end{align}
Here the ``$\mp$" signs in Eq.~(\ref{eqn7}) are for 
$\omega=\perp, \!\ \parallel$ respectively. 
The upper and lower signs
in ``$\pm$" and ``$\mp$" in the remaining of this letter shall be for  
$\omega=\perp, \!\ \parallel$ respectively. The coupling is also 
invariant under the relative gauge 
transformation together with a combination of 
changes of $\theta$, $\varphi$ and $\varphi_0$ 
under the constraint Eq.~(\ref{eqn5}), 
\begin{align}
\label{eqn8}
&\vec{\phi}_{\omega}(\theta,\varphi,\varphi_0) 
\rightarrow\mathrm{e}^{\mathrm{i}\psi} 
\vec{\phi}_{\omega}(\theta,\varphi,\varphi_0) 
\equiv \vec{\phi}_{\omega}(\theta(\psi),\varphi(\psi),\varphi_0(\psi)), \nonumber \\
&\bm{a}\rightarrow\mathrm{e}^{\mathrm{i}\psi_a}\bm{a}, 
\quad\bm{b}\rightarrow 
\mathrm{e}^{\mathrm{i}\psi_b}\bm{b}=
\mathrm{e}^{\mathrm{i}(\psi_a-\psi)}\bm{b}. 
\end{align}
Here $(\theta(\psi),\varphi(\psi),\varphi_0(\psi))$ satisfies
the constraint Eq.~(\ref{eqn5}) for an arbitrary U(1) phase 
$\psi$. 
A continuous variation of 
$(\theta(\psi),\varphi(\psi),\varphi_0(\psi))$ as a function of 
$\psi$ is shown in Fig.~4 of the 
supplementary material~\cite{SM}. 
To emphasize the  
dependence of $\vec{\phi}_{\omega}$ on the two variables of the 
Goldstone modes, we use $\vec{\phi}_{\omega}(\psi,\tilde{h},\varphi_0)$ 
instead of $\vec{\phi}_{\omega}(\theta,\varphi,\varphi_0)$  where 
$\tilde{h}$ is a massive mode defined in Eq.~(\ref{eqn5}). 
We further omit $\tilde{h}$ from the arguments of 
$\vec{\phi}_{\omega}$ in the followings. 

\textit{Coupled Josephson effects.}--- As an analogy to pure charge or spin superfluids~\cite{volovik2003,sonin2010}, 
the two Goldstone modes, $\varphi_0$ and $\psi$,  
are related to spin and charge supercurrents respectively. 
Without the exciton condensation, 
the electron and hole layers have a spin 
rotational symmetry:
\begin{equation}
\label{eqn9}
\bm{d}\rightarrow\mathrm{e}^{\mathrm{i}\varphi_d\bm{\sigma}_x}\bm{d},\quad H_d\rightarrow H_d-\hbar\partial_t\varphi_d,
\end{equation}
and a U(1) gauge symmetry:
\begin{equation}
\label{eqn10}
\bm{d}\rightarrow\mathrm{e}^{\mathrm{i}\psi_d}\bm{d},\quad V_d\rightarrow V_d-\hbar\partial_t\psi_d,
\end{equation}
where $\bm{d}=\bm{a}/\bm{b}$, $d=a/b$, $V_{a/b}$ and $H_{a/b}$ are electric 
potential and exchange field along 
$x$ in the electron/hole layer respectively. Spatial differences of $V_{a/b}/(-e)$ 
and $H_{a/b}/(-e)$ are defined to be charge voltage 
and spin voltage in the electron/hole layer, where $e$ is the unit charge.
Eqs. (\ref{eqn9}, \ref{eqn10}) in combination 
with Eqs.~(\ref{eqn7}, \ref{eqn8}) suggest that in the excitonic condensate, 
the charge and spin voltage control time dependence 
of $\psi$ and $\varphi_0$ respectively. As shown below,
the spatial differences of these two gapless phases lead to
spin-charge coupled Josephson currents.

The spin-charge coupled Josephson effects can be derived 
by a quantum-dot junction model~\cite{altland2010,SM}. The model 
comprises two 
domains and a junction between them. 
Each domain can be regarded 
as an EHDL quantum dot. 
The two domains ($j=1,2$) have exciton pairing 
$\vec{\phi}_{\omega}(\psi,\varphi_0)$ $(\omega=\perp,\parallel)$
with different values of $\psi$ and $\varphi_0$, i.e. $\psi_{j}$ and 
$\varphi_{0j}$ ($j=1,2$). The charge and spin voltages change across the junction in the electron/hole 
layer by $V_{Ca/b}$ and $V_{Sa/b}$ respectively. 
 $e|V_{Sa/b}|$ is assumed to be much smaller
than the exchange fields $|H_{a/b}|$, 
so that variations of the gapped modes 
($\rho$ and $\tilde{h}$) can be neglected. An action for 
the model is given by a functional of $V_{Cd}$, $V_{Sd}$, $\psi_{j}$ 
and $\varphi_{0j}$ ($d=a,b$, $j=1,2$),
that takes a quadratic form of the annihilation operators in the 
electron and hole layers in the two domains, 
${\bm a}_j({\vec r})$ ($j=1,2$) and 
${\bm b}_j({\vec r})$ ($j=1,2$). The action comprises of  
two parts: 
\begin{align}
\label{eqn11}
&\mathcal{S}[\bm{a}_j,\bm{a}^\dagger_j,\bm{b}_j,\bm{b}^\dagger_j,\psi_j,\varphi_{0j};V_{Cd},V_{Sd}]=\mathcal{S}_{T}[\bm{a}_j,\bm{a}^\dagger_j,\bm{b}_j,\bm{b}^\dagger_j]\nonumber\\
&\hspace{1cm} 
+\mathcal{S}_{\mathrm{mf}}[\bm{a}_j,\bm{a}^\dagger_j,\bm{b}_j,\bm{b}^\dagger_j,\psi_j,\varphi_{0j};V_{Cd},V_{Sd}],
\end{align}
where a mean-field part:
\begin{align}
\label{eqn12}
&\mathcal{S}_{\mathrm{mf}}=\int\mathrm{d}\tau  \sum_{j=1,2} \sum_{\alpha}\nonumber\\
\big\{&\bm{a}^\dagger_{j\alpha} [\hbar\partial_\tau+\bm{H}_{a\alpha}-\mu-\frac{\eta_j}{2}e(V_{Ca}+V_{Sa}\bm{\sigma}_x)]\bm{a}_{j\alpha}\nonumber\\
+&\bm{b}^\dagger_{j\alpha}[\hbar\partial_\tau+\bm{H}_{b\alpha}-\mu-\frac{\eta_j}{2}e(V_{Cb}+V_{Sb}\bm{\sigma}_x)]\bm{b}_{j\alpha} \nonumber \\
-& [\vec{\phi}_{\omega}(\psi_j,\varphi_{0j}) \cdot {\bm a}^{\dagger}_{j\alpha} 
{\vec{\bm \sigma}}{\bm b}_{j\alpha} + {\rm h.c.}] \big\},
\end{align}
and a tunneling part:
\begin{equation}
\label{eqn13}
\mathcal{S}_T=\int\mathrm{d}\tau\sum_{\alpha\beta}[\bm{a}^\dagger_{1\alpha}T^{(a)}_{\alpha\beta}\bm{a}_{2\beta}+\bm{b}^\dagger_{1\alpha}T^{(b)}_{\alpha\beta}\bm{b}_{2\beta}+\mathrm{h.c.}],
\end{equation}
with $\eta_1=-\eta_2=1$, ${\bm a}_{j}({\vec r}) \equiv \sum_{\alpha} u_{aj\alpha}({\vec r})
\!\ {\bm a}_{j\alpha}$ and ${\bm b}_{j}({\vec r}) \equiv 
\sum_{\alpha} u_{bj\alpha}({\vec r}) \!\  
{\bm b}_{j\alpha}$. Here $u_{dj\alpha}({\vec r})$ is the $\alpha$-th single-particle 
eigenstate of the kinetic energy part of 
Eq.~(\ref{eqn1}) for the electron/hole 
layer in the $j$-th domain 
region ($d=a/b$, $j=1,2$) 
with a proper boundary condition, 
together with its eigenenergy 
${\bm H}_{d\alpha} \equiv E_{d\alpha} \bm{\sigma}_0 + H_{d} \bm{\sigma}_x$. 
Tunneling matrices between the two domains are given by 
the single-particle eigenstates,  
$T^{(d)}_{\alpha\beta} \equiv \langle u_{d1\alpha}|{\cal T}^{(d)}|
u_{d2\beta} \rangle$, where ${\cal T}^{(d)}$ is the kinetic 
energy part for the electron/hole layer in the junction region ($d=a/b$). 
We assume that ${\cal T}^{(d)}$ 
is free from spin or
electron-hole mixing.

A perturbative treatment of the tunneling term in the junction model 
leads to an effective action of the Josephson junction~\cite{SM}:
\begin{align}
\label{eqn14a}
&\mathcal{S}_{\mathrm{eff}}[\tilde{\psi},\tilde{\varphi}_0,N_C,N_S;V_C,V_S]\nonumber\\
&= \int\mathrm{d}\tau 
\bigg[N_C(-\mathrm{i}\hbar\dot{\tilde{\psi}}(\tau)-eV_C)+N_S(\mp\mathrm{i}\hbar\dot{\tilde{\varphi}}_0(\tau)-eV_S) 
\nonumber \\
&
-\hbar I_0\bigg( 
\mathrm{cos}\big(\tilde{\psi} 
- \frac{e}{\hbar c} \Psi\big)
\mathrm{cos}
\tilde{\varphi}_0
+\bar{h}_\pm
\mathrm{sin}\big(\tilde{\psi}
-\frac{e}{\hbar c}\Psi\big)
\mathrm{sin} 
\tilde{\varphi}_0
\bigg)\bigg],
\end{align}
with $V_C\equiv V_{Cb}-V_{Ca}$, $V_S\equiv V_{Sb}\pm V_{Sa}$, $\tilde{\varphi}_0 \equiv \varphi_{01}-\varphi_{02}$, $\tilde{\psi} \equiv \psi_{1}-\psi_{2}$, $\dot{\tilde{\varphi}}_0\equiv\partial_\tau\tilde{\varphi_0}$ and 
$\dot{\tilde{\psi}}\equiv\partial_\tau\tilde{\psi}$. $c$ is the speed of light. $N_C$ and $N_S$ are 
differences of total charge and spin between the two
domains in the hole layer respectively~\cite{SM}. 
$\Psi$ is an external magnetic flux trapped in the junction region. 
$I_0$ and $\bar{h}_{\pm}$ are constants and 
$\bar{h}_{\pm}$ is proportional 
to a weighted average of $H_a$ and $H_b$, while $\bar{h}_{+} \ne \bar{h}_{-}$.
Spin currents are defined as differences of charge currents contributed by spin-up and 
spin-down electrons.
By analyses of current directions in the two layers, the currents have relations:
\begin{align}
\label{eqn14b}
I_C\equiv I_{Cb}=&-I_{Ca}=e\partial_t N_C,\nonumber\\
I_S\equiv I_{Sb}=&\pm I_{Sa}=e\partial_t N_S,
\end{align}
where $I_{Ca/b}$ and $I_{Sa/b}$ are charge currents and spin currents in the electron/hole layer respectively.
Josephson equations can be derived by minimization of Eq.~(\ref{eqn14a}) with Eq.~(\ref{eqn14b}) and Wick rotation ($\tau=\mathrm{i}t$)~\cite{SM}.
The first Josephson equations are: 
\begin{align}
I_C=& -eI_0[\mathrm{sin}(\tilde{\psi}-\frac{e}{\hbar c}\Psi)\mathrm{cos}\tilde{\varphi}_0 
-\bar{h}_{\pm}\mathrm{cos}(\tilde{\psi}-\frac{e}{\hbar c}\Psi)\mathrm{sin}\tilde{\varphi}_0], 
\label{eqn15} \\
\pm I_S=&  -eI_0[\mathrm{sin}\tilde{\varphi}_0\mathrm{cos}(\tilde{\psi}-\frac{e}{\hbar c}\Psi)
-\bar{h}_{\pm}\mathrm{cos}\tilde{\varphi}_0\mathrm{sin}(\tilde{\psi}-\frac{e}{\hbar c}\Psi)]. \label{eqn16}
\end{align}
The second Josephson equations are:
\begin{equation}
\label{eqn17}
\frac{\mathrm{d}\tilde{\psi}}{\mathrm{d}t}=-\frac{e}{\hbar}V_C,\quad\frac{\mathrm{d}\tilde{\varphi}_0}{\mathrm{d}t}=\mp\frac{e}{\hbar} V_S. 
\end{equation}

The Josephson equations reveal spin-charge coupled Josephson effects.
The term proportional to ${\rm sin} (\tilde{\psi}-\frac{e}{\hbar c}\Psi)$ in Eq.~(\ref{eqn15}) 
and the term proportional to ${\rm sin} (\tilde{\varphi}_0)$ in Eq.~(\ref{eqn16}) represent the well-known 
pure charge and pure spin Josephson effects~\cite{josephson1974,sonin2010,takei2017},
while they are modulated by the spin phase ($\tilde{\varphi}_0$) and the charge phase ($\tilde{\psi}$) respectively.
Moreover, the terms proportional to $\bar{h}_{\pm}$ in Eqs.~(\ref{eqn15}, \ref{eqn16})  
indicates that a pure spin (charge) phase difference can still lead to 
a charge (spin) supercurrent, as the excitons are polarized by the exchange fields.
In a trilayer ferromagnetic Josephson junction of superconductors, a relative angle between two 
ferromagnetic polarizations in two sides of a ferromagnetic junction plays a similar role to 
the spin phase~\cite{grein2009,eschrig2015}. Differently, Eq.~(\ref{eqn17}) further shows control of the spin phase 
by the spin voltage.

\begin{figure}[t]
\centering
\subfigure[ ]{
\label{fig.conversion_1}
\begin{minipage}{0.38\linewidth}
\centering
\includegraphics[width=1\linewidth]{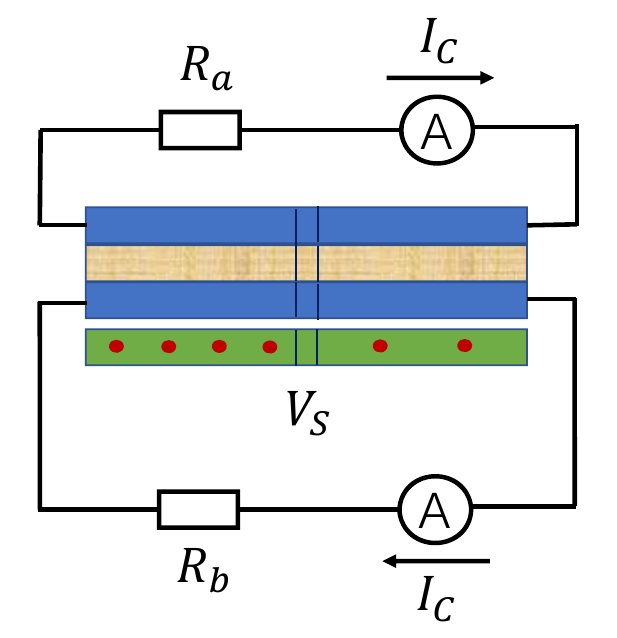}
\end{minipage}
}
\subfigure[ ]{
\label{fig.conversion_2}
\begin{minipage}{0.53\linewidth}
\centering
\includegraphics[width=1\linewidth]{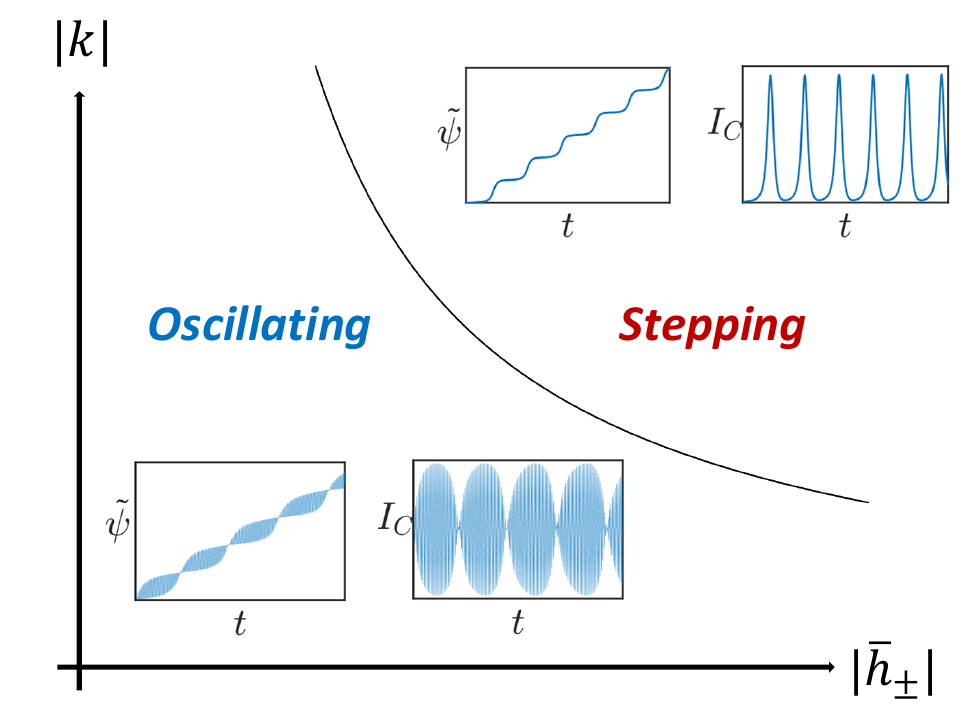}
\end{minipage}
}
\caption{\label{fig.conversion} The charge current ($I_C$) induced by the spin voltage ($V_S$). \textbf{(a)} The spin voltage is 
added at the Josephson junction in the hole layer. The charge currents can be measured by the two external 
circuits attached to the electron and hole layers respectively. \textbf{(b)} The a.c. behavior 
of $\tilde{\psi}(t)$ and $I_C(t)$ according to Eq.~(\ref{eqn18}) for small $|\bar{h}_{\pm}|$. 
$\tilde{\psi}(t)$ shows an oscillating behavior for $|\bar{h}_{\pm}k|<1$ and a stepping behavior for
$|\bar{h}_{\pm}k|>1$.}
\end{figure}

\textit{Device setup.}---
To propose the spin-charge conversion in 
a feasible experimental setup, we consider to put
two magnetic substrates with different magnetizations along the 
same ($x$) direction under the hole layer (Fig.~\ref{fig.conversion_1}).
The two substrates introduce the two domains in the EHDL system, 
whose hole layers experience the magnetic exchange 
fields through the proximity effect. The difference of the exchange fields results 
in a finite d.c. spin voltage $V_S$ across the junction in the 
hole layer. The d.c. spin voltage 
results in a linear increase of $\tilde{\varphi}_0$,
$\tilde{\varphi}_0=\mp\frac{e}{\hbar} V_S t$ ($\tilde{\varphi}_0=0$ 
at $t=0$ is taken without loss of generality). The time dependence 
of $\tilde{\varphi}_0$ gives rise to a.c. electric currents 
in counter-propagating directions in the electron and hole 
layers respectively. The electric currents induce the a.c. 
charge voltages across the junction in the electron and hole 
layers as $I_{Ca} R_a$ and $I_{Cb} R_b$, where 
$R_a$ and $R_b$ are external resistances (Fig.~\ref{fig.conversion_1}).
The exciton U(1) phase $\tilde{\psi}$ couples only with 
the difference between the charge voltages in 
the two layer, $V_C=I_{Ca}R_a - I_{Cb}R_b$.
Thus, Eqs.~(\ref{eqn15}, \ref{eqn17}) 
give an equation of motion  (EOM) for $\tilde{\psi}$:
\begin{align}
\label{eqn18}
&I_C(s)\frac{R}{V_S}=\frac{\mathrm{d}\tilde{\psi}}{\mathrm{d}s}=-k[\mathrm{sin}(\tilde{\psi})\mathrm{cos}(s) 
\pm \bar{h}_{\pm}\mathrm{cos}(\tilde{\psi})\mathrm{sin}( s)], 
\end{align}
with $R\equiv R_a+R_b$, 
a normalized time $s\equiv eV_S t/\hbar$, two 
dimensionless parameters, $k \equiv eI_0R/V_S$ and 
$\overline{h}_{\pm}$. 

Solutions of the EOM  are obtained numerically
in the supplementary material~\cite{SM}. $\tilde{\psi}(s)$
shows an oscillating behavior with a double-sine form 
for $|\bar{h}_{\pm}|<1/|k|$, and a stepping
behaviour for $|\bar{h}_{\pm}|>1/|k|$ (Fig.~\ref{fig.conversion_2}).
$\tilde{\psi}(s)$ has an oscillatory component with $2\pi$ periodicity in $s$.
The $\bar{h}_{\pm}$ term with $\mp k\bar{h}_{\pm} > 0 \!\ (<0)$
gives rise to a component of $\tilde{\psi}(s)$ that
increases (decreases) linearly in the time $s$ and
an additional longer oscillatory periodicity
over which $\tilde\psi(s)$ increases (decreases) by $\pi$.
When the longer periodicity approaches the shorter periodicity, 
the oscillating behavior shows a crossover to the stepping behavior.   
The double-sine form appears because the electric charge current 
is induced not only by a sine of the charge phase $\tilde{\psi}$ but also 
by another sine of the spin phase $\tilde{\varphi}_0$.
The spin voltage $V_S$ can be measured from the (short) 
period of the a.c. electric current (Fig.~\ref{fig.conversion_2}).


\textit{Spin-orbit coupling.}--- 
A 
semiconductor hetetrostructure of the 2D EHDL systems breaks a spatial inversion symmetry, 
causing an effective Rashba SOC in the electron layer~\cite{winkler2003,liu2008,liu2010,pikulin2014}. 
The Rashba SOC endows the excitonic pseudospin polarizations with a nonzero 
momentum $K$ in a 
direction perpendicular to the exchange fields; $\varphi_0$ in Eqs.~(\ref{eqn3}, \ref{eqn4}) 
is replaced by $\varphi_0-Ky$~\cite{chen2019,SM}. The condensate with the broken 
traslational symmetry also have the relative U(1) phase ($\psi$) and the spin rotational phase ($\varphi_0$) 
as low-energy Goldstone modes. The spin rotational phase $\varphi_0$ appears 
together with the spatial coordinate ($y$), so that it is also a 
translational phase (phason). 
The gapless $\varphi_0$ phase originates purely 
from the spin-rotational symmetry in the hole layer. Charge and (hole-layer) spin voltages 
control these two gapless modes respectively (Eq.~(\ref{eqn17})), while spatial gradients of these 
two modes generate charge and spin currents as well~\cite{SM}. Accordingly, the dissipationless 
spin-charge conversion property is robust against the presence of the Rashba SOC.

\textit{Summary.}---
In this letter, we clarify the spin-charge coupled Josephson effects 
in the EHDL exciton system under magnetic exchange fields, where the charge Josephson 
current can be a response to the spin voltage. The spin-charge coupling effects
provide a dissipationless way of the spin-charge conversion in a feasible device  
setup. 

\begin{acknowledgments}
Y. Z. and R. S. thanks fruitful discussions with Junren Shi, Rui-Rui Du, Xi Lin, 
Ke Chen, Zhenyu Xiao and Lingxian Kong. The work is supported by 
the National Basic Research Programs of China (No. 2019YFA0308401) and 
by National Natural Science Foundation of China (No.11674011 and No. 12074008). 
\end{acknowledgments}

\bibliography{spin_superfluid}

\clearpage
\begin{widetext}
\numberwithin{equation}{subsection}
\section*{Supplementary Material for ``Dissipationless Spin-Charge Conversion in Excitonic Pseudospin Superfluid"} 

In the main text, the $\phi^4$-type effective Lagrangian is introduced for the 
four-component excitonic fields.
The $\phi^4$-type Lagrangian is derived from Eq.~(1) perturbatively in the 
exchange fields.   
The Lagrangian is minimized in terms of a classical solution 
of the excitonic fields, leading to the prediction of 
transverse and longitudinal phases. 
Both of them are excitonic pseudospin superfluid phases. Using a 
coupled quantum dots model or the effective Lagrangian, 
we derive spin-charge coupled 
Josephson equations in these excitonic pseudospin superfluid phases.
We also discuss the robustness of the spin-charge conversion against the Rashba 
spin-orbit coupling (SOC) in an electron layer. In this supplementary material, we explain 
these derivations and minimizations as well as related details. 

The structure of this supplementary material is as follows. In the next section, 
we review possible forms of relativistic SOCs in 2D semiconductor 
heterostructures for the EHDL systems. In Sec.~\ref{sec1}, we give the perturbative derivation 
of the $\phi^4$-type effective Lagrangian with the Rashba SOC. In Sec.~\ref{sec2}, we describe the minimization 
of the Lagrangian without the SOC, where the classical ground-state phase diagram of transverse 
and longitudinal phases are presented. In Sec.~\ref{sec3} and \ref{sec4}, we clarify what global 
continuous symmetries are broken in the transverse and longitudinal phases, and 
we associate the broken symmetries with the gapless Goldstone modes in these 
phases. In Sec.~\ref{sec5}, we derive the spin-charge coupled Josephson equations based on 
the coupled quantum dots model.
In Sec.~\ref{sec6}, we describe solutions of the Josephson 
equation under a physical experimental setup. In Sec.~\ref{sec7}, we describe the minimization of 
the Lagrangian with the Rashba SOC in the electron layer, 
where a classical ground-state phase diagram of helicoidal and helical 
phases are presented. In Sec.~\ref{sec8}, we derive the spin-charge 
coupled Josephson equation with the Rashba SOC, using Noether's theorem for the $\phi^4$-type 
effective Lagrangian.
In Sec.~\ref{sec10} and Sec.~\ref{sec11}, we give 
supplementary details of Sec.~\ref{sec5} and Sec.~\ref{sec7} respectively.

There are some notation simplifications in this supplementary material.
We take $\hbar=c=e=1$ unless dictated otherwise (we recover these fundamental physical constants 
in the very last expressions of important physical equations).

\subsection{\label{sec0} Spin-orbit couplings (SOCs) in 2D semiconductor heterostructure systems}
For a 2D semiconductor heterostructure of the EHDL systems, an effective confinement 
potential along $z$ direction may break spatial inversion symmetry. This kind of the broken spatial 
inversion symmetry is called structural inversion asymmetry (SIA). The SIA results in an effective 
electric field along $z$ direction, which leads to Rashba SOC in the electron layer;
\begin{equation}
\label{eqn0-0}
\hat{H}_R=\xi_e\int\mathrm{d}^2\vec{r}\bm{a}^\dagger( \vec{r})(-\mathrm{i}\partial_y\bm{\sigma}_x+\mathrm{i}\partial_x\bm{\sigma}_y)\bm{a}(\vec{r}),
\end{equation}
where $\xi_e$ is a strength of the Rashba coupling. The SIA in the electron-hole 
double-layer semiconductor systems results not only in the Rashba coupling in the 
electron layer with $S_z=\pm 1/2$, but also Rashba coupling 
in the hole layer with $J_z=\pm 3/2$. As the Pauli matrices ${\bm \sigma}_x$ and ${\bm \sigma}_y$ 
connect between the $J_z=\pm 3/2$ Kramers doublet in the hole layer, the SOC for the hole layer
is proportional to a cubic in the momentum $k$ and atomic spin-orbit interaction strength and it 
is therefore negligibly small especially around the $\Gamma$ point. In addition to the SIA, 
the bulk crystal structure itself may break spatial inversion symmetry. 
The bulk inversion asymmetry (BIA) leads to Dirac-type SOC term in the hole 
layer of the EHDL systems~\cite{chen2019,liu2008,liu2010}:
\begin{align}
    \hat{H}_{D} = \Delta_h \int d^2 \vec{r} {\bm b}^{\dagger}(\vec{r}) \big(-\mathrm{i}\partial_x {\bm \sigma}_x 
    - \mathrm{i}\partial_y {\bm \sigma}_y\big){\bm b}(\vec{r}). \label{eqn0-1}
\end{align}
For a typical semiconductor heterostructure system, however, 
the Dirac-type SOC is still quantitatively negligible, $\Delta_h\sim 0.01\xi_e$\cite{liu2013,pikulin2014}.
In this paper, we thus consider only the Rashba SOC in the electron layer and study 
how it changes the nature of the excitonic condensate (Sec.~\ref{sec7}) and whether it influences the  
dissipationless spin-charge conversion property or not (Sec.~\ref{sec8}).

\subsection{\label{sec1}Derivation of $\phi^4$ type effective Lagrangian}
In this section, we describe a derivation of the effective 
Lagrangian from the Hamiltonian Eq.~(1) in the main text with the Rashba SOC, Eq.~(\ref{eqn0-0}).
The derivation is perturbative in the exchange 
fields in Eq.~(1) and in the Rashba coupling in Eq.~(\ref{eqn0-0}). 
Since both the exchange fields and the Rashba coupling are often much 
smaller than the typical energy scale of the electron and hole bands, 
we include only their first-order perturbation effect. 

The partition function of the four-components exciton pairing fields $\vec{\phi}_{\mu} = \frac{g}{2}\langle {\bm b}^{\dagger} {\bm \sigma}_{\mu}{\bm a}\rangle$ ($\mu=0,x,y,z$) can be derived
from Eq.~(1) together with the Rashba coupling
in the electron band (Eq.~(\ref{eqn0-0}));
\begin{equation}
\label{eqn1-1}
Z=\int\mathcal{D}\phi^\dagger\mathcal{D}\phi\mathrm{exp}\{-\frac{2}{g}|\vec{\phi}|^2+\mathrm{Tr}[G_0G^{-1}_RG_0\Psi G_0\Psi+G_0G^{-1}_HG_0\Psi G_0\Psi-\frac{1}{2}G_0\Psi G_0\Psi-\frac{1}{4}G_0\Psi G_0\Psi G_0\Psi G_0\Psi]\},
\end{equation}
where
\begin{equation}
\label{eqn1-2}
G^{-1}_0(q)\equiv\left(\begin{array}{cc}(-\mathrm{i}\omega_n+\mathcal{E}_a(\vec{k})-\mu)\bm{\sigma}_0 & 0 \\ 0 & (-\mathrm{i}\omega_n+\mathcal{E}_b(\vec{k})-\mu)\bm{\sigma}_0\end{array}\right) 
\equiv \left(\begin{array}{cc} {g^{0}_a}^{-1}(k) & 0 \\
0 & {g^{0}_b}^{-1}(k) \\
\end{array}\right), 
\end{equation}
\begin{equation}
\label{eqn1-3}
G^{-1}_R(q)\equiv\left(\begin{array}{cc}\xi_e(k_y\bm{\sigma}_x-k_x\bm{\sigma}_y)& 0 \\ 0 & 0\end{array}\right),\qquad G^{-1}_H(q)\equiv\left(\begin{array}{cc}H\bm{\sigma}_x& 0 \\ 0 & H\bm{\sigma}_x\end{array}\right),
\end{equation}
\begin{equation}
\label{eqn1-4}
\Psi(q)=\frac{1}{\sqrt{\beta V}}\left(\begin{array}{cc}0 & -\vec{\phi}(-k)\cdot\vec{\sigma} \\ -\vec{\phi}^{*}(k)\cdot\vec{\sigma} & 0 \end{array}\right),
\end{equation}
with $\mathcal{E}_a(\vec{k})\equiv\frac{\hbar^2\vec{k}^2}{2m_e}-E_g$, $\mathcal{E}_b(\vec{k})\equiv-\frac{\hbar^2\vec{k}^2}{2m_h}+E_g$,  $q\equiv(\mathrm{i}\omega_n,\vec{k})$ and $\vec{k}=(k_x,k_y)$. 
The expansion is perturbative in the excitonic field $\Psi(q)$, 
the Rashba coupling $G^{-1}_R(q)$ and the exchange 
fields $G^{-1}_{H}(q)$. From the gauge symmetry, the expansion 
contains only the even order in $\Psi(q)$. For the 2nd order 
in $\Psi(q)$, we expand the exchange fields and 
the Rashba coupling up to the first order. 
We ignore their effect in the quartic order 
in $\Psi(q)$. 

The expansion has been previously 
carried out only for triplet-components 
excitonic field in Ref.~\cite{chen2019}. 
In the following, we describe the expansion for 
singlet-component as well.  
$\mathrm{Tr}'[...]$ denotes an additional 
contribution from the singlet-component $\phi_0\bm{\sigma}_0$, 
\begin{equation}
    {\rm Tr}^{\prime}\big[\cdots\big] \equiv  
    {\rm Tr}\big[\cdots\big] - {\rm Tr}\big[\cdots\big]_{\phi_0=0},
    \label{eqn1-5a}
\end{equation}
and $\hat{\phi}$ and $\hat{\bm{\sigma}}$ denote 
the three-component (spin-triplet) vectors. 
The leading-order terms in the 
expansion are as follows:
\begin{equation}
\label{eqn1-5}
\mathrm{Tr}'[G_0G_R^{-1}G_0\Psi G_0\Psi]=-\mathrm{i}D\int \mathrm{d}\tau \mathrm{d}^2 \vec{r} \!\  
\vec{e}_z\cdot[(\hat{\phi}^*\times\nabla)\phi_0-\phi_0^*(\nabla\times\hat{\phi})],
\end{equation}
\begin{equation}
\label{eqn1-6}
\mathrm{Tr}'[G_0G_H^{-1}G_0\Psi G_0\Psi]=-h'\int \mathrm{d}\tau \mathrm{d}^2 \vec{r} \!\    
\vec{e}_x\cdot(\phi_0^*\hat{\phi}+\phi_0\hat{\phi}^*),
\end{equation}
\begin{equation}
\label{eqn1-7}
\mathrm{Tr}'[-\frac{1}{4}(G_0\Psi)^4]=\gamma\int \mathrm{d}\tau \mathrm{d}^2 \vec{r} \!\    
[|\phi_0|^4+4|\phi_0|^2|\hat{\phi}|^2+(\phi_0^*)^2(\hat{\phi})^2+(\phi_0)^2(\hat{\phi}^*)^2],
\end{equation}
\begin{equation}
\label{eqn1-8}
\mathrm{Tr}'[-\frac{1}{2}G_0\Psi G_0\Psi]=\alpha\int \mathrm{d}\tau \mathrm{d}^2 \vec{r} \!\    
|\phi_0|^2+\eta\int \mathrm{d}\tau \mathrm{d}^2 \vec{r} \!\ 
\phi_0^*\partial_\tau\phi_0+\lambda\int \mathrm{d}\tau \mathrm{d}^2 \vec{r} \!\ \phi_0^*\nabla^2\phi_0. 
\end{equation}
These terms share the same coefficients as 
those in the leading-order terms in Ref.~\cite{chen2019}, 
except for $h'$. For the 
later convenience, 
we give the expressions of $D$, $h$ and $h^{\prime}$ 
as follows:
\begin{equation}
\label{eqn1-9a}
D=-\frac{2\xi_e}{\beta V} 
\frac{\hbar^2}{m_b} \sum_{k} k^2_x g^{0}_a(k)^2 g^0_b(k)^2,  
\end{equation}
\begin{equation}
\label{eqn1-9b}
h=\frac{2}{\beta V}\sum_{k} g^{0}_a(k) g^0_b(k) (H_b g^{0}_b(k) - H_a g^0_a(k)),  
\end{equation}
\begin{equation}
\label{eqn1-9c}
h'=\frac{2}{\beta V} \sum_{k} g^0_a(k) g^0_b(k) \big[H_a g^0_a(k)+ H_b g^0_b(k)\big]
= -\frac{2}{V}\sum_{\vec{k}}\frac{\beta}{\mathcal{E}_a-\mathcal{E}_b}[\frac{H_a}{2+2\mathrm{cosh}\beta(\mathcal{E}_a-\mu)}-\frac{H_b}{2+2\mathrm{cosh}\beta(\mathcal{E}_b-\mu)}].
\end{equation}

Putting Eqs.~(\ref{eqn1-5}--\ref{eqn1-8}) into Eq.~(\ref{eqn1-1}) and add them 
into the triplet component (Eq.~(5) of Ref.~\cite{chen2019}), 
we obtain 
the $\phi^4$-type effective Lagrangian 
for the four-components excitonic field:
\begin{align}
\label{eqn1-11}
S&=\int 
\mathrm{d}\tau \mathrm{d}^2 \vec{r} \!\  
\{-\eta\vec{\phi}^\dagger\partial_\tau\vec{\phi}+\lambda|\nabla\vec{\phi}|^2-(\alpha-\frac{2}{g})|\vec{\phi}|^2-\gamma[2|\hat{\phi}|^4-(\hat{\phi}^*)^2(\hat{\phi})^2+|\hat{\phi}_0|^4+4|\phi_0|^2|\hat{\phi}|^2+(\phi_0^*)^2(\hat{\phi})^2+(\phi_0)^2(\hat{\phi}^*)^2]\nonumber\\
&-D[\vec{e}_y\cdot(\hat{\phi}^*\times\partial_x\hat{\phi})-\vec{e}_x\cdot(\hat{\phi}^*\times\partial_y\hat{\phi})-\mathrm{i}\vec{e}_z\cdot((\hat{\phi}^*\times\nabla)\phi_0-\phi_0^*(\nabla\times\hat{\phi}))]+\mathrm{i}h\vec{e}_x\cdot(\hat{\phi}^*\times\hat{\phi})+h'\vec{e}_x\cdot(\phi_0^*\hat{\phi}+\phi_0\hat{\phi}^*)\},
\end{align}
where $\vec{\phi}\equiv(\phi_0,\hat{\phi})=(\phi_0,\phi_x,\phi_y,\phi_z)$, and $\nabla=(\partial_x,\partial_y,0)$. In terms of  $\vec{\Phi}\equiv(-\mathrm{i}\phi_0,\hat{\phi})\equiv\vec{\Phi}'+\mathrm{i}\vec{\Phi}''$, 
the Lagrangian takes a more symmetric form:
\begin{align}
\label{eqn1-12}
S&=\int_0^\beta\mathrm{d}\tau\int\mathrm{d}^2\vec{r}\{-\eta\vec{\Phi}^\dagger\partial_\tau\vec{\Phi}-(\alpha-\frac{2}{g})|\vec{\Phi}|^2-\gamma[(\vec{\Phi}'^2)^2+(\vec{\Phi}''^2)^2+6\vec{\Phi}'^2\vec{\Phi}''^2-4(\vec{\Phi}'\cdot\vec{\Phi}'')^2]+\lambda[(\nabla\vec{\Phi}')^2+(\nabla\vec{\Phi}'')^2]\nonumber\\
&-D(\Phi'_z\partial_x\Phi'_x-\Phi'_x\partial_x\Phi'_z+\Phi'_z\partial_y\Phi'_y-\Phi'_y\partial_y\Phi'_z)-D(\Phi'_0\partial_x\Phi'_y-\Phi'_y\partial_x\Phi'_0+\Phi'_x\partial_y\Phi'_0-\Phi'_0\partial_y\Phi'_x)\nonumber\\
&-D(\Phi''_z\partial_x\Phi''_x-\Phi''_x\partial_x\Phi''_z+\Phi''_z\partial_y\Phi''_y-\Phi''_y\partial_y\Phi''_z)-D(\Phi''_0\partial_x\Phi''_y-\Phi''_y\partial_x\Phi''_0+\Phi''_x\partial_y\Phi''_0-\Phi''_0\partial_y\Phi''_x)\nonumber\\
&-2h(\Phi'_y\Phi''_z-\Phi'_z\Phi''_y)+2h'(\Phi'_0\Phi''_x-\Phi'_x\Phi''_0)\}+\mathcal{O}(\xi_{e}^2,H^2,\xi_{e}H),
\end{align}
with $|\vec{\Phi}|^2 \equiv \vec{\Phi}'^2 + \vec{\Phi}''^2$. In absence of the Rashba term ($D=0$), 
this reduces to Eq.~(2) in the main text. 

Before closing this section, we like to mention a relation between $\lambda$ and $D$ and that between $h$ and $h^{\prime}$. 
$\lambda$, $\alpha$ and $\eta$ in the action comes from an expansion of the 
bare polarization function in frequency and momentum,
\begin{equation}
\label{eqn1-13}
\alpha_q\equiv -\frac{2}{\beta V}\sum_k g_b^0(k-\frac{q}{2})g_a^0(k+\frac{q}{2})=\alpha_q^{(0)}+\alpha_q^{(1)}\mathrm{i}\omega_m+\alpha_q^{(2)}\vec{q}^2+...
\end{equation}
$\alpha$ is the zero-th order term in the expansion;
\begin{align}
\alpha \equiv \alpha^{(0)}_{q} = -\frac{2}{\beta V} g^{0}_a(k) g^{0}_b(k) = -\frac{2}{V} 
\sum_{\bm k} \frac{1}{\mathcal{E}_a-\mathcal{E}_b} \Big\{\frac{1}{e^{\beta (\mathcal{E}_a-\mu)}+1}- 
\frac{1}{e^{\beta (\mathcal{E}_b-\mu)}+1}\Big\} >0.  
\end{align}
$\alpha$ is positive and it increases on lowering the temperature.  In terms of a relation, 
\begin{equation}
\label{eqn1-14}
\sum_k g^0_b(k-\frac{q_x}{2})g^0_a(k+\frac{q_x}{2})=\sum_k g^0_b(k-q_x)g^0_a(k)=\sum_k g^0_b(k)g^0_a(k+q_x),
\end{equation}
$\lambda$ is calculated as follows:
\begin{equation}
\label{eqn1-15}
\lambda=-\alpha_q^{(2)}=-\frac{1}{\beta V}\sum_k g^0_b(k)'g^0_a(k)'=
\frac{1}{\beta V }\frac{\hbar^4}{m_a m_b}\sum_k k_x^2 g^0_a(k)^2 g^0_b(k)^2,
\end{equation}
with $g^0_a(k)^{\prime} \equiv \partial_{k_x}g^0_a(k)$. A comparison 
to Eq.~(\ref{eqn1-9a}) gives the relation between $\lambda$ and $D$ as
\begin{equation}
\label{eqn1-18}
K \equiv \frac{D}{2\lambda}=-\frac{\xi_e m_a}{\hbar^2},
\end{equation}
where we recover $\hbar$ in Eq.~(\ref{eqn1-18}) by substitutions $1/m_{a/b}\rightarrow\hbar^2/m_{a/b}$ in 
Eqs.~(\ref{eqn1-9a}, \ref{eqn1-15}, \ref{eqn1-18}). The ratio between $h'$ and $h$ are determined by $m_b/m_a$ and $\beta E_g$. Using 
Eqs.~(\ref{eqn1-9b}, \ref{eqn1-9c}), we calculate $h^{\prime}/h$ as a function of 
$\beta E_g$ and $m_b/m_a$ (Fig.~\ref{fig.proportion_1}),
where we take the chemical potential $\mu$ at intersections of the electron band and the hole bands.
In Sec.~\ref{sec2}, we show that when $|h^{\prime}|>|h|$ / $|h|>|h|^{\prime}$, 
the classical effective Lagrangian 
Eq.~(2) is minimized by the longitudinal/transverse 
phase (Fig.~\ref{fig.phases_1}). We combine 
Fig.~\ref{fig.proportion_2} and 
Fig.~\ref{fig.phases_1}, to have a finite-temperature 
phase diagram, Fig.~\ref{fig.proportion_2}. 
Note that the phase 
diagram is valid for the case with
$\xi_e=0$. In the case with
$\xi_e\ne 0$, the transverse and longitudinal 
phases are replaced by helicoidal and helical 
phases respectively (see Sec.~\ref{sec7}). Note also that the 
zero-temperature limit of the phase diagram 
($\beta \rightarrow \infty$) indicates the 
classical ground state of the 2D EHDL system 
under the exchange fields
is the transverse phase. 

\begin{figure}[t]
\centering
\subfigure[ ]{
\label{fig.proportion_1}
\begin{minipage}{0.45\textwidth}
\centering
\includegraphics[width=\textwidth]{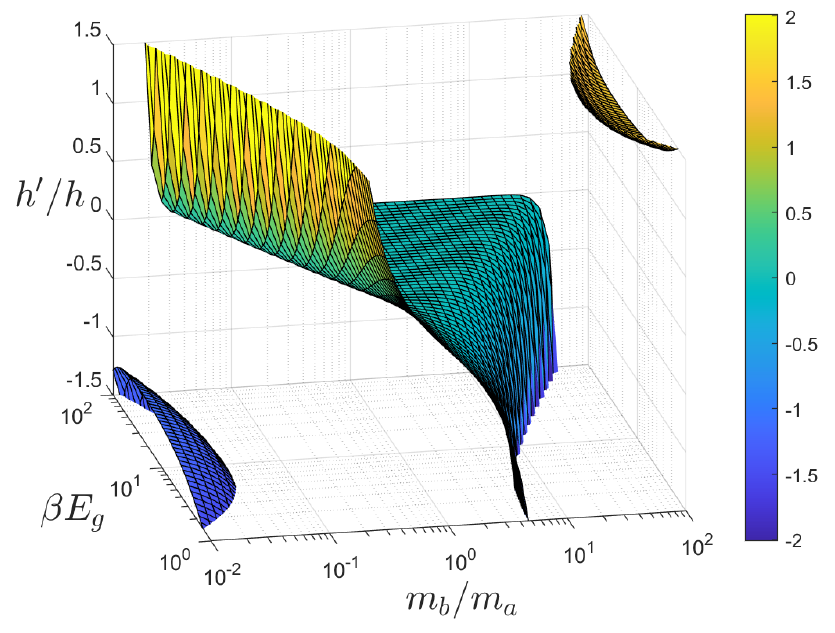}
\end{minipage}
}
\subfigure[ ]{
\label{fig.proportion_2}
\begin{minipage}{0.47\textwidth}
\centering
\includegraphics[width=\textwidth]{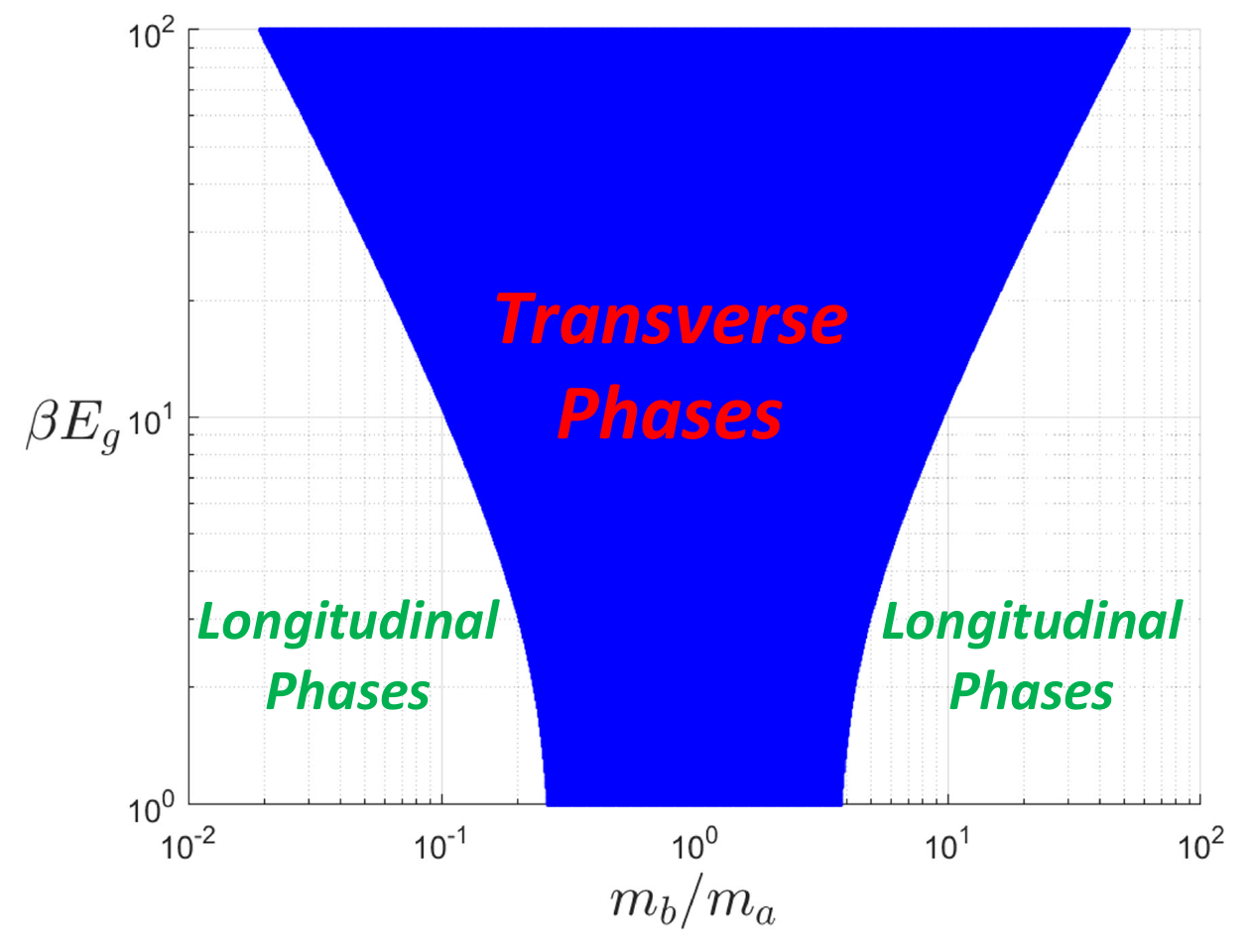}
\end{minipage}
}
\caption{\label{fig.proportion} \textbf{(a)} A ratio between $h'$ and $h$ 
depends on a ratio between the two effective mass ($m_a$ and $m_b$), temperature, band inversion parameter $E_g$ and chemical potential.
The ratio $h'/h$ is plotted as a function 
of $m_b/m_a$, and the band inversion parameter 
normalized by the temperature. A chemical potential at intersections of the two band is considered.
\textbf{(b)} When $|h'/h|<1$ ($|h'/h|>1$), the classical ground state is transverse (longitudinal) phases. 
Combining this with Fig.~(a), we show the 
phase diagram as a function of $m_b/m_a$ and 
$\beta E_g$.} 
\end{figure}

\subsection{\label{sec2}Derivation of classical ground-state phase diagram without Rashba coupling}

In this section, we describe the minimization of Eq.~(2) of the main text, while  
we describe the minimization of Eq.~(\ref{eqn1-12}) in Sec.~\ref{sec7}. 
Note first that spatial derivative term in Eq.~(2) is positive definite,  $\lambda|\nabla\vec{\Phi}|^2\geq 0$. Thus, only a spatially uniform solution of 
$\vec{\Phi}$ minimizes the action,  
\begin{align}
\label{eqn2-1}
\mathcal{L}&=A(\Phi'^2+\Phi''^2)+B[\Phi'^4+\Phi''^4+6\Phi'^2\Phi''^2]\nonumber\\
&-4B\Phi'^2\Phi''^2(\mathrm{cos}\eta_1\mathrm{cos}\eta_2\mathrm{cos}\alpha_1+\mathrm{sin}\eta_1\mathrm{sin}\eta_2\mathrm{cos}\alpha_2)^2+2h'\Phi'\Phi''\mathrm{cos}\eta_1\mathrm{cos}\eta_2\mathrm{sin}\alpha_1-2h\Phi'\Phi''\mathrm{sin}\eta_1\mathrm{sin}\eta_2\mathrm{sin}\alpha_2, \nonumber \\
&\equiv A(\Phi'^2+\Phi''^2)+B[\Phi'^4+\Phi''^4+6\Phi'^2\Phi''^2] 
- 2 \Phi^{\prime} \Phi^{\prime\prime} g(\eta_1,\eta_2,\alpha_1,\alpha_2),
\end{align}
with $A\equiv -(\alpha-2/g)<0$ and $B=-\gamma>0$. 
Here $\Phi^{\prime}$ and $\Phi^{\prime\prime}$ are the norm of 
the four-component vector fields $\vec{\Phi}^{\prime}$ and $\vec{\Phi}^{\prime\prime}$ 
respectively. $\eta_1$, $\eta_2$, $\alpha_1$ and $\alpha_2$ 
define relative angles among $\vec{\Phi}^{\prime}$, 
$\vec{\Phi}^{\prime\prime}$ and a $0x$ plane subtended by $\vec{e}_x$ and 
$\vec{e}_0$ (Figs.~\ref{fig.minima_1}--\ref{fig.minima_4}). $\eta_1$ ($\eta_2$) is an angle between $\vec{\Phi}^{\prime}$
($\vec{\Phi}^{\prime\prime}$) and the $0x$ plane (Figs.~\ref{fig.minima_1}, \ref{fig.minima_2}). To define $\alpha_1$ and 
$\alpha_2$, we decompose $\vec{\Phi}^{\prime}$ and 
$\vec{\Phi}^{\prime\prime}$ into a component parallel 
to the $0x$ plane and the other, $\vec{\Phi}^{\prime}=\vec{\Phi}^{\prime}_{0x}+\vec{\Phi}^{\prime}_{yz}$, 
$\vec{\Phi}^{\prime\prime}=\vec{\Phi}^{\prime\prime}_{0x}
+\vec{\Phi}^{\prime\prime}_{yz}$. $\alpha_1$ is an angle between 
$\vec{\Phi}^{\prime}_{0x}$ and $\vec{\Phi}^{\prime\prime}_{0x}$ and 
$\alpha_2$ is an angle between $\vec{\Phi}^{\prime}_{yz}$ and $\vec{\Phi}^{\prime\prime}_{yz}$ (Figs.~\ref{fig.minima_3}, \ref{fig.minima_4}). 

\begin{figure}[t]
\centering
\renewcommand{\thesubfigure}{(a)}
\subfigure[ ]{
\label{fig.minima_1}
\begin{minipage}{0.25\textwidth}
\centering
\includegraphics[width=\textwidth]{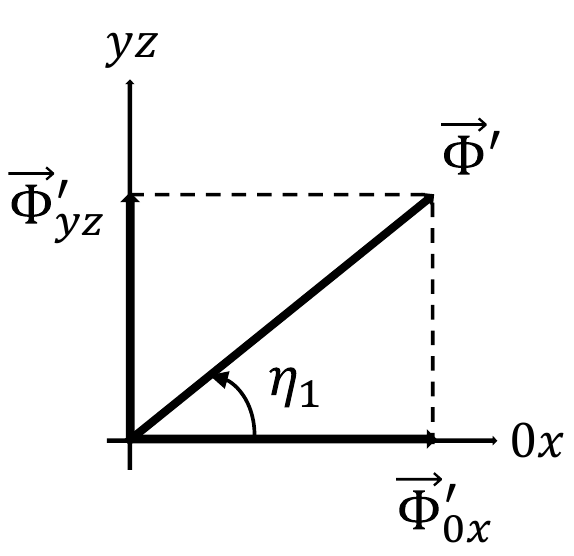}
\end{minipage}
}
\renewcommand{\thesubfigure}{(c)}
\subfigure[ ]{
\label{fig.minima_3}
\begin{minipage}{0.25\textwidth}
\centering
\includegraphics[width=\textwidth]{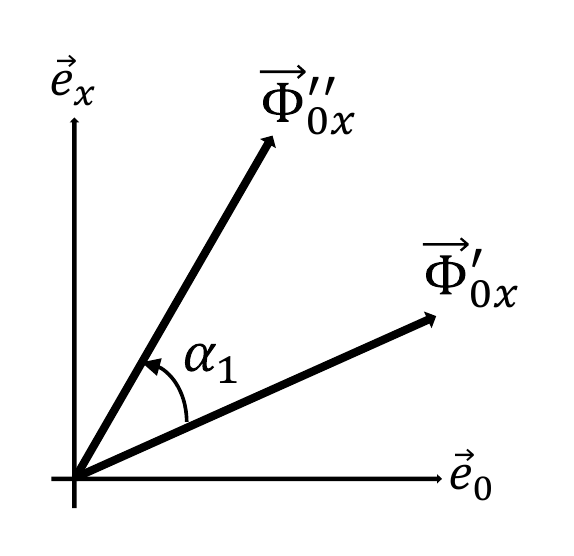}
\end{minipage}
}
\renewcommand{\thesubfigure}{(e)}
\subfigure[ ]{
\label{fig.minima_5}
\begin{minipage}{0.25\textwidth}
\centering
\includegraphics[width=\textwidth]{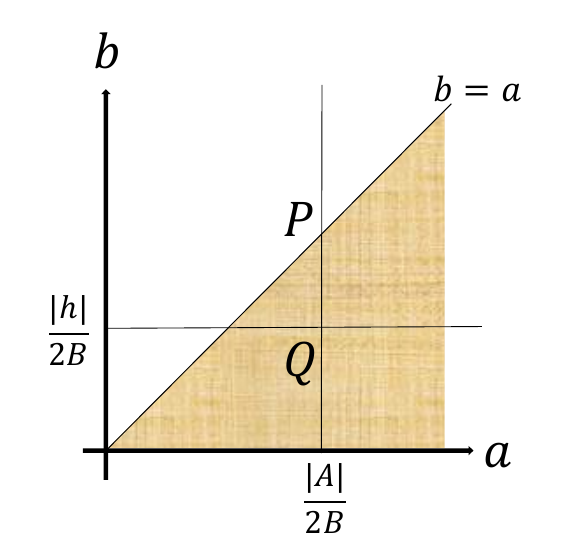}
\end{minipage}
}

\renewcommand{\thesubfigure}{(b)}
\subfigure[ ]{
\label{fig.minima_2}
\begin{minipage}{0.25\textwidth}
\centering
\includegraphics[width=\textwidth]{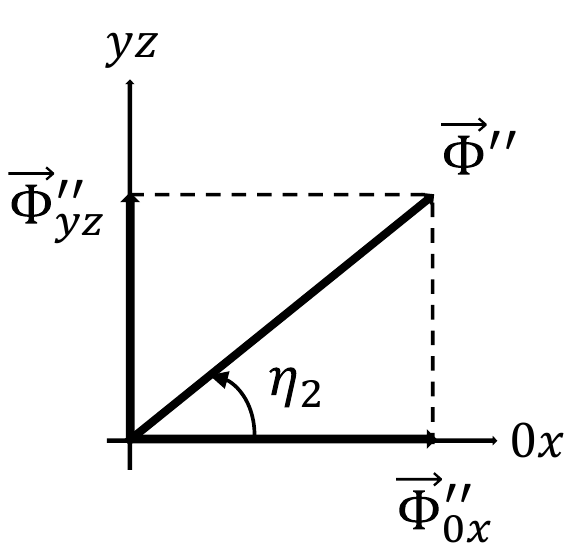}
\end{minipage}
}
\renewcommand{\thesubfigure}{(d)}
\subfigure[ ]{
\label{fig.minima_4}
\begin{minipage}{0.25\textwidth}
\centering
\includegraphics[width=\textwidth]{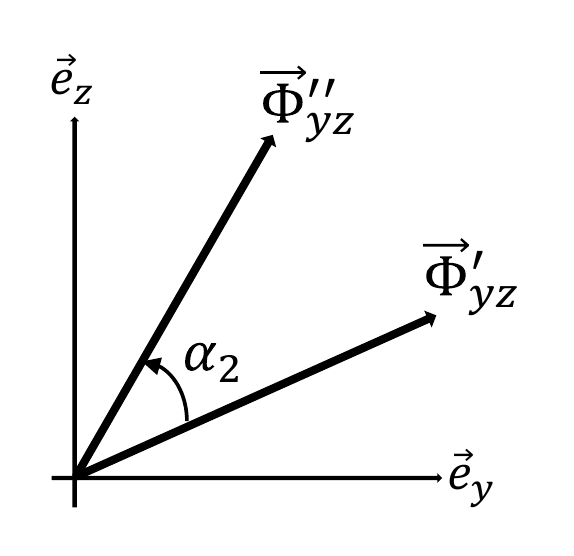}
\end{minipage}
}
\renewcommand{\thesubfigure}{(f)}
\subfigure[ ]{
\label{fig.minima_6}
\begin{minipage}{0.25\textwidth}
\centering
\includegraphics[width=\textwidth]{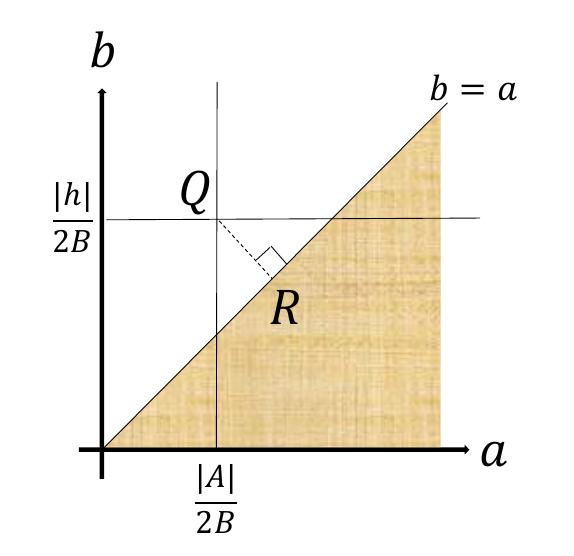}
\end{minipage}
}
\caption{\label{fig.minima} \textbf{(a--d)} Definitions of $\eta_1$, $\eta_2$, $\alpha_1$ and $\alpha_2$. \textbf{(e, f)} Minimization of 
${\cal L}_1(a)+{\cal L}_{2}(b)$ in a domain of $0 \le b\le a$. ${\cal L}_1(a)$ is minimized along a line of $a=|A|/2B$. ${\cal L}_2(b)$ is minimized in a region of $b\ge |h|/2B$. When $|A|>|h|$, ${\cal L}_1(a)+{\cal L}_2(b)$ is minimized along a finite length of line: $a=|A|/2B$ and $|A|/2B \ge b\ge |h|/2B$ (a line of PQ in Fig.~\textbf{(e)}). When $|A|<|h|$, ${\cal L}_1(a)+{\cal L}_2(b)$ is minimized at a point on the domain boundary: $a=b=\frac{1}{2}(\frac{|h|}{2B}+\frac{|A|}{2B})$ (a point of R in Fig.~\textbf{(f)}).} 
\end{figure}

We first minimize the third term of Eq.~(\ref{eqn2-1}) that depends on $\eta_1$, $\eta_2$, $\alpha_1$ and $\alpha_2$ for fixed $\Phi^{\prime}$ and $\Phi^{\prime\prime}$. Namely, 
we maximize the following function for fixed $\Phi^{\prime}$ and $\Phi^{\prime\prime}$, 
\begin{equation}
\label{eqn2-2}
g(\eta_1,\eta_2,\alpha_1,\alpha_2)= 2B \Phi'\Phi''
(\mathrm{cos}\eta_1\mathrm{cos}\eta_2\mathrm{cos}\alpha_1+\mathrm{sin}\eta_1\mathrm{sin}\eta_2\mathrm{cos}\alpha_2)^2-h'\mathrm{cos}\eta_1\mathrm{cos}\eta_2\mathrm{sin}\alpha_1+h\mathrm{sin}\eta_1\mathrm{sin}\eta_2\mathrm{sin}\alpha_2. 
\end{equation}
The function is a sum of 
quadratic functions of $x\equiv\mathrm{cos}(\eta_1-\eta_2)$ and $y\equiv\mathrm{cos}(\eta_1+\eta_2)$, 
\begin{equation}
\label{eqn2-3}
g=C(\frac{\mathrm{cos}\alpha_1+\mathrm{cos}\alpha_2}{2}x+\frac{\mathrm{cos}\alpha_1-\mathrm{cos}\alpha_2}{2}y)^2+\frac{h\mathrm{sin}\alpha_2-h'\mathrm{sin}\alpha_1}{2}x-\frac{h\mathrm{sin}\alpha_2+h'\mathrm{sin}\alpha_1}{2}y,
\end{equation}
with $C\equiv 2B\Phi'\Phi''>0$, 
$\frac{\partial^2 g}{\partial x^2}\geq 0$ and
$\frac{\partial^2 g}{\partial y^2}\geq 0$. Since a domain of $x$ and $y$ is bounded by  
$(x,y)\in [-1,1]\times [-1,1]$, the function takes a maximum 
value at either one of the four 
corners of the domain; $(x,y) =  \{(-1,-1),(-1,1),(1,-1),(1,1)\}$. 
By definition, 
$(\alpha_1,\alpha_2,x,y)$ and $(\alpha_1+\pi,\alpha_2+\pi,-x,-y)$ represent 
the same vectors. 
Thus, we take $(x,y)$ at $(1,1)$ or at $(1,-1)$ 
and maximize $g$ with respect to $\alpha_1$ and $\alpha_2$. 
When $x=y=1$, $\eta_1=\eta_2=0$, and 
$g=-C\mathrm{sin}^2\alpha_1-h'\mathrm{sin}\alpha_1+C$; 
When $x=-y=1$, $\eta_1=\eta_2=\frac{\pi}{2}$, and  $g=-C\mathrm{sin}^2\alpha_2+h\mathrm{sin}\alpha_2+C$. Such $g$ 
is maximized with respect to $\alpha_1$ and/or 
$\alpha_2$ at the following points, 
\begin{align}
    \left\{\begin{array}{cllll} 
    \eta_1=\eta_2=\frac{\pi}{2}, & \alpha_1 \!\ {\rm undefined}, 
    & \sin\alpha_2 = \frac{h}{2C}, & g = C + \frac{h^2}{4C}, & 
    (|h^{\prime}|<|h|<2C), \\
     \eta_1=\eta_2=0, & \sin\alpha_1 = -\frac{h^{\prime}}{2C}, 
     & \alpha_2 \!\ {\rm undefined}, & g = C + \frac{{h^{\prime}}^2}{4C}, & 
    (|h|<|h^{\prime}|<2C), \\ 
    \eta_1=\eta_2=\frac{\pi}{2}, & \alpha_1 \!\ {\rm undefined}, 
    & \sin\alpha_2 = \frac{h}{|h|}, & g = |h|, & 
    (|h^{\prime}|<|h|, 2C<|h|),  \\
    \eta_1=\eta_2=0, & \sin\alpha_1 = -\frac{h^{\prime}}{|h^{\prime}|}, 
    & \alpha_2 \!\ {\rm undefined}, & g = |h^{\prime}|, & 
    (|h|<|h^{\prime}|,2C<|h^{\prime}|). \\ 
    \end{array}\right.
    \label{eqn2-4-7}
\end{align}
Eq.~(\ref{eqn2-4-7}) is symmetric with respect to an exchange between $h$ and $-h^{\prime}$ and between $\alpha_1$ and $\alpha_2$. 
We consider a case of $|h|>|h^{\prime}|$ first, where 
$\vec{\Phi}^{\prime}$ and $\vec{\Phi}^{\prime\prime}$ are 
on a $yz$ plane subtended 
by $\vec{e}_y$ and $\vec{e}_z$ ($\eta_1=\eta_2=\pi/2$) 
and an angle between $\vec{\Phi}^{\prime}$ and $\vec{\Phi}^{\prime\prime}$ 
is $\alpha_2$ (Figs.~\ref{fig.minima_1}, \ref{fig.minima_2}, \ref{fig.minima_4}). $g$ in Eq.~(\ref{eqn2-4-7}) is substituted 
into Eq.~(\ref{eqn2-1}), 
\begin{align}
    \left\{\begin{array}{lc} 
    {\cal L} = -|A| ({\Phi^{\prime}}^2+{\Phi^{\prime\prime}}^2) 
    + B ({\Phi^{\prime}}^2+{\Phi^{\prime\prime}}^2)^2 - \frac{h^2}{4B} & 
    (2\Phi^{\prime} \Phi^{\prime\prime} \ge \frac{|h|}{2B}), \\
    {\cal L} = -|A| ({\Phi^{\prime}}^2+{\Phi^{\prime\prime}}^2) 
    + B ({\Phi^{\prime}}^2+{\Phi^{\prime\prime}}^2)^2 + 4 B  {\Phi^{\prime}}^2{\Phi^{\prime\prime}}^2 - 2|h|{\Phi^{\prime}}{\Phi^{\prime\prime}}  & 
    (2\Phi^{\prime} \Phi^{\prime\prime} \le \frac{|h|}{2B}). \\ 
    \end{array}\right.  \label{eqn2-10-11}
\end{align}

The Lagrangian in Eq.~(\ref{eqn2-10-11}) is 
further minimized in $\Phi^{\prime}$ and 
$\Phi^{\prime\prime}$. First, ${\cal L}$ is decomposed  
into a function of $a \equiv {\Phi^{\prime}}^2+{\Phi^{\prime\prime}}^2$ and a 
function of 
$b \equiv 2\Phi^{\prime} \Phi^{\prime\prime}$,
each of which can be separately minimized;
\begin{align}
    {\cal L} \equiv {\cal L}_1(a) + {\cal L}_2(b), \ \ 
    {\cal L}_1(a) = -|A| a + B a^2, \ \ 
    {\cal L}_{2}(b) = \left\{\begin{array} {lc}
    - \frac{h^2}{4B} & (b\ge \frac{|h|}{2B}), \\
    B b^2 - |h| b & (b \le \frac{|h|}{2B}). \\
\end{array}\right. 
\end{align}
${\cal L}_1(a)$ and ${\cal L}_2(b)$ 
take respective minimum at the following point or 
region, 
\begin{align} 
    a \equiv {\Phi^{\prime}}^2+{\Phi^{\prime\prime}}^2 = 
    \frac{|A|}{2B}, \ \ \ b \equiv  2\Phi^{\prime} \Phi^{\prime\prime} 
    \equiv a \sin 2\theta  \ge \frac{|h|}{2B}. \label{eqn2-12} 
\end{align}
with $(\Phi^{\prime},\Phi^{\prime\prime}) \equiv \sqrt{a} (\cos\theta,\sin\theta)$ 
and $0<\theta<\pi/2$. Noting that a domain of 
$a$ and $b$ is limited by $0<b<a$ together with $|A| \equiv h_c$, $B \equiv |\gamma|$, we 
complete the minimization of the Lagrangian with a 
help of Figs.~\ref{fig.minima_5}, \ref{fig.minima_6}.

{\bf Case 1: $h_c \ge |h|$ and $|h|>|h^{\prime}|$,} 
the global minimum of the action 
is achieved on a finite length of a line defined as:
\begin{align}
  a \sin\alpha_2 \sin 2\theta = \frac{h}{2|\gamma|}, 
  \ \  \ 
 a \equiv {\Phi^{\prime}}^2+{\Phi^{\prime\prime}}^2 =  \frac{h_c}{2|\gamma|}, 
 \ \ \  \frac{|h|}{2|\gamma|} \le b \equiv a \sin 2\theta 
 \le \frac{h_c}{2|\gamma|}. \label{eqn2-13} 
\end{align}

{\bf Case 2: $h_c \le |h|$ and $|h|>|h^{\prime}|$,} 
the global minimum in the domain is achieved at a point 
on the domain boundary, $a=b=\frac{a+b}{2}=\frac{h_c+|h|}{4|\gamma|}$:
\begin{align}
\sin\alpha_2 = \frac{h}{|h|}, \ \ 
 a \equiv {\Phi^{\prime}}^2+{\Phi^{\prime\prime}}^2   
= \frac{h_c+|h|}{4|\gamma|}, \ \ \theta = \frac{\pi}{4}. \label{eqn2-14} 
\end{align}

In the case of $|h|<|h^{\prime}|$, $\vec{\Phi}^{\prime}$ and 
$\vec{\Phi}^{\prime\prime}$ are on the $0x$ plane subtended by 
$\vec{e}_0$ and $\vec{e}_x$ ($\eta_1=\eta_2=0$) and the angle 
between $\vec{\Phi}^{\prime}$ and 
$\vec{\Phi}^{\prime\prime}$ is $\alpha_1$ (Figs.~\ref{fig.minima_1}, \ref{fig.minima_2}, \ref{fig.minima_3}). 
Following the same argument, we obtain the the other two cases. 

{\bf Case 3: $h_c \ge |h^{\prime}|$ and $|h|<|h^{\prime}|$,} 
the global minimum of the action 
is achieved on a finite length of a line given by:
\begin{align}
  a \sin\alpha_1 \sin 2\theta = -\frac{h^{\prime}}{2|\gamma|}, 
  \ \  \ 
 a \equiv {\Phi^{\prime}}^2+{\Phi^{\prime\prime}}^2 =  \frac{h_c}{2|\gamma|}, 
 \ \ \  \frac{|h^{\prime}|}{2|\gamma|} \le b \equiv a \sin 2\theta 
 \le \frac{h_c}{2|\gamma|}. \label{eqn2-15} 
\end{align}

{\bf Case 4: $h_c \le |h^{\prime}|$ and $|h|<|h^{\prime}|$,} 
the global minimum in the domain is achieved at a point 
on the domain boundary, $a=b=\frac{a+b}{2}=\frac{h_c+|h|}{4|\gamma|}$:
\begin{align}
\sin\alpha_1 = -\frac{h^{\prime}}{|h^{\prime}|}, \ \ 
 a \equiv {\Phi^{\prime}}^2+{\Phi^{\prime\prime}}^2   
= \frac{h_c+|h^{\prime}|}{4|\gamma|}, \ \ \theta = \frac{\pi}{4}. \label{eqn2-15d} 
\end{align}


To summarize these four cases, we have the following four phases.\\
{\bf For $|h'|<|h|<h_c$ (regular transverse phase: Case 1 with $\alpha_2=\varphi$)}:
\begin{align}
\label{eqn2-16}
&\vec{\phi}=\rho\mathrm{cos}\theta(\mathrm{cos}\varphi_0\vec{e}_y+\mathrm{sin}\varphi_0\vec{e}_z)+\mathrm{i}\rho\mathrm{sin}\theta[\mathrm{cos}(\varphi+\varphi_0)\vec{e}_y+\mathrm{sin}(\varphi+\varphi_0)\vec{e}_z],\nonumber\\
&\rho=\sqrt{\frac{h_c}{2|\gamma|}},\quad\mathrm{sin}\varphi\mathrm{sin}2\theta=\frac{h}{h_c}.
\end{align}
{\bf For $|h|<|h'|<h_c$ (regular longitudinal phase: case 3 with $\alpha_1 = \varphi$)}:
\begin{align}
\label{eqn2-17}
&\vec{\phi}=\rho[-\mathrm{sin}\theta\mathrm{cos}(\varphi+\varphi_0)\vec{e}_0+\mathrm{cos}\theta\mathrm{sin}\varphi_0\vec{e}_x]+\mathrm{i}\rho[\mathrm{cos}\theta\mathrm{cos}\varphi_0\vec{e}_0+\mathrm{sin}\theta\mathrm{sin}(\varphi+\varphi_0)\vec{e}_x],\nonumber\\
&\rho=\sqrt{\frac{h_c}{2|\gamma|}},\quad\mathrm{sin}\varphi\mathrm{sin}2\theta=-\frac{h'}{h_c}.
\end{align}
{\bf For $|h'|<|h|$, $h_c<|h|$ (saturated transverse phase: case 2 with 
$\alpha_2 = {\rm sgn}(h)\frac{\pi}{2}$)}:
\begin{equation}
\label{eqn2-18}
\vec{\phi}=\rho(\mathrm{cos}\varphi_0\vec{e}_y+\mathrm{sin}\varphi_0\vec{e}_z)-\mathrm{i}\rho\mathrm{sgn}(h)[\mathrm{sin}\varphi_0\vec{e}_y-\mathrm{cos}\varphi_0\vec{e}_z],\quad \rho=\sqrt{\frac{h_c+|h|}{8|\gamma|}}.
\end{equation}
{\bf For $|h|<|h'|$, $h_c<|h'|$ (saturated longitudinal phase: case 4 with 
$\alpha_1 = - {\rm sgn}(h^{\prime})\frac{\pi}{2}$)}:
\begin{equation}
\label{eqn2-19}
\vec{\phi}=\rho[\mathrm{sgn}(-h')\mathrm{sin}\varphi_0\vec{e}_0+\mathrm{sin}\varphi_0\vec{e}_x]+\mathrm{i}\rho[\mathrm{cos}\varphi_0\vec{e}_0+\mathrm{sgn}(-h')\mathrm{cos}\varphi_0\vec{e}_x],\quad\rho=\sqrt{\frac{h_c+|h'|}{8|\gamma|}}.
\end{equation}
From Eqs.~(\ref{eqn2-16}--\ref{eqn2-19}), we obtain a classical ground-state 
phase diagram at $D=0$ (Fig.~\ref{fig.phases_1}). 

The phase boundaries at $|h|=|h^{\prime}|$ 
are of the first order. To be more specific, Eq.~(2) in the main text at $h=\pm h^{\prime}$
is invariant under the following 
SO(2) rotation in the four-component vector space of $\vec{\Phi}\equiv 
\vec{\Phi}^{\prime}+{\rm i}\vec{\Phi}^{\prime\prime} \equiv (-{\rm i}\phi_0,\phi_x,\phi_y,\phi_z)$, 
\begin{align}
\left(\begin{array}{c}
-{\rm i}\phi_0 \\
\phi_x \\
\phi_y \\
\phi_z \\
\end{array}\right) \rightarrow \left(\begin{array}{cccc}
\cos\Theta & 0 & \sin \Theta & 0 \\ 
0 & \cos \Theta & 0 & \mp \sin  \Theta \\
-\sin \Theta & 0 & \cos\Theta & 0 \\
0 & \pm \sin \Theta & 0 & \cos\Theta \\
\end{array}\right) \left(\begin{array}{c}
-{\rm i}\phi_0 \\
\phi_x \\
\phi_y \\
\phi_z \\
\end{array}\right),  
\end{align} 
with real-valued U(1) phase $\Theta$. The SO(2) rotation interpolates between 
the transverse configuration and longitudinal configuration. Thus, the general classical solution 
at $|h|=|h^{\prime}|$ is given by a linear superposition of the two configurations. The phase boundaries at 
$|h|=h_c$ and at $|h^{\prime}|=h_c$ are of the second order.

At $h=h^{\prime}=0$, Eq.~(2) in the main text 
is given only by an amplitude of the complex-valued four-components vector $\vec{\Phi}\equiv 
\vec{\Phi}^{\prime}+{\rm i}\vec{\Phi}^{\prime\prime}$ ($\rho$), an angle between 
$\vec{\Phi}^{\prime}$ and $\vec{\Phi}^{\prime\prime}$ ($\varphi$), and an amplitude ratio between 
$\vec{\Phi}^{\prime}$ and $\vec{\Phi}^{\prime\prime}$ ($\theta$), 
\begin{align}
{\cal L} = -h_c \rho^2 + |\gamma| \Big(\rho^4 + \rho^4 \big(\sin 2\theta \sin \varphi\big)^2\Big),  
\end{align}
with 
\begin{align}
(|\vec{\Phi}^{\prime}|,|\vec{\Phi}^{\prime\prime}|) \equiv \rho(\cos\theta,\sin\theta), \ \ 
\vec{\Phi}^{\prime}\cdot \vec{\Phi}^{\prime\prime} \equiv |\vec{\Phi}^{\prime}||\vec{\Phi}^{\prime\prime}| 
\cos\varphi \equiv \frac{\rho^2}{2} \sin 2\theta \cos\varphi.
\end{align}
Here $\vec{\Phi}^{\prime}$ and $\vec{\Phi}^{\prime\prime}$ are real and imginary part of 
the four-components vector $\vec{\Phi}$ respectively, and they are real-valued four-components 
vectors. Thus, 
for $h_c>0$, the classical solution at $h=h^{\prime}=0$ is given by 
\begin{align}
\rho^2 = \frac{h_c}{2|\gamma|}  \  \cap \ \sin 2\theta \sin \varphi =0,
\end{align}
or equivalently,
\begin{align}
\vec{\Phi} = \rho \vec{n} e^{{\rm i}\theta},
\end{align}
with an arbitrary phase $\theta$ and an arbitrary 4-components real-valued unit vector $\vec{n}$.

As the exchange fields $H_a$ and $H_b$ are supposed to be small, in the main text (Eqs.~(3--5)) we 
focus on the regular transverse phase and the regular longitudinal phase.

\begin{figure}[t]
\centering
\subfigure[ ]{
\label{fig.phases_1}
\begin{minipage}{0.4\textwidth}
\centering
\includegraphics[width=\textwidth]{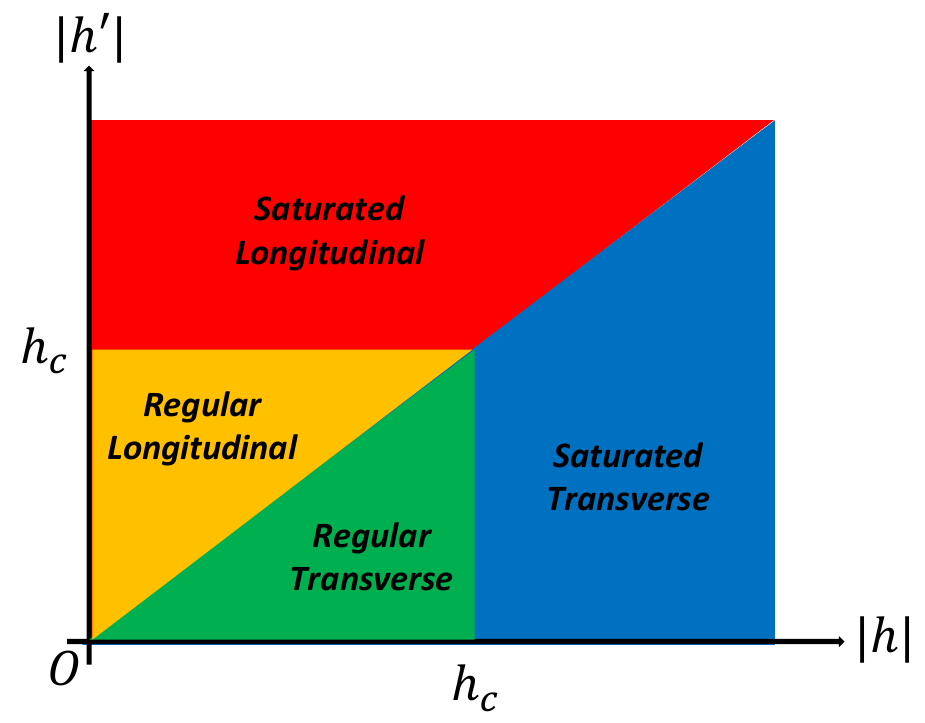}
\end{minipage}
}
\subfigure[ ]{
\label{fig.phases_2}
\begin{minipage}{0.4\textwidth}
\centering
\includegraphics[width=\textwidth]{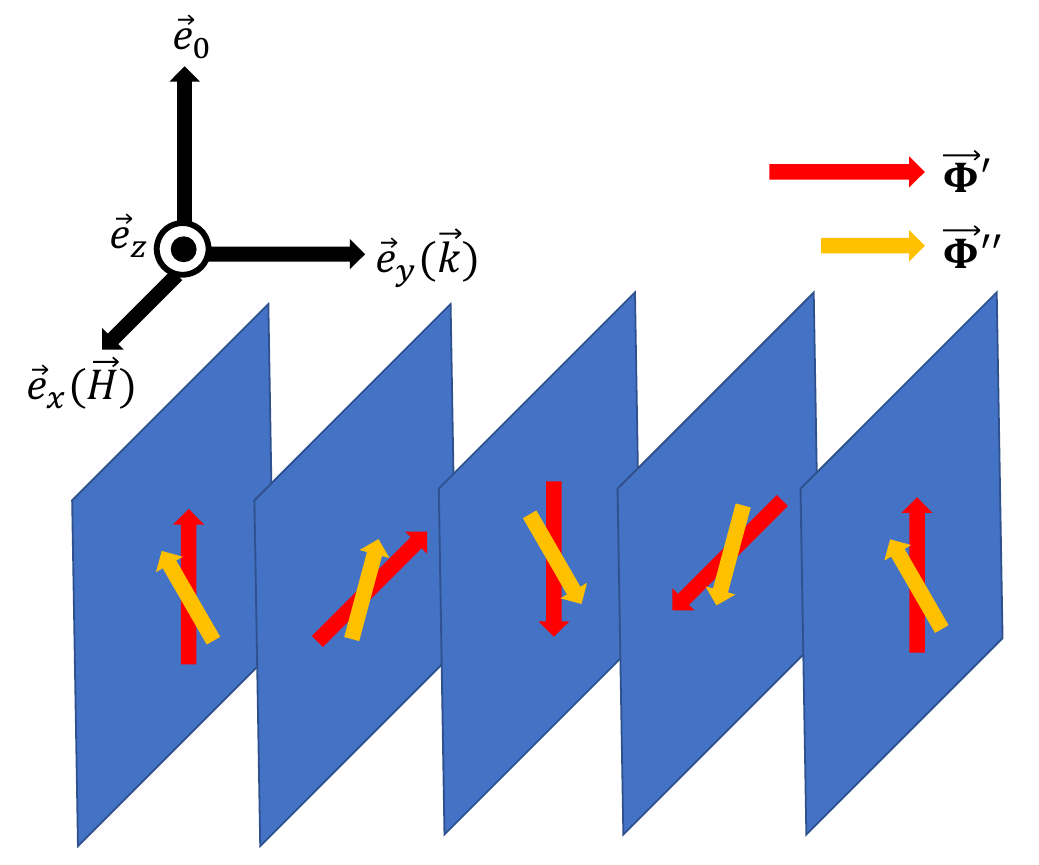}
\end{minipage}
}
\caption{\label{fig.phases} \textbf{(a)} Classical ground-state phase diagram of the EHDL 
excitons under magnetic exchange fields without Rashba coupling. The phase diagram remains unchanged with Rashba coupling, except the transverse/longitudinal phases are substituted by corresponding helicoidal/helical phases. \textbf{(b)} A schematic picture of the helical structure of 
condensed excitons. The real ($\vec{\Phi}'$) and imaginary ($\vec{\Phi}''$) parts are depicted by red and yellow arrows respectively. The propagation direction $\vec{k}(\vec{e}_y)$ is along the the in-plane direction perpendicular to the magnetic field, and $\vec{\Phi}'$ and $\vec{\Phi}''$ rotate in the plane (depicted by blue planes) subtended by a direction of spin singlet ($\vec{e}_0$) and a direction of magnetic field $\vec{H}(\vec{e}_x)$. An angle between $\vec{\Phi}'$ and $\vec{\Phi}''$ is 
acute for the regular helical phase. The angle becomes $0$ for $h'=0$ and $\pi/2$ for 
$|h'|\geq |h|$ (saturated helical phase). The length of $\vec{\Phi}'$ and that of $\vec{\Phi}''$ 
become identical to each other for the saturated helical phase.} 
\end{figure}

\subsection{\label{sec3}Spin rotational symmetry of the excitonic condensate system}
In this section, we clarify what continuous spin-rotational symmetry is broken in the 
transverse and longitudinal phases, i.e. Eq.~(6) in the main text.  
Let us begin with the longitudinal phase:
\begin{align}
\label{eqn3-1}
&\vec{\phi}_{\parallel}(\theta,\varphi,\varphi_0) \cdot 
\vec{\bm{\sigma}} \equiv 
-\rho\mathrm{sin}\theta[\bm{\sigma}_0\mathrm{cos}(\varphi+\varphi_0) 
-\mathrm{i}\bm{\sigma}_x\mathrm{sin}(\varphi+\varphi_0)]+
\mathrm{i}\rho\mathrm{cos}\theta[\bm{\sigma}_0\mathrm{cos}\varphi_0 
-\mathrm{i}\bm{\sigma}_x\mathrm{sin}\varphi_0]\nonumber\\ 
=&-\rho\mathrm{sin}\theta\mathrm{e}^{-\mathrm{i}(\varphi+\varphi_0)\bm{\sigma}_x}
+\mathrm{i}\rho\mathrm{cos}\theta\mathrm{e}^{-\mathrm{i}\varphi_0\bm{\sigma_x}}
=[-\rho\mathrm{sin}\theta\mathrm{e}^{-\mathrm{i}\varphi\bm{\sigma}_x}
+\mathrm{i}\rho\mathrm{cos}\theta]\mathrm{e}^{-\mathrm{i}\varphi_0\bm{\sigma_x}}
= \vec{\phi}_{\parallel}(\theta,\varphi,0) \cdot 
\vec{\bm{\sigma}}
\!\ \mathrm{e}^{-\mathrm{i}\varphi_0\bm{\sigma_x}}. 
\end{align}
A change of $\varphi_0$ by $\delta \varphi_0$ 
in Eq.~(\ref{eqn3-1}) leads to a  
pseudospin rotation of the four-component excitonic pseudospin vector $\vec{\Phi}
\equiv \vec{\Phi}^{\prime}+{\rm i}\vec{\Phi}^{\prime\prime}\equiv (-{\rm i}\phi_0,\phi_x,\phi_y,\phi_z)$. 
The rotation is within the $0x$ plane subtended by $\vec{e}_0$ (singlet component) 
and $\vec{e}_x$ ($x$-component of the triplet pairing field);
\begin{align}
\left(\begin{array}{c}
-{\rm i}\phi_0 \\
\phi_x \\
\phi_y \\
\phi_z \\
\end{array}\right)_{|\varphi_0+\delta\varphi_{0}} = 
\left(\begin{array}{cccc}
 \cos\delta \varphi_0 & -\sin \delta \varphi_0 & &  \\
\sin \delta \varphi_0 & \cos\delta \varphi_0   & & \\
& & 1 & 0 \\
& & 0 & 1 \\
\end{array}\right) \left(\begin{array}{c}
-{\rm i}\phi_0 \\
\phi_x \\
\phi_y \\
\phi_z \\
\end{array}\right)_{|\varphi_0}. 
\end{align} 
The pseudospin rotation within the $0x$ plane can be absorbed 
by spin rotations around the $x$ axis in the electron and hole layers through a mean-field coupling term,
\begin{equation}
\label{eqn3-2}
\vec{\phi}_{\parallel}(\varphi_0+\delta\varphi_0)\cdot\bm{a}^\dagger\vec{\bm{\sigma}}\bm{b}=\vec{\phi}_{\parallel}(\varphi_0)\cdot\bm{a}^\dagger\vec{\bm{\sigma}}\mathrm{e}^{-\mathrm{i}\delta\varphi_0\bm{\sigma_x}}\bm{b}=\vec{\phi}_{\parallel}(\varphi_0)\cdot\bm{a}^\dagger\mathrm{e}^{-\mathrm{i}\delta\varphi_0\bm{\sigma_x}}\vec{\bm{\sigma}}\bm{b}.
\end{equation}
Namely, the mean-field term $\vec{\phi}_{\parallel}\cdot {\bm a}^{\dagger} \vec{\bm \sigma}{\bm b}$ 
is invariant under the followings;  
\begin{equation}
\label{eqn3-3}
\bm{a}\rightarrow\mathrm{e}^{\mathrm{i}\varphi_a {\bm \sigma}_x}\bm{a},\quad  
\bm{b}\rightarrow\mathrm{e}^{\mathrm{i}\varphi_b{\bm \sigma}_x}\bm{b}, \quad 
\vec{\phi}_{\parallel}(\varphi_0)\rightarrow\vec{\phi}_{\parallel}(\varphi_0+\varphi_b-\varphi_a).
\end{equation}

Similarly, the transverse phase is given by the following  classical configuration:
\begin{align}
\label{eqn3-4}
&\vec{\phi}_{\perp}(\theta,\varphi,\varphi_0) 
\cdot \vec{\bm{\sigma}} \equiv 
\rho \cos\theta {\bm \sigma}_y e^{{\rm i}\varphi_0 {\bm \sigma}_x} 
+ {\rm i}\rho \sin\theta {\bm \sigma}_y e^{{\rm i}(\varphi+\varphi_0){\bm \sigma}_x} \nonumber \\
&= \vec{\phi}_{\perp}(\theta,\varphi,0) 
\cdot \vec{\bm{\sigma}}
\!\ \mathrm{e}^{\mathrm{i}\varphi_0\bm{\sigma_x}} =  \mathrm{e}^{-\mathrm{i}\varphi_0\bm{\sigma_x}}\!\  
\vec{\phi}_{\perp}(\theta,\varphi,0) 
\cdot \vec{\bm{\sigma}}.
\end{align}
A variation of $\varphi_0$ by $\delta \varphi_0$ 
in Eq.~(\ref{eqn3-4}) leads to
a pseudospin rotation of $\vec{\Phi} \equiv (-{\rm i}\phi_0,\phi_x,\phi_y,\phi_z)$ 
within the $yz$ plane subtended by $\vec{e}_y$ ($y$-component of the triplet 
pairing field) and $\vec{e}_z$ ($z$-component of the triplet pairing field);
\begin{align}
\left(\begin{array}{c}
-{\rm i}\phi_0 \\
\phi_x \\
\phi_y \\
\phi_z \\
\end{array}\right)_{|\varphi_0+\delta\varphi_{0}} = 
\left(\begin{array}{cccc}
1 & 0 & & \\
0 & 1 & & \\
& & \cos\delta \varphi_0 & -\sin \delta \varphi_0 \\
& & \sin \delta \varphi_0 & \cos\delta \varphi_0 \\
\end{array}\right) \left(\begin{array}{c}
-{\rm i}\phi_0 \\
\phi_x \\
\phi_y \\
\phi_z \\
\end{array}\right)_{|\varphi_0}. 
\end{align} 
The pseudospin rotation within the $yz$ plane
transforms the mean-field coupling term as 
\begin{equation}
\label{eqn3-5}
\vec{\phi}_{\perp}(\varphi_0+\delta\varphi_0)\cdot\bm{a}^\dagger\vec{\bm{\sigma}}\bm{b}=\vec{\phi}_{\perp}(\varphi_0)\cdot\bm{a}^\dagger\vec{\bm{\sigma}}\mathrm{e}^{\mathrm{i}\delta\varphi_0\bm{\sigma_x}}\bm{b}=\vec{\phi}_{\perp}(\varphi_0)\cdot\bm{a}^\dagger\mathrm{e}^{-\mathrm{i}\delta\varphi_0\bm{\sigma_x}}\vec{\bm{\sigma}}\bm{b}.
\end{equation}
The variation can be absorbed 
by the following spin rotations around 
the $x$ axis in the electron and hole layers,
\begin{equation}
\label{eqn3-6}
\bm{a}\rightarrow\mathrm{e}^{\mathrm{i}\varphi_a {\bm \sigma}_x}\bm{a},
\quad\bm{b}\rightarrow\mathrm{e}^{\mathrm{i}\varphi_b {\bm \sigma}_x}\bm{b},
\quad\vec{\phi}_{\perp}(\varphi_0)\rightarrow\vec{\phi}_{\perp}(\varphi_0-\varphi_b-\varphi_a).
\end{equation}
Eqs.~(\ref{eqn3-3}, \ref{eqn3-6}) are equivalent to Eq.~(6) in the main text.

\subsection{\label{sec4}Relation between the Goldstone modes and the $U(1)$ gauge symmetry}

In the main text, the transverse configuration and longitudinal configuration are described by 
three phase variables, $\theta$, $\varphi$ and $\varphi_0$. As shown in Sec.~\ref{sec3}, 
$\varphi_0$ is a rotational angle of the spin rotation of the four-component exciton's 
pseudospin vector $\vec{\Phi}\equiv \vec{\Phi}^{\prime}+{\rm i}\vec{\Phi}^{\prime\prime} 
\equiv (-{\rm i}\phi_0,\phi_x,\phi_y,\phi_z)$. On the other hand, $\theta$ and $\varphi$ define 
an amplitude ratio between $\vec{\Phi}^{\prime}$ and $\vec{\Phi}^{\prime\prime}$ and an angle 
between $\vec{\Phi}^{\prime}$ and $\vec{\Phi}^{\prime\prime}$: 
\begin{align}
\big(|\vec{\Phi}^{\prime}|,|\vec{\Phi}^{\prime\prime}|\big) \equiv \rho (\cos\theta,\sin\theta), \ \ 
\vec{\Phi}^{\prime}\cdot \vec{\Phi}^{\prime\prime} = |\vec{\Phi}^{\prime}| |\vec{\Phi}^{\prime\prime}| 
\cos\varphi. 
\end{align}
Here $\vec{\Phi}^{\prime}$ and $\vec{\Phi}^{\prime\prime}$ are real and imaginary parts 
of $\vec{\Phi} \equiv (-{\rm i}\phi_0,\phi_x,\phi_y,\phi_z)$ respectively, and they are four-component 
real-valued vectors. Roughly speaking, a combination of $\varphi$ (angle) and $\theta$ (amplitude ratio) 
can be regarded as the relative U(1) phase degree of freedom between the two layers. To show this, 
consider the U(1) gauge transformations in the two layers, that induces a U(1) gauge transformation 
of the excitonic order parameters 
$\vec{\Phi} \equiv \vec{\Phi}^{\prime}+{\rm i}\vec{\Phi}^{\prime\prime}$, 
\begin{align}
&{\bm a} \rightarrow e^{{\rm i}\frac{\psi}{2}} {\bm a}, \ \ 
{\bm b} \rightarrow e^{-{\rm i}\frac{\psi}{2}} {\bm b}, \nonumber \\
&\vec{\Phi} \rightarrow e^{{\rm i}\psi} \!\ \vec{\Phi}, \ \ \ 
\vec{\Phi}^{\prime} \rightarrow \cos \psi \!\ \vec{\Phi}^{\prime} - \sin\psi \!\ \vec{\Phi}^{\prime\prime}, \ \ 
\vec{\Phi}^{\prime\prime} \rightarrow  \sin \psi \!\ \vec{\Phi}^{\prime} + \cos\psi \!\  \vec{\Phi}^{\prime\prime}. 
\end{align} 
Under the transformation, the angle and amplitude ratio are transformed as follows, 
\begin{align}
\left(\begin{array}{c} 
|\vec{\Phi}^{\prime}|^2 - |\vec{\Phi}^{\prime\prime}|^2 \\
2\vec{\Phi}^{\prime}\cdot \vec{\Phi}^{\prime\prime} \\
\end{array}\right) \equiv 
\rho^2 \left(\begin{array}{c}
\cos2\theta \\
\sin2\theta \cos\varphi \\
\end{array}\right) \rightarrow \rho^2 \left(\begin{array}{cc}
\cos2\psi & -\sin 2\psi \\
\sin 2\psi & \cos 2\psi \\
\end{array}\right) \left(\begin{array}{c}
\cos2\theta \\
\sin2\theta \cos\varphi \\
\end{array}\right). 
\end{align}
This suggests that $\sin2\theta\sin\varphi$ is 
invariant under the relative U(1) gauge transformation, and a simultaneous change of the angle ($\varphi$) 
and the amplitude ratio ($\theta$)  along a loop of $\sin2\theta\sin\varphi \!\ =$ 
constant can be regarded as the relative U(1) 
phase degree of freedom between the two layers. To be more precise, the relative U(1) gauge 
transformation not only induces the change of the angle ($\varphi$) and the amplitude ratio ($\theta$), 
but also induces a pseudospin rotation of the 
four-component excitonic pseudospin vector $\vec{\Phi}$ ($\varphi_0$). 

To see this, let us 
show that the relative U(1) gauge transformation can be 
absorbed into a combination of changes of $\theta$, $\varphi$ and $\varphi_0$ that 
satisfies the constraint Eq.~(5); 
\begin{align}
\left\{\begin{array}{l}
    e^{i\psi} \vec{\phi}_{\lambda}(\theta,\varphi,\varphi_0) = 
    \vec{\phi}_{\lambda}(\theta(\psi),\varphi(\psi),\varphi_0(\psi)),  \\
    \sin 2\theta(\psi) \sin \varphi(\psi) =  \sin 2\theta \sin \varphi =  {\sf{h}} \equiv \left\{\begin{array}{cc}
    \frac{h}{h_c} & (\lambda=\perp), \\
    -\frac{h^{\prime}}{h_c} & (\lambda=\parallel), \\
    \end{array}\right. \\
    \end{array}\right. \label{eqn4-0}
\end{align}
for both the transverse phase ($\lambda=\perp$) and the longitudinal phase ($\lambda=\parallel$). 
When there is the Rashba coupling, we can generalize the argument into the helical and helicoidal phase by replacing 
$\varphi_0$ by $\varphi_0-Ky$ (see Sec.~\ref{sec7}). In the following, we only 
sketch the argument for the transverse phase, while the argument for the longitudinal 
phase goes as well. Without loss of generality, 
we take $\varphi_0$ to be $-\varphi$ in Eq.~(3) of the main text and apply the 
gauge transformation on Eq.~(3) as, 
\begin{align}
\vec{\phi}_{\perp}(\theta,\varphi,\varphi_0)&=\rho(\mathrm{cos}\theta\mathrm{cos}\varphi+\mathrm{i}\mathrm{sin}\theta)\vec{e}_y-\rho\mathrm{cos}\theta\mathrm{sin}\varphi\vec{e}_z, \label{eqn4-1} \\
e^{i\psi}\vec{\phi}_{\perp}(\theta,\varphi,\varphi_0)&=\rho(\mathrm{cos}\theta\mathrm{cos}\varphi\mathrm{cos}\psi-\mathrm{sin}\theta\mathrm{sin}\psi)\vec{e}_y+\mathrm{i}\rho(\mathrm{sin}\theta\mathrm{cos}\psi+\mathrm{cos}\theta\mathrm{cos}\varphi\mathrm{sin}\psi)\vec{e}_y\nonumber \\ 
&-\rho\mathrm{cos}\theta\mathrm{sin}\varphi\mathrm{cos}\psi\vec{e}_z-\mathrm{i}\rho\mathrm{cos}\theta\mathrm{sin}\varphi\mathrm{sin}\psi\vec{e}_z. \label{eqn4-2}
\end{align}
In terms of $\theta^{\prime}$, $\varphi^{\prime}$, 
$\varphi^{\prime}_0$ ($\varphi^{\prime\prime}_0\equiv \varphi^{\prime} +\varphi^{\prime}_0$), Eq.~(\ref{eqn4-2}) is equated to  
\begin{equation}
\label{eqn4-3}
\vec{\phi}_{\perp} (\theta^{\prime},\varphi^{\prime},\varphi^{\prime}_0)
=\rho\mathrm{cos}\theta'\mathrm{cos}(\varphi'-\varphi''_0)\vec{e}_y+\mathrm{i}\rho\mathrm{sin}\theta'\mathrm{cos}\varphi''_0\vec{e}_y-\rho\mathrm{cos}\theta'\mathrm{sin}(\varphi'-\varphi''_0)\vec{e}_z+\mathrm{i}\rho\mathrm{sin}\theta'\mathrm{sin}\varphi''_0\vec{e}_z. 
\end{equation}
The comparison of (\ref{eqn4-3}) with (\ref{eqn4-2}) gives 
four equations:
\begin{equation}
\label{eqn4-4}
\mathrm{cos}\theta'\mathrm{cos}(\varphi'-\varphi''_0)=\mathrm{cos}\theta\mathrm{cos}\varphi\mathrm{cos}\psi-\mathrm{sin}\theta\mathrm{sin}\psi,
\end{equation}
\begin{equation}
\label{eqn4-5}
\mathrm{sin}\theta'\mathrm{cos}\varphi''_0=\mathrm{sin}\theta\mathrm{cos}\psi+\mathrm{cos}\theta\mathrm{cos}\varphi\mathrm{sin}\psi,
\end{equation}
\begin{equation}
\label{eqn4-6}
\mathrm{cos}\theta'\mathrm{sin}(\varphi'-\varphi''_0)=\mathrm{cos}\theta\mathrm{sin}\varphi\mathrm{cos}\psi,
\end{equation}
\begin{equation}
\label{eqn4-7}
-\mathrm{sin}\theta'\mathrm{sin}\varphi''_0=\mathrm{cos}\theta\mathrm{sin}\varphi\mathrm{sin}\psi.
\end{equation}
Note first that 
$\mathrm{(\ref{eqn4-4})}^2+\mathrm{(\ref{eqn4-5})}^2+\mathrm{(\ref{eqn4-6})}^2+\mathrm{(\ref{eqn4-7})}^2$ are trivially satisfied, 
so that Eqs.~(\ref{eqn4-4}--\ref{eqn4-7}) have only three 
independent equations. Three unknown variables $(\theta',\varphi',\varphi'_0)$  can be solved in favor 
for $(\theta,\varphi,\varphi_0=-\varphi)$ and $\psi$. 

\begin{figure}[t]
\centering
\subfigure[ ]{
\label{fig.angles_1}
\begin{minipage}{0.4\textwidth}
\centering
\includegraphics[width=\textwidth]{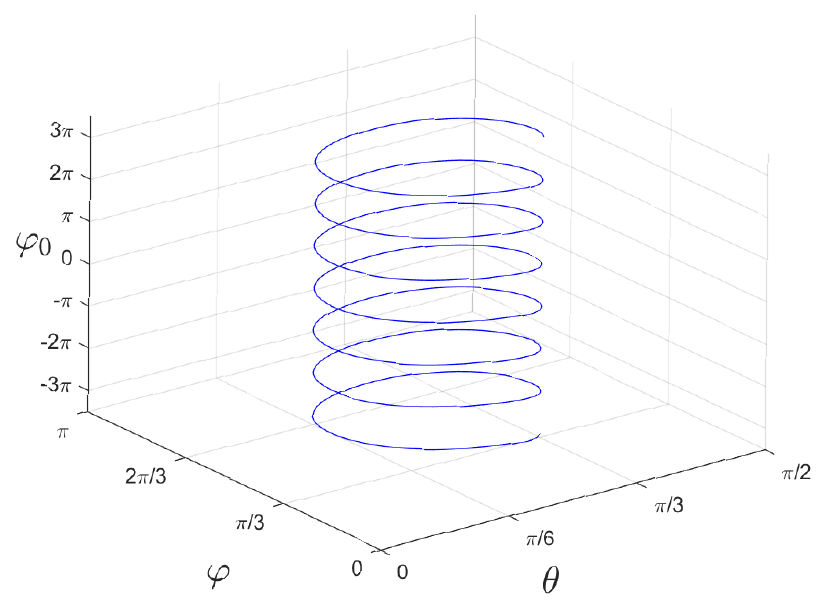}
\end{minipage}
}
\subfigure[ ]{
\label{fig.angles_2}
\begin{minipage}{0.5\textwidth}
\centering
\includegraphics[width=\textwidth]{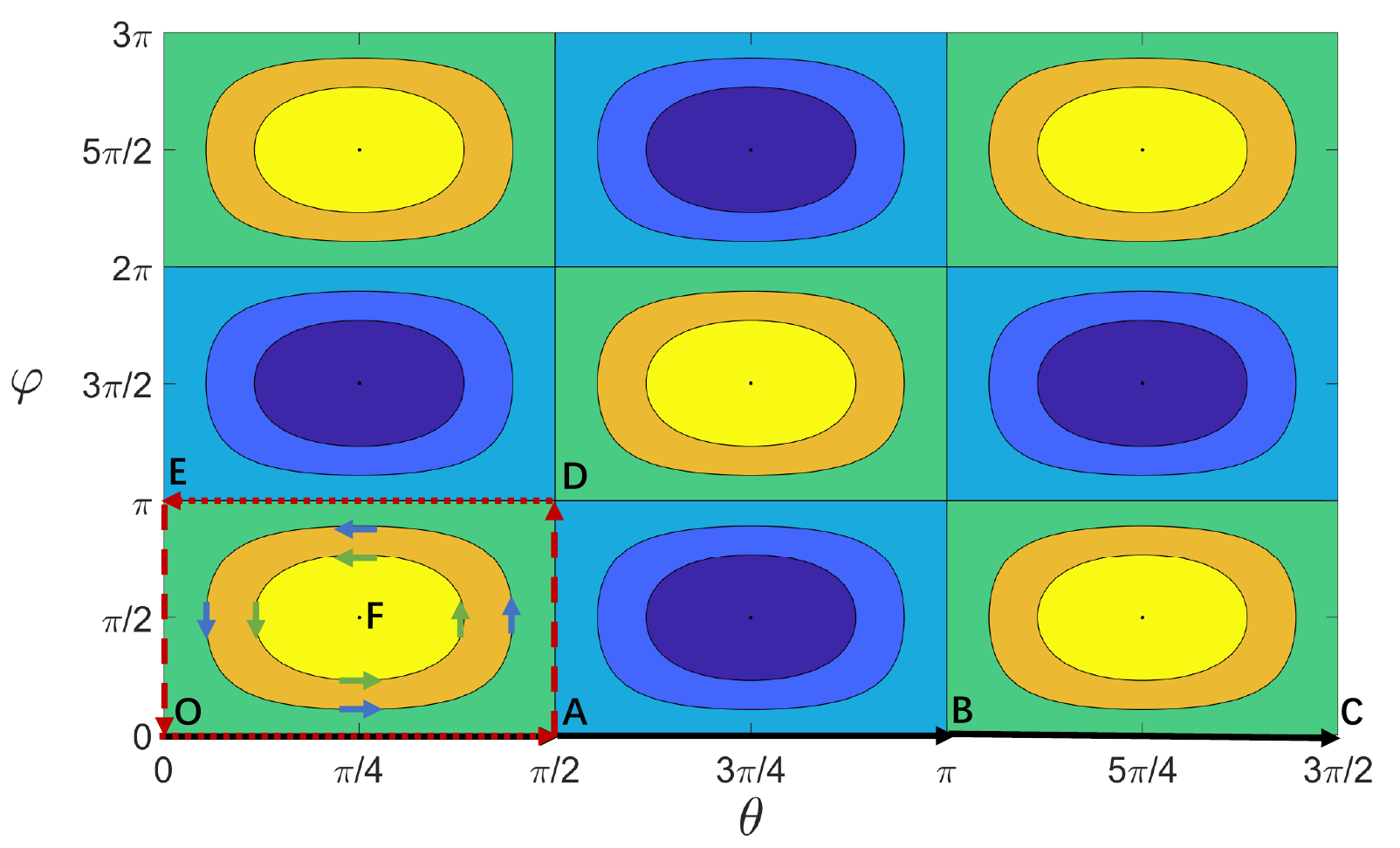}
\end{minipage}
}
\caption{\label{fig.angles} 
\textbf{(a)} A parameter plot of $(\theta(\psi)$, $\varphi(\psi)$, $\varphi_0(\psi))$ where the initial point 
$(\theta(0)$, $\varphi(0)$, $\varphi_0(0))$ satisfy  
${\sf{h}}=\sin 2\theta(0)\sin \varphi(0)=0.75$, $\sin\varphi(0)=0.8$, and 
$\varphi_0(0)=0$. When $\psi$ increases by $\pi$, $\varphi$ decreases by $\pi$, while $(\theta,\varphi)$ goes along a closed curve ($\tilde{h}\equiv\sin 2\theta\sin\varphi=\sf{h}$) at one time. \textbf{(b)} The contour plot of $\tilde{h}={\sf{h}}$ for different values of ${\sf{h}}$. When ${\sf{h}}=1$, the projection becomes a point (F). 
When ${\sf{h}}=0$, the projection tends to a rectangule (O$\rightarrow$ A $\rightarrow$ D $\rightarrow$ E $\rightarrow$ O), but it is also equivalent to go along a straight line (O$\rightarrow$ A $\rightarrow$ B $\rightarrow$ C $\rightarrow$ O), as different values of $(\psi,\theta,\varphi,\varphi_0)$ 
may be equivalent in the special case of $\tilde{h}=0$.}
\end{figure}

From Eqs.~(\ref{eqn4-5}, \ref{eqn4-7}), we get:
\begin{equation}
\label{eqn4-9}
-\mathrm{tan}\varphi''_0=\frac{\mathrm{cos}\theta\mathrm{sin}\varphi\mathrm{sin}\psi}{\mathrm{sin}\theta\mathrm{cos}\psi+\mathrm{cos}\theta\mathrm{cos}\varphi\mathrm{sin}\psi},
\end{equation}
From Eqs.~(\ref{eqn4-4}, \ref{eqn4-6}), we get:
\begin{equation}
\label{eqn4-10}
\mathrm{tan}(\varphi'-\varphi''_0)=\frac{\mathrm{cos}\theta\mathrm{sin}\varphi\mathrm{cos}\psi}{\mathrm{cos}\theta\mathrm{cos}\varphi\mathrm{cos}\psi-\mathrm{sin}\theta\mathrm{sin}\psi},
\end{equation}
Using Eqs.~(\ref{eqn4-9}, \ref{eqn4-10}) together with 
$\sin2\theta \sin\varphi={\sf{h}}$, we have 
\begin{equation}
\label{eqn4-11}
\mathrm{tan}\varphi'=\frac{\mathrm{tan}(\varphi'-\varphi''_0)+\mathrm{tan}\varphi''_0}{1-\mathrm{tan}(\varphi'-\varphi''_0)\mathrm{tan}\varphi''_0}
=\frac{{\sf{h}}}{\mathrm{sin}2\theta\mathrm{cos}\varphi\mathrm{cos}2\psi+\mathrm{cos}2\theta\mathrm{sin}2\psi}.
\end{equation}
From Eqs.~(\ref{eqn4-4}, \ref{eqn4-6}), we get:
\begin{equation}
\label{eqn4-12}
\mathrm{cos}^2\theta'=\frac{1}{2}(1+\mathrm{cos}2\theta\mathrm{cos}2\psi-\mathrm{sin}2\theta\mathrm{sin}2\psi\mathrm{cos}\varphi),
\end{equation}
From Eqs.~(\ref{eqn4-5}, \ref{eqn4-7}), we get:
\begin{equation}
\label{eqn4-13}
\mathrm{sin}^2\theta'=\frac{1}{2}(1-\mathrm{cos}2\theta\mathrm{cos}2\psi+\mathrm{sin}2\theta\mathrm{sin}2\psi\mathrm{cos}\varphi).
\end{equation}
Eqs.~(\ref{eqn4-9}, \ref{eqn4-11}--\ref{eqn4-13}) 
are nothing but the solutions of 
$\theta^{\prime}$, $\varphi^{\prime}$, 
$\varphi^{\prime}_0$ in favor for $\theta$, $\varphi$, 
$\varphi_0=-\varphi$ and $\psi$. To see that such $\theta^{\prime}$ 
and $\varphi^{\prime}$ satisfy the same condition as 
$\theta$ and $\varphi$, i.e. 
$\sin2\theta^{\prime} \sin\varphi^{\prime}={\sf{h}}$, 
we multiply 
Eq.~(\ref{eqn4-12}) by Eq.~(\ref{eqn4-13}), to have 
\begin{equation}
\label{eqn4-14}
\mathrm{sin}^2 2\theta^{\prime}=4\mathrm{sin}^2\theta'\mathrm{cos}^2\theta'=1-(\mathrm{cos}2\theta\mathrm{cos}2\psi-\mathrm{sin}2\theta\mathrm{sin}2\psi\mathrm{cos}\varphi)^2,
\end{equation}
and we square Eq.~(\ref{eqn4-11}), to have 
\begin{equation}
\label{eqn4-15}
\mathrm{sin}^2\varphi'=\frac{\mathrm{tan}^2\varphi'}{1+\mathrm{tan}^2\varphi'}=\frac{{\sf{h}}^2}{(\mathrm{sin}2\theta\mathrm{cos}\varphi\mathrm{cos}2\psi+\mathrm{cos}2\theta\mathrm{sin}2\psi)^2+{\sf{h}}^2}.
\end{equation}
Combining these two, we get:
\begin{equation}
\label{eqn4-16}
\mathrm{sin}^2 2\theta'\mathrm{sin}^2\varphi'
={\sf{h}}^2\frac{1-(\mathrm{cos}2\theta\mathrm{cos}2\psi-\mathrm{sin}2\theta\mathrm{sin}2\psi\mathrm{cos}\varphi)^2}{(\mathrm{sin}2\theta\mathrm{cos}\varphi\mathrm{cos}2\psi+\mathrm{cos}2\theta\mathrm{sin}2\psi)^2
+{\sf{h}}^2}={\sf{h}}^2.
\end{equation}
Because $\theta^{\prime}$ and $\varphi^{\prime}$ 
can be regarded as smooth functions of $\psi$ that reduce 
to $\theta$ and $\varphi$ at $\psi=0$ respectively, we can 
conclude that $\mathrm{sin}2\theta'\mathrm{sin}\varphi' = {\sf{h}}$. 
This completes the proof of Eq.~(\ref{eqn4-0}) for the 
transverse phase. In other words, the gapless Goldstone mode 
associated with the symmetry breaking of the relative gauge symmetry 
is given by a combination of the $\varphi_0$ 
mode and a variation of $\theta$ and $\varphi$ within the 
constraint of Eq.~(5) in the main text. 


A parameter-plot of  
$(\theta(\psi),\varphi(\psi),\varphi_0(\psi))$ is given in Fig.~\ref{fig.angles_1} for a given 
$(\theta(0),\varphi(0),\varphi_0(0))$.
The plot takes a form of a helical curve in the $(\theta,\varphi,\varphi_0)$ space. A projection of the curve onto the $(\theta,\varphi)$ plane 
is a circle that is defined by  $\tilde{h}\equiv\sin 2\theta \sin\varphi={\sf{h}}$. 
When $\psi$ changes by $\pi$, $(\theta,\varphi)$ goes around the circle once and $\varphi_0$ changes by $-\pi$. When ${\sf{h}}= 1 \!\ (h= h_c)$, 
the circle reduces to a point of $\theta=\pi/4+n\pi/2$ and 
$\varphi=\pi/2+n\pi$, and 
the helical curve reduces to 
a straight line of $\psi=-\varphi_0$. When ${\sf{h}}=0$, the 
parameter plot of $(\theta(\psi),\varphi(\psi),\varphi_0(\psi))$ 
still preserves the periodicity in 
a trickly way. To see this, we take an initial point at $\psi=0$ 
as $(\theta,\varphi,\varphi_0)=(0,0,0)$. When $\psi$ changes from $0$ to $\pi/{2}$, $(\theta(\psi),\varphi(\psi),\varphi_0(\psi))=(\psi,0,0)$. 
At $\psi=\pi/2$, $\varphi(\psi)$ jumps from $0$ to $\pi$. When $\psi$ changes from $\pi/2$ to $\pi$, $(\theta(\psi),\varphi(\psi),\varphi_0(\psi))=(\pi-\psi,\pi,0)$. At $\psi=\pi$, $\varphi(\psi)$ jumps from $\pi$ to $0$, and 
$\varphi_0(\psi)$ jumps from $0$ to $-\pi$. Thus, the periodicity is still true: when $\psi$ changes by $\pi$, $\varphi_0$ changes by $-\pi$, and $(\theta,\varphi)$ comes back to the same point. 

\subsection{\label{sec5}Derivation of the spin-charge coupled Josephson equations without Rashba coupling}
In this section, we derive the spin-charge coupled Josephson equation for the transverse  
and longitudinal phases. In the main text, we introduced a quantum-dot junction model 
(Eqs.~(10--12)). Applying local (time-dependent) gauge transformations in the electron  
and hole layers, we obtain
\begin{align}
    \mathcal{S}_{\rm mf} = \int d\tau \sum_{j=1,2} \sum_{\alpha} 
    \Big\{ {\bm a}^{\dagger}_{j\alpha} \big[\partial_{\tau} + {\bm H}_{a\alpha} 
    - \mu\big] {\bm a}_{j\alpha} + {\bm b}^{\dagger}_{j\alpha} \big[\partial_{\tau} + {\bm H}_{b\alpha} 
    - \mu -  \frac{\eta_{j}}{2} \big(V_C+V_S {\bm \sigma}_x\big)\big] {\bm b}_{j\alpha} 
    - \vec{\phi}_{\omega}(\psi_j,\varphi_{0j}) \cdot {\bm a}^{\dagger}_{j\alpha}
    \vec{\bm \sigma} {\bm b}_{j\alpha} + {\rm h.c.}\Big\} \label{eqn5-0}
\end{align}
with $\omega=\perp,\parallel$ and $\eta_1=-\eta_2=1$. Here 
$V_C$ is charge voltage 
difference between the electron 
and hole layers respectively, 
while $V_S$ is the sum of (difference between) the spin voltage in the electron layer and 
the spin voltage in the hole layer for the transverse (longitudinal) phase; 
\begin{align}
V_C = V_{Cb} - V_{Ca}, \quad V_S = V_{Sb}\pm V_{Sa}. \label{eqn5-0a}
\end{align}
Namely, ``$+$" is for transverse $(\omega=\perp)$ and ``$-$" 
for longitudinal 
$(\omega=\parallel)$. Note that we treat $V_C$ and $V_S$ as external fields.
The excitonic mean fields in the two regions are identical to each other except for the two gapless U(1) phase variables;
\begin{align}
    \vec{\phi}_{\omega}(\psi_j,\varphi_{0j})\cdot \vec{\bm \sigma} = 
    \vec{\phi}_{\omega}(\psi=0,\varphi_0=0) \cdot \vec{\bm \sigma} 
    \mathrm{e}^{{\rm i}\psi_j \pm {\rm i}\varphi_{0j}{\bm \sigma}_x}  
\end{align}
with $\pm$ for $\omega=yz,0x$ respectively. Note that 
$V_C$, $V_S$, and $\vec{\phi}_{\omega}(\psi=0,\varphi_0=0)\equiv 
\vec{\phi}_{\omega}$ are treated as given (e.g. external) static 
variables, and the gapless U(1) phase variables, $\psi_{j}$ and 
$\varphi_{0j}$ $(j=1,2)$, are treated as dynamical variables.  
In terms of a global gauge transformation in the 
hole layer, $\mathrm{e}^{{\rm i}\psi_j\pm {\rm i}\varphi_{0j}{\bm \sigma}_x}{\bm b}_{j\alpha}
\rightarrow {\bm b}_{j\alpha}$, the dependence on the gapless phase variables can be 
removed from the mean field coupling. After the transformation, the phase variables appear in the tunneling 
part $S_T$; accordingly, we have 
\begin{equation}
\label{eqn5-1}
\mathcal{S}[\bm{\Psi},\bm{\Psi}^\dagger,\psi_j,\varphi_{0j};V_C,V_S] \equiv 
\mathcal{S}_{\rm mf}[\bm{\Psi},\bm{\Psi}^\dagger,\psi_j,\varphi_{0j};V_C,V_S] 
+ \mathcal{S}_{T}[\bm{\Psi},\bm{\Psi}^\dagger,\psi_j,\varphi_{0j}] = 
\int\mathrm{d}\tau\sum_{\alpha\beta}\bm{\Psi}_{\alpha}^\dagger(\bm{\mathcal{G}}^{-1})_{\alpha\beta}\bm{\Psi}_{\beta},
\end{equation}
\begin{equation}
\label{eqn5-3}
(\bm{\mathcal{G}}^{-1})_{\alpha\beta}=\left(\begin{array}{cccc}
\bm{G}^{-1}_{a\alpha}\delta_{\alpha\beta} & -\vec{\phi}_\omega\cdot\vec{\bm{\sigma}}\delta_{\alpha\beta} & T^{(a)}_{\alpha\beta} & 0 \\
-\vec{\phi}_\omega^*\cdot\vec{\bm{\sigma}}\delta_{\alpha\beta} & ({\bm G}^{-1}_{b\alpha}+{\bm \Delta}_1)\delta_{\alpha\beta} & 0 & T^{(b)}_{\alpha\beta}\mathrm{e}^{\mathrm{i}(\tilde{\psi}\pm\tilde{\varphi}_0\bm{\sigma}_x)} \\
T_{\beta\alpha}^{(a)*} & 0 & {\bm G}^{-1}_{a\alpha}\delta_{\alpha\beta} & -\vec{\phi}_\omega\cdot\vec{\bm{\sigma}}\delta_{\alpha\beta} \\
0 & T_{\beta\alpha}^{(b)*}\mathrm{e}^{-\mathrm{i}(\tilde{\psi}\pm\tilde{\varphi}_0\bm{\sigma}_x)} & -\vec{\phi}_\omega^*\cdot\vec{\bm{\sigma}}\delta_{\alpha\beta} & (
{\bm G}^{-1}_{b\alpha}+{\bm \Delta}_2)\delta_{\alpha\beta}
\end{array}\right),
\end{equation}
where $\vec{\phi}_{\omega} \equiv \vec{\phi}_{\omega}(\psi=0,\varphi_0=0)$, $\tilde{\psi} \equiv \psi_1-\psi_2$, $\tilde{\varphi}_0 \equiv \varphi_{01}-\varphi_{02}$, 
and 
\begin{equation}
\label{eqn5-4}
{\bm G}^{-1}_{a\alpha}\equiv\partial_\tau+\bm{H}_{a\alpha}-\mu,\quad {\bm G}^{-1}_{b\alpha}\equiv\partial_\tau+\bm{H}_{b\alpha}-\mu,
\end{equation}
\begin{equation}
\label{eqn5-5}
{\bm \Delta}_j\equiv-(\mathrm{i}\dot{\psi}_{j}+\eta_j\frac{V_C}{2})-(\pm\mathrm{i}\dot{\varphi}_{0j}+\eta_j\frac{V_S}{2})\bm{\sigma}_x,
\end{equation}
with $\dot\psi_{j} \equiv \partial_{\tau}\psi_j$ and 
$\dot{\varphi}_{0j} \equiv \partial_{\tau}\varphi_{0j}$ 
$(j=1,2)$. 
The multiple signs in the tunneling matrix element in the hole layer 
are chosen as ``$+$" for the transverse phase and ``$-$" 
for the longitudinal phase. 
 $\bm{\Psi}_{\alpha}\equiv(\bm{a}_{1\alpha},\bm{b}_{1\alpha},\bm{a}_{2\alpha},\bm{b}_{2\alpha})^T$ is an eight-components vectors with the domain ($j=1,2$), the layer ($a,b$), and the spin ($\uparrow,\downarrow$) indices. 
The phase variables are decomposed into their average parts  
($\bar{\psi}\equiv \frac{\psi_1+\psi_2}{2}$ and $\bar{\varphi}_0 \equiv\frac{\varphi_{01}+\varphi_{02}}{2}$) and their 
difference parts ($\tilde{\psi}\equiv \psi_1-\psi_2$ and $\tilde{\varphi}_0 \equiv\varphi_{01}-\varphi_{02}$), i.e.  
\begin{equation}
\label{eqn5-6}
\psi_j=\bar{\psi}+\eta_j\frac{\tilde{\psi}}{2},\quad\varphi_{0j}=\bar{\varphi}_0+\eta_j\frac{\tilde{\varphi}_0}{2}.
\end{equation}

The difference parts, $\tilde{\psi}$ and $\tilde{\varphi}_0$, 
together with $V_C$ and $V_S$, are coupled with charge and spin density differences $N_C$ and $N_S$ respectively; 
\begin{align}
\label{eqn5-6-1}
    N_C \equiv \frac{1}{2}\sum_{\alpha} \Big[{\bm b}^{\dagger}_{1\alpha} 
    {\bm b}_{1\alpha}-{\bm b}^{\dagger}_{2\alpha}{\bm b}_{2\alpha}\Big], \quad 
    N_S \equiv \frac{1}{2}\sum_{\alpha} \Big[{\bm b}^{\dagger}_{1\alpha}{\bm \sigma}_x {\bm b}_{1\alpha}-{\bm b}^{\dagger}_{2\alpha}{\bm \sigma}_x{\bm b}_{2\alpha}\Big].
\end{align}
To see this coupling, we follow a standard procedure 
and introduce $N_C$ and $N_S$ and their canonical conjugate variables 
$\mu_C$ and $\mu_S$, 
\begin{align}
\label{eqn5-7}
&\mathcal{Z}[V_{C},V_{S}]\equiv\int\mathcal{D}\psi_j\mathcal{D}\varphi_{0j}\mathcal{D}\bm{\Psi}^\dagger\mathcal{D}\bm{\Psi}\mathrm{e}^{-\mathcal{S}[\bm{\Psi},\bm{\Psi}^\dagger,\psi_j,\varphi_{0j};V_C,V_S]}\nonumber\\
=&\int\mathcal{D}N_{C}\mathcal{D}N_{S}\mathcal{D}\psi_j\mathcal{D}\varphi_{0j}\mathcal{D}\bm{\Psi}^\dagger\mathcal{D}\bm{\Psi}\delta(N_{C}-\sum_{j\alpha}\eta_j\bm{b}_{j\alpha}^\dagger\bm{b}_{j\alpha}/2)\delta(N_{S}-\sum_{j\alpha}\eta_j\bm{b}_{j\alpha}^\dagger\bm{\sigma}_x\bm{b}_{j\alpha}/2)\mathrm{e}^{-\mathcal{S}[\bm{\Psi},\bm{\Psi}^\dagger,\psi_j,\varphi_{0j};V_C,V_S]}\nonumber\\
=&\int\mathcal{D}\mu_{C}\mathcal{D}\mu_{S}\mathcal{D}N_{C}\mathcal{D}N_{S}\mathcal{D}\psi_j\mathcal{D}\varphi_{0j}\mathcal{D}\bm{\Psi}^\dagger\mathcal{D}\bm{\Psi}\mathrm{e}^{\mathrm{i}\int\mathrm{d}\tau[\mu_{C}(N_{C}-\sum_{j\alpha}\eta_j\bm{b}_{j\alpha}^\dagger\bm{b}_{j\alpha}/2)+\mu_{S}(N_{S}-\sum_{j\alpha}\eta_j\bm{b}_{j\alpha}^\dagger\bm{\sigma}_x\bm{b}_{j\alpha}/2)]-\mathcal{S}[\bm{\Psi},\bm{\Psi}^\dagger,\psi_j,\varphi_{0j};V_C,V_S]}\nonumber\\
=&\int\mathcal{D}\mu_{C}\mathcal{D}\mu_{S}\mathcal{D}N_{C}\mathcal{D}N_{S}\mathcal{D}\psi_j\mathcal{D}\varphi_{0j}\mathcal{D}\bm{\Psi}^\dagger\mathcal{D}\bm{\Psi}\mathrm{e}^{\int\mathrm{d}\tau({\rm i}\mu_{C}N_{C}+{\rm i}\mu_{S}N_{S})-\mathcal{S}[\bm{\Psi},\bm{\Psi}^\dagger,\psi_j,\varphi_{0j};V_C-\mathrm{i}\mu_{C},V_S-\mathrm{i}\mu_{S}]},
\end{align}
where $\delta(N_{C}-\sum_{j\alpha}\eta_j\bm{b}_{j\alpha}^\dagger\bm{b}_{j\alpha}/2)$ and $\delta(N_{S}-\sum_{j\alpha}\eta_j\bm{b}_{j\alpha}^\dagger\bm{\sigma}_x\bm{b}_{j\alpha}/2)$ are delta functions defined between real numbers and bilinear Grassmann numbers, whose definition and properties are discussed in Sec.~\ref{sec10}. An integration over $\bm{\Psi}_\alpha$ and $\bm{\Psi}^\dagger_\alpha$ leads to 
an effective action of the dynamical variables, $\psi_j$, $\varphi_{0j}$, $N_{C}$, $N_{S}$, $\mu_{C}$, and $\mu_{S}$,
\begin{equation}
\label{eqn5-8}
\mathcal{Z}[V_{C},V_{S}]=\int\mathcal{D}\mu_{C}\mathcal{D}\mu_{S}\mathcal{D}N_{C}\mathcal{D}N_{S}\mathcal{D}\psi_j\mathcal{D}\varphi_{0j}\mathrm{e}^{\mathrm{i}\int\mathrm{d}\tau(\mu_{C}N_{C}+\mu_{S}N_{S})+\mathrm{Tr}\mathrm{ln}\bm{\mathcal{G}}_{\mu}^{-1}},
\end{equation}
where the subscript ``$\mu$"  in ${\cal G}^{-1}_{\mu}$ represents that $V_C$ and $V_S$ in ${\cal G}^{-1}$ in Eq.~(\ref{eqn5-3}) are replaced by $V_C-\mathrm{i}\mu_C$  and $V_S-\mathrm{i}\mu_S$ in ${\cal G}^{-1}_{\mu}$. 
$\mathrm{Tr}$ includes the integral over the time and the summation over single-particle energy-levels as well as a trace of $8 \times 8$ matrix associated with domain, layer and spin indices. The $8 \times 8$ matrix-formed $\bm{\mathcal{G}}_{\mu}^{-1}$ can be decomposed into four parts: 
\begin{equation}
\label{eqn5-9}
\bm{\mathcal{G}}_{\mu}^{-1}=\tilde{\bm{\mathcal{G}}}_{\mu  0}^{-1}+\bm{\mathcal{T}},\quad\tilde{\bm{\mathcal{G}}}_{\mu 0}^{-1}=\bm{\mathcal{G}}_0^{-1}+\bm{\Phi}+\bm{\Delta}_{\mu},
\end{equation}
\begin{equation}
\label{eqn5-10}
(\bm{\mathcal{G}}^{-1}_0)_{\alpha\beta}=\delta_{\alpha\beta}\left(\begin{array}{cccc}
\bm{G}^{-1}_{a\alpha} & 0 & 0 & 0 \\
0 & \bm{G}^{-1}_{b\alpha} & 0 & 0 \\
0 & 0 & \bm{G}^{-1}_{a\alpha} & 0 \\
0 & 0 & 0 & \bm{G}^{-1}_{b\alpha}
\end{array}\right),
\end{equation}
\begin{equation}
\label{eqn5-11}
\bm{\Phi}_{\alpha\beta}=\delta_{\alpha\beta}\left(\begin{array}{cccc}
0 & -\vec{\phi}_\omega\cdot\vec{\bm{\sigma}} & 0 & 0 \\
-\vec{\phi}_\omega^*\cdot\vec{\bm{\sigma}} & 0 & 0 & 0 \\
0 & 0 & 0 & -\vec{\phi}_\omega\cdot\vec{\bm{\sigma}} \\
0 & 0 & -\vec{\phi}_\omega^*\cdot\vec{\bm{\sigma}} & 0
\end{array}\right),
\end{equation}
\begin{equation}
\label{eqn5-12}
(\bm{\Delta}_\mu)_{\alpha\beta}=\delta_{\alpha\beta}\left(\begin{array}{cccc}
0 & 0 & 0 & 0 \\
0 & \bm{\Delta}_{\mu 1} & 0 & 0 \\
0 & 0 & 0 & 0 \\
0 & 0 & 0 & \bm{\Delta}_{\mu 2}
\end{array}\right),
\end{equation}
\begin{equation}
\label{eqn5-13}
\bm{\mathcal{T}}_{\alpha\beta}=\left(\begin{array}{cccc}
0 & 0 & T^{(a)}_{\alpha\beta} & 0 \\
0 & 0 & 0 & T^{(b)}_{\alpha\beta}\mathrm{e}^{\mathrm{i}(\tilde{\psi}\pm\tilde{\varphi}_0\bm{\sigma}_x)} \\
T_{\beta\alpha}^{(a)*} & 0 & 0 & 0 \\
0 & T_{\beta\alpha}^{(b)*}\mathrm{e}^{-\mathrm{i}(\tilde{\psi}\pm\tilde{\varphi}_0\bm{\sigma}_x)} & 0 & 0
\end{array}\right), 
\end{equation}
where 
\begin{align}
    \Delta_{\mu j} \equiv -\Big({\rm i}\dot{\psi}_{j} + \eta_j \frac{V_C-{\rm i}\mu_C}{2}\Big) 
    - \Big(\pm {\rm i} \dot{\varphi}_{0j} +
    \eta_j \frac{V_S-{\rm i}\mu_S}{2} \Big) {\bm \sigma}_x, 
\end{align}
for $j=1,2$. 
As the tunneling matrix elements $\bm{\mathcal{T}}$ 
are small quantities, the effective action can be expanded in ${\bm {\mathcal{T}}}$:
\begin{equation}
\label{eqn5-14}
-\mathrm{Tr}\mathrm{ln}\bm{\mathcal{G}}_{\mu}^{-1}=-\mathrm{Tr}\mathrm{ln}\tilde{\bm{\mathcal{G}}}_{\mu 0}^{-1}-\mathrm{Tr}\mathrm{ln}(1+\tilde{\bm{\mathcal{G}}}_{\mu 0}\bm{\mathcal{T}})=-\mathrm{Tr}\mathrm{ln}\tilde{\bm{\mathcal{G}}}_{\mu 0}^{-1}+\frac{1}{2}\mathrm{Tr}(\tilde{\bm{\mathcal{G}}}_{\mu 0}\bm{\mathcal{T}})^2+\frac{1}{4}\mathrm{Tr}(\tilde{\bm{\mathcal{G}}}_{\mu 0}\bm{\mathcal{T}})^4+...
\end{equation}
In the expansion, only the even-order terms remains, as 
$\bm{\mathcal{T}}$ is off-diagonal in the domain index ($i=1,2$). 

The couplings between $N_C$, $N_S$, $\mathrm{i}\partial_{\tau}\tilde{\psi}+V_C$ and  
$\pm \mathrm{i}\partial_{\tau}\tilde{\varphi}_{0}+V_S$ are encoded in the zeroth-order term 
in $\bm{\mathcal{T}}$ in Eq.~(\ref{eqn5-14}). To see this coupling, we further expand the zero-order term in 
${\bm \Delta}_{\mu}$,  
\begin{align}
\label{eqn5-15}
&-\mathrm{Tr}\mathrm{ln}\tilde{\bm{\mathcal{G}}}_{\mu 0}^{-1}=-\mathrm{Tr}\mathrm{ln}(\bm{\mathcal{G}}^{-1}_0+\bm{\Phi})-\mathrm{Tr}\mathrm{ln}[1+\bm{(\mathcal{G}}^{-1}_0+\bm{\Phi})^{-1}\bm{\Delta}_\mu)]\nonumber\\
=&-\mathrm{Tr}\mathrm{ln}(\bm{\mathcal{G}}^{-1}_0+\bm{\Phi})-\mathrm{Tr}[(\bm{\mathcal{G}}^{-1}_0+\bm{\Phi})^{-1}\bm{\Delta}_\mu]+\frac{1}{2}\mathrm{Tr}[(\bm{\mathcal{G}}^{-1}_0+\bm{\Phi})^{-1}\bm{\Delta}_\mu(\bm{\mathcal{G}}^{-1}_0+\bm{\Phi})^{-1}\bm{\Delta}_\mu]+...
\end{align}
The first term is constant in variables. 
The second term is proportional to $\int d\tau ({\bm \Delta}_{\mu 1}(\tau) + {\bm \Delta}_{\mu 2}(\tau))=-2{\rm i}\int d\tau 
(\partial_{\tau} \bar{\psi} + \partial_{\tau}\bar{\varphi}_{0})$, that vanishes 
after the integral over the time. The third term forms 
a saddle point in a space of $\partial_{\tau} \bar{\psi}$, 
$\partial_{\tau} \bar{\varphi}_0$, $\partial_{\tau} \tilde{\psi} -\mathrm{i} V_C-\mu_C$, 
and $\pm \partial_{\tau} \tilde{\varphi}_0-\mathrm{i}V_S-\mu_S$,  
\begin{align}
& \mathrm{Tr}[(\bm{\mathcal{G}}^{-1}_0+\bm{\Phi})^{-1}\bm{\Delta}_\mu(\bm{\mathcal{G}}^{-1}_0+\bm{\Phi})^{-1}\bm{\Delta}_\mu] = \nonumber \\ 
& \int \int d\tau_1 d\tau_2 \!\ \chi_{00}(\tau_1-\tau_2) \!\  
    \Big( \big(\dot{\bar{\psi}}\big)_{\tau_1} \big(\dot{\bar{\psi}}\big)_{\tau_2} 
    + \frac{1}{4} \big(\dot{\tilde{\psi}}-\mathrm{i}V_C-\mu_C\big)_{\tau_1} 
    \big(\dot{\tilde{\psi}}-\mathrm{i}V_C-\mu_C\big)_{\tau_2} \Big) \nonumber \\
 + &\int \int d\tau_1 d\tau_2 \!\ \chi_{xx}(\tau_1-\tau_2) \!\  
    \Big( \big(\dot{\bar{\varphi}}_0\big)_{\tau_1} \big(\dot{\bar{\varphi}}_0\big)_{\tau_2} 
    + \frac{1}{4} \big(\pm \dot{\tilde{\varphi}}_0-\mathrm{i}V_S-\mu_S\big)_{\tau_1} 
    \big(\pm \dot{\tilde{\varphi}}_0-\mathrm{i}V_S-\mu_S\big)_{\tau_2} \Big) \nonumber \\
 + &\int \int d\tau_1 d\tau_2 \!\ \chi_{0x}(\tau_1-\tau_2) \!\  
    \Big( \big(\dot{\bar{\psi}}\big)_{\tau_1} \big(\dot{\bar{\varphi}}_0\big)_{\tau_2} 
    + \frac{1}{4} \big(\dot{\tilde{\psi}}-\mathrm{i}V_C-\mu_C\big)_{\tau_1} 
    \big(\pm \dot{\tilde{\varphi}}_0-\mathrm{i}V_S-\mu_S\big)_{\tau_2} \Big) \nonumber \\ 
 + &\int \int d\tau_1 d\tau_2 \!\ \chi_{x0}(\tau_1-\tau_2) \!\  
    \Big( \big(\dot{\bar{\varphi}}_0\big)_{\tau_1} \big(\dot{\bar{\psi}}\big)_{\tau_2} 
    + \frac{1}{4} \big(\pm \dot{\tilde{\varphi}}_0-\mathrm{i}V_S-\mu_S\big)_{\tau_1} 
    \big(\dot{\tilde{\psi}}-\mathrm{i}V_C-\mu_C\big)_{\tau_2} \Big).
\end{align}
As the higher-order expansion terms in Eq.~(\ref{eqn5-15}) do not change this saddle-point 
structure, we can fairly 
conclude that $-\mathrm{Tr}\mathrm{ln}\tilde{\bm{\mathcal{G}}}_{\mu 0}^{-1}$ 
has the following saddle point, 
\begin{align}
    -\mu_C + \dot{\tilde{\psi}} -\mathrm{i} V_C=0, \quad 
    -\mu_S \pm \dot{\tilde{\varphi}}_0 -\mathrm{i} V_S=0, \quad 
    \dot{\bar{\psi}} = 0, \quad \dot{\bar{\varphi}}_0=0. \label{eqn5-17b}
\end{align}
Due to a term of $\mu_C N_C+\mu_S N_S$ in the action, the saddle point of 
the whole action in Eq.~(\ref{eqn5-8}) is deviated from Eq.~(\ref{eqn5-17b}) 
by $\mathcal{O}(N_C,N_S)$. Given $N_C(0)=N_S(0)=0$, however, the deviation 
is on the order $\mathcal{O}(\bm{\mathcal{T}}^2)$, so that the 
correction term results in higher-order effects in Josephson equations and we can ignore them legitimately.  

Then an integration over $\mu_C$, $\mu_S$, $\bar{\psi}$ and $\bar{\varphi}_0$ in Eq.~(\ref{eqn5-8}) 
under the saddle-point approximation leads to the following effective action for $\tilde{\psi}$, $\tilde{\varphi}_0$, $N_C$ and $N_S$; 
\begin{equation}
\label{eqn5-18}
\mathcal{Z}[V_{C},V_{S}]=\int\mathcal{D}N_{C}\mathcal{D}N_{S}\mathcal{D}\tilde{\psi}\mathcal{D}\tilde{\varphi}_{0}\mathrm{e}^{\mathrm{i}\int\mathrm{d}\tau[N_C(\dot{\tilde{\psi}}-\mathrm{i}V_C)+N_S(\pm\dot{\tilde{\varphi}}_0-\mathrm{i}V_S)]-\frac{1}{2}\mathrm{Tr}
[((\bm{\mathcal{G}}_0^{-1}+\bm{\Phi})^{-1}\bm{\mathcal{T}})^2] 
- \frac{1}{4}\mathrm{Tr}
[((\bm{\mathcal{G}}_0^{-1}+\bm{\Phi})^{-1}\bm{\mathcal{T}})^4] + \cdots },
\end{equation}
where $\mu_C$ and $\mu_S$ in Eq.~(\ref{eqn5-8}) were replaced by 
$\partial_{\tau} \tilde{\psi} -\mathrm{i}V_C$ and $\pm \partial_{\tau} \tilde{\psi} -\mathrm{i}V_S$ respectively, and $\tilde{\bm{\mathcal{G}}}_{\mu 0}^{-1}$ in Eq.~(\ref{eqn5-14}) was replaced by $\bm{\mathcal{G}}_0^{-1}+\bm{\Phi}$.
In Eq.~(\ref{eqn5-18}), one can clearly see that 
$\partial_{\tau} \tilde{\psi}-\mathrm{i}V_C$ and $\partial_{\tau} \tilde{\varphi}_0-\mathrm{i}V_S$ 
are coupled only with $N_C$ and $N_S$ respectively. The couplings result 
in the second Josephson equations.  

The first Josephson equation comes from the
the second-order term in the tunneling part, $\bm{\mathcal{T}}$ in Eq.~(\ref{eqn5-14}), which can be further 
expanded in ${\bm \Phi}$; 
\begin{align}
\label{eqn5-19}
&\frac{1}{2}\mathrm{Tr}(\tilde{\bm{\mathcal{G}}}_{\mu 0}\bm{\mathcal{T}})^2=\frac{1}{2}\mathrm{Tr}[(\bm{\mathcal{G}}^{-1}_0+\bm{\Phi})^{-1}\bm{\mathcal{T}}(\bm{\mathcal{G}}^{-1}_0+\bm{\Phi})^{-1}\bm{\mathcal{T}}]\nonumber\\
=&\frac{1}{2}\mathrm{Tr}[(\bm{\mathcal{G}}_0-\bm{\mathcal{G}}_0\bm{\Phi}\bm{\mathcal{G}}_0+(\bm{\mathcal{G}}_0\bm{\Phi})^2\bm{\mathcal{G}}_0-(\bm{\mathcal{G}}_0\bm{\Phi})^3\bm{\mathcal{G}}_0+...)\bm{\mathcal{T}}(\bm{\mathcal{G}}_0-\bm{\mathcal{G}}_0\bm{\Phi}\bm{\mathcal{G}}_0+(\bm{\mathcal{G}}_0\bm{\Phi})^2\bm{\mathcal{G}}_0-(\bm{\mathcal{G}}_0\bm{\Phi})^3\bm{\mathcal{G}}_0+...)\bm{\mathcal{T}}]\nonumber\\
=&\frac{1}{2}\mathrm{Tr}\bm{\mathcal{G}}_0\bm{\mathcal{T}}\bm{\mathcal{G}}_0\bm{\mathcal{T}}+\mathrm{Tr}\bm{\mathcal{G}}_0\bm{\Phi}\bm{\mathcal{G}}_0\bm{\Phi}\bm{\mathcal{G}}_0\bm{\mathcal{T}}\bm{\mathcal{G}}_0\bm{\mathcal{T}}+\frac{1}{2}\mathrm{Tr}\bm{\mathcal{G}}_0\bm{\Phi}\bm{\mathcal{G}}_0\bm{\mathcal{T}}\bm{\mathcal{G}}_0\bm{\Phi}\bm{\mathcal{G}}_0\bm{\mathcal{T}}+...
\end{align}
The first two terms do not depend on $\tilde{\psi}$ and $\tilde{\varphi}_0$, 
when dissipation effect is neglected from the Josephson equation. Namely, 
${\bm G}^{-1}_b$ commutes with ${\bm \sigma}_x$ and the dissipation effect 
comes from the time-dependence of $\tilde{\psi}$ and $\tilde{\varphi}_0$. 
The (dissipationless) Josephson equation comes from the third term 
in the right hand side,
in which the spin-charge coupled nature of the Josephson equations are encoded;
\begin{align}
\label{eqn5-21}
&\mathrm{Tr}[(\bm{\mathcal{G}}_0\bm{\Phi})(\bm{\mathcal{G}}_0\bm{\mathcal{T}})(\bm{\mathcal{G}}_0\bm{\Phi})(\bm{\mathcal{G}}_0\bm{\mathcal{T}})]\nonumber\\
&=\mathrm{tr}[\bm{G}_{a\alpha}(\vec{\phi}_\omega\cdot\bm{\vec{\sigma}})\bm{G}_{b\alpha}T^{(b)}_{\alpha\beta}\mathrm{e}^{\mathrm{i}(\tilde{\psi}\pm\tilde{\varphi}_0\bm{\sigma_x})}\bm{G}_{b\beta}(\vec{\phi}_\omega^*\cdot\vec{\bm{\sigma}})\bm{G}_{a\beta}T_{\alpha\beta}^{(a)*}]+\mathrm{Tr}[\bm{G}_{b\alpha}(\vec{\phi}_\omega^*\cdot\vec{\bm{\sigma}})\bm{G}_{a\alpha}T^{(a)}_{\alpha\beta}\bm{G}_{a\beta}(\vec{\phi}_\omega\cdot\vec{\bm{\sigma}})\bm{G}_{b\beta}T^{(b)*}_{\alpha\beta}\mathrm{e}^{-\mathrm{i}(\tilde{\psi}\pm\tilde{\varphi}_0\bm{\sigma_x})}]\nonumber\\
&+\mathrm{Tr}[\bm{G}_{a\alpha}(\vec{\phi}_\omega\cdot\vec{\bm{\sigma}})\bm{G}_{b\alpha}T^{(b)*}_{\beta\alpha}\mathrm{e}^{-\mathrm{i}(\tilde{\psi}\pm\tilde{\varphi}_0\bm{\sigma_x})}\bm{G}_{b\beta}(\vec{\phi}_\omega^*\cdot\vec{\bm{\sigma}})\bm{G}_{a\beta}T^{(a)}_{\beta\alpha}]+\mathrm{Tr}[\bm{G}_{b\alpha}(\vec{\phi}_\omega^*\cdot\vec{\bm{\sigma}})\bm{G}_{a\alpha}T^{(a)*}_{\beta\alpha}\bm{G}_{a\beta}(\vec{\phi}_\omega\cdot\vec{\bm{\sigma}})\bm{G}_{b\beta}T^{(b)}_{\beta\alpha}\mathrm{e}^{\mathrm{i}(\tilde{\psi}\pm\tilde{\varphi}_0\bm{\sigma_x})}]\nonumber\\
&=2T^{(b)}_{\beta\alpha}T^{(a)*}_{\beta\alpha}
\mathrm{Tr}[(\vec{\phi}_\omega\cdot\vec{\bm{\sigma}})
e^{-i(\tilde{\psi} \pm \tilde{\varphi}_0 {\bm \sigma}_x)} (\vec{\phi}_\omega^*\cdot\vec{\bm{\sigma}}) \bar{\bm G}_{b\alpha}
\bm{G}_{a\alpha}\bm{G}_{a\beta}\bar{\bm G}_{b\beta}]  
+ {\rm c.c.}  
\end{align}
Here ${\bm G}^{-1}_{d\alpha}\equiv \partial_{\tau}+E_{d\alpha}{\bm \sigma}_0
+H{\bm \sigma}_x -\mu$ and $\bar{\bm G}^{-1}_{d\alpha}\equiv \partial_{\tau}+E_{d\alpha}{\bm \sigma}_0
\mp H_d{\bm \sigma}_x -\mu$ for $d=a,b$, and the upper and 
lower signs of the multiple sign 
are for the transvere ($\omega=\perp$) and longitudinal ($\omega=\parallel$) 
phases respectively. In Eq.~(\ref{eqn5-21}), a product between two 
tunneling matrix element picks up an external magnetic flux $\Psi$ that 
penetrates through the junction area in the transversal direction; 
\begin{align}
T^{(b)}_{\beta\alpha}T^{(a)*}_{\beta\alpha}=
|T_{\beta\alpha}|^2 e^{\mathrm{i}\Psi}. \label{eqn5-21a}
\end{align}
Using Eq.~(\ref{eqn5-21a}) together with 
$\bar{\bm G}_{b\beta}\bm{G}_{a\beta}\bm{G}_{a\alpha}\bar{\bm G}_{b\alpha}=
\bar{\bm G}_{b\alpha} \bm{G}_{a\alpha}\bm{G}_{a\beta}\bar{\bm G}_{b\beta}$, 
we obtain a tunneling term ${\cal S}_{\rm tun}$ in the effective action as; 
\begin{align}
\label{eqn5-21b}
{\cal S}_{\rm tun}&\equiv \frac{1}{2} 
\mathrm{Tr}\Big[(\bm{\mathcal{G}}_0\bm{\Phi})(\bm{\mathcal{G}}_0\bm{\mathcal{T}})(\bm{\mathcal{G}}_0\bm{\Phi})(\bm{\mathcal{G}}_0\bm{\mathcal{T}})]\nonumber\\
&=2|T_{\beta\alpha}|^2 
\mathrm{Tr}[\big(\vec{\phi}_\omega\cdot\vec{\bm{\sigma}}\big)
\big(\cos\big(\tilde{\psi}-\Psi\big) 
\cos\tilde{\varphi}_{0} \mp \sin\big(\tilde{\psi}-\Psi\big) 
\sin\tilde{\varphi}_{0} {\bm \sigma}_x \big) 
\big(\vec{\phi}_\omega^*\cdot\vec{\bm{\sigma}}\big) \bar{\bm G}_{b\alpha}
\bm{G}_{a\alpha}\bm{G}_{a\beta}\bar{\bm G}_{b\beta}\Big].
\end{align}
Note that 
\begin{align}
&(\vec{\phi}_\omega\cdot\vec{\bm{\sigma}})
(\cos\big(\tilde{\psi}-\Psi\big) 
\cos\tilde{\varphi}_{0} \mp \sin\big(\tilde{\psi}-\Psi\big) 
\sin\tilde{\varphi}_{0} {\bm \sigma}_x ) 
(\vec{\phi}_\omega^*\cdot\vec{\bm{\sigma}}) \nonumber \\
& \ \ = 
(\vec{\phi}_\omega\cdot\vec{\bm{\sigma}}) 
(\vec{\phi}_\omega^*\cdot\vec{\bm{\sigma}}) 
(\cos\big(\tilde{\psi}-\Psi\big) 
\cos\tilde{\varphi}_{0} + \sin\big(\tilde{\psi}-\Psi\big) 
\sin\tilde{\varphi}_{0} {\bm \sigma}_x ) \nonumber \\
& \ \ = \big( {\bm \sigma}_0 + \tilde{h} {\bm \sigma}_x \big) 
\Big(\cos\big(\tilde{\psi}(\tau)-\Psi\big) 
\cos(\tilde{\varphi}_{0}(\tau)) + \sin\big(\tilde{\psi}(\tau)-\Psi\big) 
\sin(\tilde{\varphi}_{0}(\tau)) \!\ {\bm \sigma}_x \Big) 
\equiv {\bm{\mathcal{F}}}_{\lambda}(\tau),  \label{eqn5-23a} 
\end{align}
where $\tilde{h}={\sf{h}}\equiv h/h_c$ for $\omega=\perp$ and $\tilde{h}={\sf{h}}\equiv-h^{\prime}/h_c$ 
for $\omega=\parallel$. Since we do not include the dissipation effect 
in the Josephson equation, we take the 
zero Matsubara frequency component 
of ${\bm{\mathcal{F}}}_{\omega}(\tau)$. This leads to 
\begin{align}
\label{eqn5-24}
{\cal S}_{\rm tun}& =\int\mathrm{d}\tau\mathrm{tr}\big[\bm{\mathcal{F}}_\omega(\tau)\sum_{\alpha,\beta}
2|T_{\beta\alpha}|^2\frac{1}{\beta}\sum_{{\rm  i}\omega_n} 
\bar{\bm{G}}_{b\alpha}(\mathrm{i}\omega_n)\bm{G}_{a\alpha}(\mathrm{i}\omega_n)\bm{G}_{a\beta}(\mathrm{i}\omega_n)\bar{\bm{G}}_{b\beta}(\mathrm{i}\omega_n)\big]
\equiv\int\mathrm{d}\tau\mathrm{tr} 
\big[\bm{\mathcal{F}}_\omega(\tau) \!\ \bm{\mathcal{G}}
\big], 
\end{align} 
where ${\rm tr}$ is an trace of 2 by 2 matrices associated with the spin index. A $2$ by $2$ matrix $\bm{\mathcal{G}}$ can be evaluated up to the first 
order in the exchange fields, 
\begin{align}
\label{eqn5-25}
\bm{\mathcal{G}}&=\frac{1}{\beta}\sum_{{\rm i}\omega_n}\sum_{\alpha,\beta} 2|T_{\beta\alpha}|^2 g_{b\alpha}(1\pm g_{b\alpha}H_b\bm{\sigma}_x)g_{a\alpha}(1-g_{a\alpha}H_a\bm{\sigma}_x)g_{a\beta}(1-g_{a\beta}H_a\bm{\sigma}_x)g_{b\beta}(1\pm g_{b\beta}H_b\bm{\sigma}_x) + {\cal O}(H_a^2,H_b^2) \nonumber\\ 
&=\frac{1}{\beta}\sum_{{\rm i}\omega_n}\sum_{\alpha,\beta}g_{a\alpha}g_{b\alpha}g_{a\beta}g_{b\beta} |T_{\beta\alpha}|^2 - 
\frac{\bm{\sigma}_x}{\beta}\sum_{{\rm i}\omega_n} 
\sum_{\alpha,\beta} (H_a g_{a\alpha}\mp H_b g_{b\alpha}+H_a g_{a\beta}\mp H_b g_{b\beta})g_{a\alpha}g_{b\alpha}g_{a\beta}g_{b\beta} 
|T_{\beta\alpha}|^2+ {\cal O}(H_a^2,H_b^2) \nonumber \\
&\equiv A_0 {\bm \sigma}_0 + A_x {\sf{h}} {\bm \sigma}_x + {\cal O}(H_a^2,H_b^2),  
\end{align}
with 
\begin{equation}
\label{eqn5-27}
A_0\equiv \frac{1}{\beta} 
\sum_{{\rm i}\omega_n} 
\sum_{\alpha,\beta} 
g_{a\alpha}g_{b\alpha}g_{a\beta}g_{b\beta} |T_{\beta\alpha}|^2,\quad  A_x\equiv-\frac{1}{{\sf{h}}}\frac{1}{\beta}\sum_{{\rm i}\omega_n} 
\sum_{\alpha,\beta}(H_a g_{a\alpha}\mp H_b g_{b\alpha}+H_a g_{a\beta} \mp H_b g_{b\beta})g_{a\alpha}g_{b\alpha}g_{a\beta}g_{b\beta}|T_{\beta\alpha}|^2. 
\end{equation}
Substituting Eqs.~(\ref{eqn5-25}, \ref{eqn5-23a}) into Eq.~(\ref{eqn5-24}), 
we obtain the following spin-charge coupled potential term from
the second order expansion in $\bm{\mathcal{T}}$:
\begin{equation}
\label{eqn5-28}
\mathcal{S}_{\mathrm{tun}}=\int\mathrm{d}\tau\mathrm{tr}[\bm{\mathcal{F}}(\tau) \!\ 
(A_0+A_x{\sf{h}}\bm{\sigma}_x)]=-I_0\int\mathrm{d}
\tau \bigg(\mathrm{cos}\Big(\tilde{\psi}({\tau})-\Psi\Big)
\mathrm{cos}\big(\tilde{\varphi}_0(\tau)\big)+
\bar{h} \!\ \mathrm{sin}\Big(\tilde{\psi}(\tau)-\Psi\Big)
\mathrm{sin}\big(\tilde{\varphi}_0(\tau)\big)\bigg),
\end{equation}
with
\begin{equation}
\label{eqn5-29}
I_0\equiv -\rho^2 A_0,\quad \bar{h}\equiv{\sf{h}}(1+\frac{A_x}{A_0}).
\end{equation}
$\bar{h}$ is different for the transverse phase and the longitudinal phase, so we write $\bar{h}_\pm$ in the main text to explicitly show the difference.

Substituting Eq.~(\ref{eqn5-28}) into Eq.~(\ref{eqn5-18}), 
we finally obtain the effective action for $N_C$, $N_S$, $\tilde{\psi}$ and 
$\tilde{\varphi}_0$:
\begin{align}
\label{eqn5-32}
\mathcal{S}_{\mathrm{eff}}[\tilde{\psi},\tilde{\varphi}_0,N_C,N_S;V_C,V_S]= & 
\int\mathrm{d}\tau 
\bigg[N_C(-\mathrm{i}\hbar\dot{\tilde{\psi}}(\tau)-eV_C)+N_S(\mp\mathrm{i}\hbar\dot{\tilde{\varphi}}_0(\tau)-eV_S) 
\nonumber \\
&\hspace{-0.4cm} 
-\hbar I_0\bigg( 
\mathrm{cos}\big(\tilde{\psi}(\tau) - \frac{e}{\hbar c} \Psi\big)
\mathrm{cos}\big(\tilde{\varphi}_0(\tau)\big)+\bar{h} \!\ 
\mathrm{sin}\big(\tilde{\psi}(\tau)-\frac{e}{\hbar c}\Psi\big)
\mathrm{sin}\big(\tilde{\varphi}_0(\tau)\big)\bigg)\bigg],
\end{align}
where we recover $\hbar$ as base unit of the action, unit charge $e$
in front of $V_C$ and $V_S$, and the inverse of magnetic flux unit $e/\hbar c$.
This is exactly Eq.~(13) in the main text.
A variation of the effective action with respect to these variables lead to 
the spin-charge coupled Josephson equations:
\begin{equation}
\label{eqn5-33}
\mathrm{i}\dot{\tilde{\psi}}=-\frac{e}{\hbar}V_C,\quad \mathrm{i}\dot{\tilde{\varphi}}_0=\mp\frac{e}{\hbar} V_S,
\end{equation}
\begin{equation}
\label{eqn5-34}
\partial_\tau N_C=\mathrm{i}I_0\Big(\mathrm{sin}\big(\tilde{\psi}-\frac{e}{\hbar c}\Psi\big)\mathrm{cos} 
\big(\tilde{\varphi}_0\big)-\bar{h} \!\ 
\mathrm{cos}\big(\tilde{\psi}-\frac{e}{\hbar c}\Psi\big) 
\mathrm{sin}\big(\tilde{\varphi}_0\big)\Big),
\end{equation}
\begin{equation}
\label{eqn5-35}
\partial_\tau N_S=\pm\mathrm{i}I_0\Big(\mathrm{sin}\big(\tilde{\varphi}_0\big)
\mathrm{cos}\big(\tilde{\psi}-\frac{e}{\hbar c}\Psi\big) 
-\bar{h}\!\ \mathrm{cos}\big(\tilde{\varphi}_0\big)\mathrm{sin}
\big(\tilde{\psi}-\frac{e}{\hbar c}\Psi\big)\Big).
\end{equation}
Under the Wick rotation from the imaginary time to the real time,
\begin{equation}
\label{eqn5-36}
\tau=\mathrm{i}t,\quad I_C=(-e)(-\partial_t N_C)=\mathrm{i}e\partial_\tau N_C,\quad I_S=(-e)(-\partial_t N_S)=\mathrm{i}e\partial_\tau N_S, 
\end{equation}
we obtain 
\begin{equation}
\label{eqn5-37}
I_C=-eI_0\Big(\mathrm{sin}\big(\tilde{\psi}-\frac{e}{\hbar c}\Psi\big)\mathrm{cos} 
\big(\tilde{\varphi}_0\big)-\bar{h} \!\ 
\mathrm{cos}\big(\tilde{\psi}-\frac{e}{\hbar c}\Psi\big) 
\mathrm{sin}\big(\tilde{\varphi}_0\big)\Big),
\end{equation}
\begin{equation}
\label{eqn5-38}
I_S=\mp eI_0 \Big(\mathrm{sin}\big(\tilde{\varphi}_0\big)
\mathrm{cos}\big(\tilde{\psi}-\frac{e}{\hbar c}\Psi\big) 
-\bar{h}\!\ \mathrm{cos}\big(\tilde{\varphi}_0\big)\mathrm{sin}
\big(\tilde{\psi}-\frac{e}{\hbar c}\Psi\big)\Big),
\end{equation}
\begin{equation}
\label{eqn5-39}
\frac{\mathrm{d}\tilde{\psi}}{\mathrm{d}t}=-\frac{e}{\hbar}V_C,\quad \frac{\mathrm{d}\tilde{\varphi}_0}{\mathrm{d}t}=\mp\frac{e}{\hbar} V_S.
\end{equation}
$I_C$ and $I_S$ are the charge and spin currents in the hole layer. The charge current 
in the electron layer must be along in the opposite direction to that in the hole layer, 
\begin{align}
\label{eqn5-39-a}
I_{Ca}=-I_{Cb}=-I_C.
\end{align}
The spin current $I_S$ is defined as a difference between the charge 
current of the hole layer with up spin (along $+x$) and the charge current of the hole layer 
with down spin, 
\begin{align}
\label{eqn5-39-b}
I_S = e \partial_t N_S =  \frac{\partial}{\partial t} \Big\langle \frac{e}{2} \sum_{\alpha} 
\big({\bm b}^{\dagger}_{1\alpha} {\bm \sigma}_x {\bm b}_{1\alpha} 
-{\bm b}^{\dagger}_{2\alpha} {\bm \sigma}_x {\bm b}_{2\alpha} \big) \Big\rangle. 
\end{align}
Thus, $I_S$ is always equal to the $+x$-component spin current in the hole layer, i.e.
$I_S = I_{Sb}$. The $+x$-component spin current in the electron layer can have the 
same sign as or opposite sign to $I_S$, depending on whether the pseudospin superfluid 
phase is either the transverse phase or the longitudinal phase. The transverse phase breaks 
the continuous symmetry of the spin rotation that rotates spin in the electron layer and spin 
in the hole layer in the same direction around the $x$ axis in the spin space. Accordingly, 
\begin{align}
I_{Sa}=I_{Sb}=I_S, \label{eqn5-40-a}
\end{align} 
for the transverse phase. The longitudinal phase breaks the continuous symmetry of the spin 
rotation that rotates spins in the electron layer and spins in the hole layer in the opposite direction 
around the $x$ axis. Thus,  
\begin{align}
-I_{Sa}=I_{Sb}=I_S, \label{eqn5-40-b}
\end{align}
for the longitudinal phase.
Eqs.~(\ref{eqn5-40-a}, \ref{eqn5-40-b}) are consistent with our intuition. Namely, 
the electron with spin polarized along $+x$ and the hole with spin polarized along 
$\mp x$ form a excitonic pairing in the transverse/longitudinal phases, where $I_{Sb}$ must 
have the same sign as / opposite sign to $I_{Sa}$ respectively.  In conclusion, we obtain 
Eqs.~(14--17) in the main text from Eqs.~(\ref{eqn5-37}--\ref{eqn5-40-b}).  

Suppose that charge or spin voltages are applied across the two dots 
in both the electron and hole layers. Then, we can decompose the charge and spin 
voltages, $V_{Ca}$, $V_{Cb}$, $V_{Sa}$, and $V_{Sb}$, into the two 
components, $V_{Ca}\pm V_{Cb}$ and $V_{Sa}\pm V_{Sb}$. According to 
Eq.~(\ref{eqn5-0a}),  only one out of the two induces the a.c. Josephson 
currents. We summarize these voltage components in Fig.~\ref{fig.devices}.

\begin{figure}[t]
\centering
\subfigure[ ]{
\label{fig.devices_1}
\begin{minipage}{0.2\textwidth}
\centering
\includegraphics[width=\textwidth]{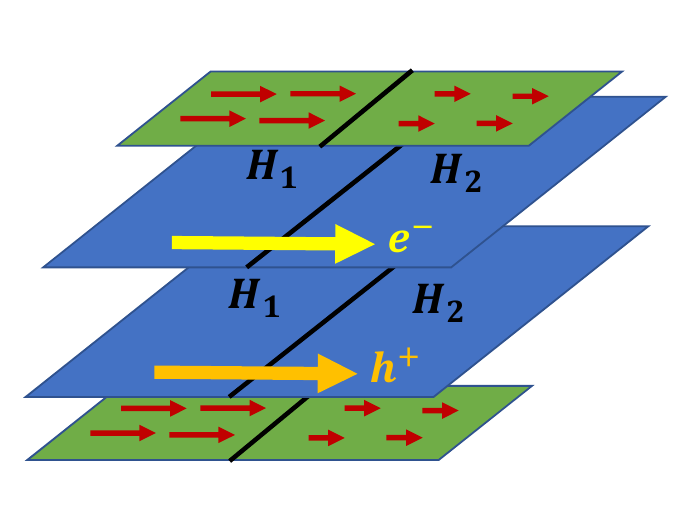}
\end{minipage}
}
\subfigure[ ]{
\label{fig.devices_2}
\begin{minipage}{0.2\textwidth}
\centering
\includegraphics[width=\textwidth]{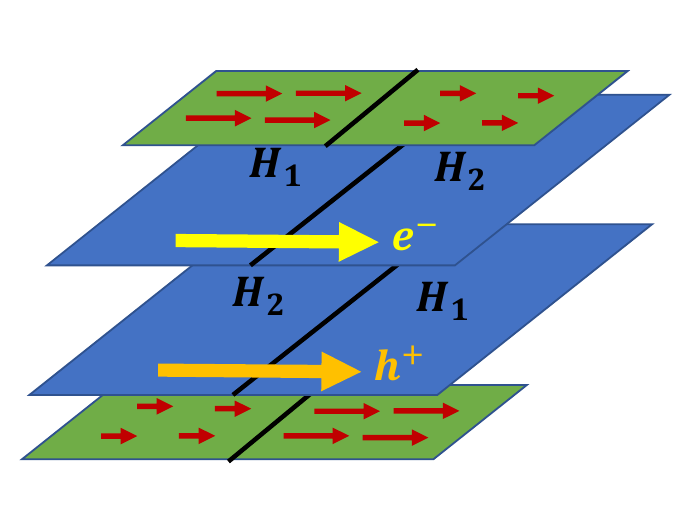}
\end{minipage}
}
\subfigure[ ]{
\label{fig.devices_3}
\begin{minipage}{0.22\textwidth}
\centering
\includegraphics[width=\textwidth]{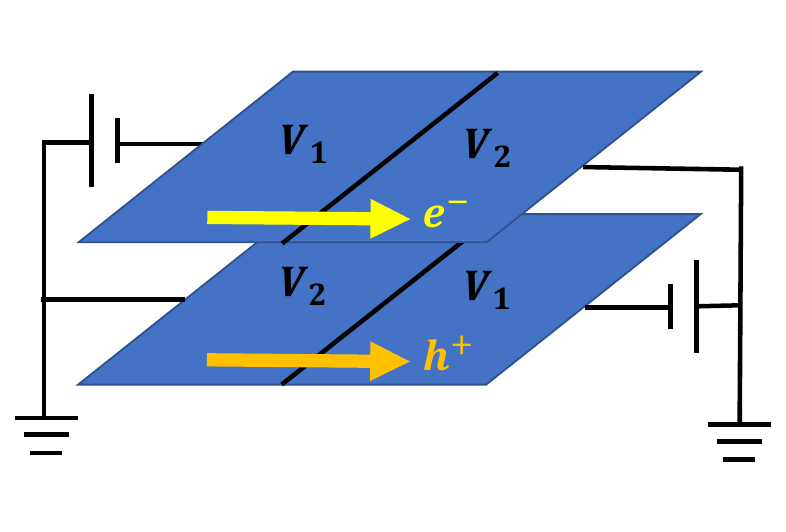}
\end{minipage}
}
\subfigure[ ]{
\label{fig.devices_4}
\begin{minipage}{0.2\textwidth}
\centering
\includegraphics[width=\textwidth]{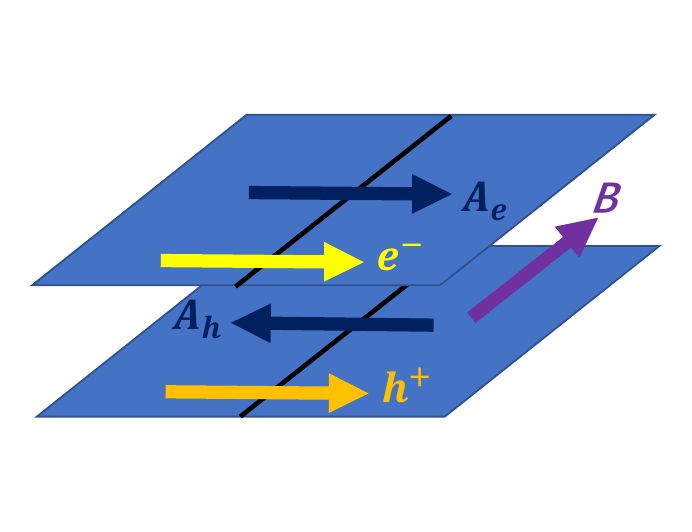}
\end{minipage}
}
\caption{\label{fig.devices} Four ways to induce the counterflow charge Josephson currents. 
\textbf{(a)} By a spin voltage across the junction, $V_S=V_{Sb}+V_{Sa}$ in the transverse phase. 
\textbf{(b)} By a spin voltage across the junction, $V_S=V_{Sb}-V_{Sa}$ in the longitudinal phase. 
\textbf{(c)} By a charge voltage across the junction, $V_C = V_{Cb}-V_{Ca}$. \textbf{(d)} By 
the transverse magnetic flux through the junction, $\Psi$.}
\end{figure}


\subsection{\label{sec6}Solutions of the spin-charge coupled Josephson equations: $V_S$-$I_C$ conversion}

\begin{figure}[t]
\centering
\subfigure[ ]{
\label{fig.solver_phases_1}
\begin{minipage}{0.3\textwidth}
\centering
\includegraphics[width=\textwidth]{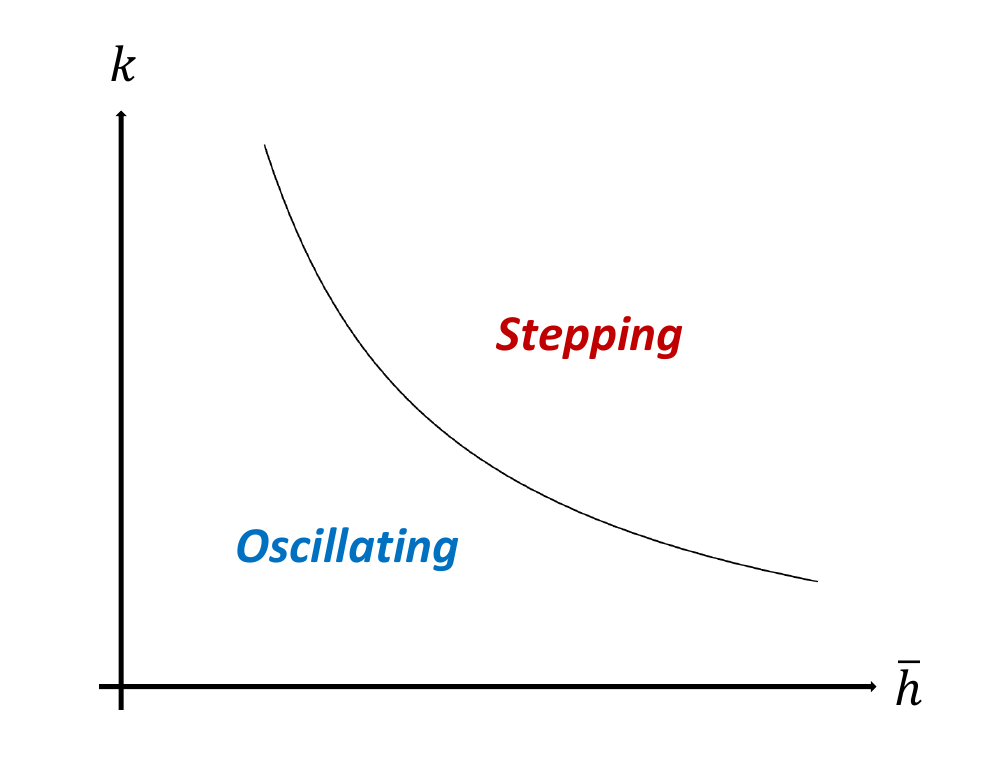}
\end{minipage}
}
\subfigure[ ]{
\label{fig.solver_phases_2}
\begin{minipage}{0.3\textwidth}
\centering
\includegraphics[width=\textwidth]{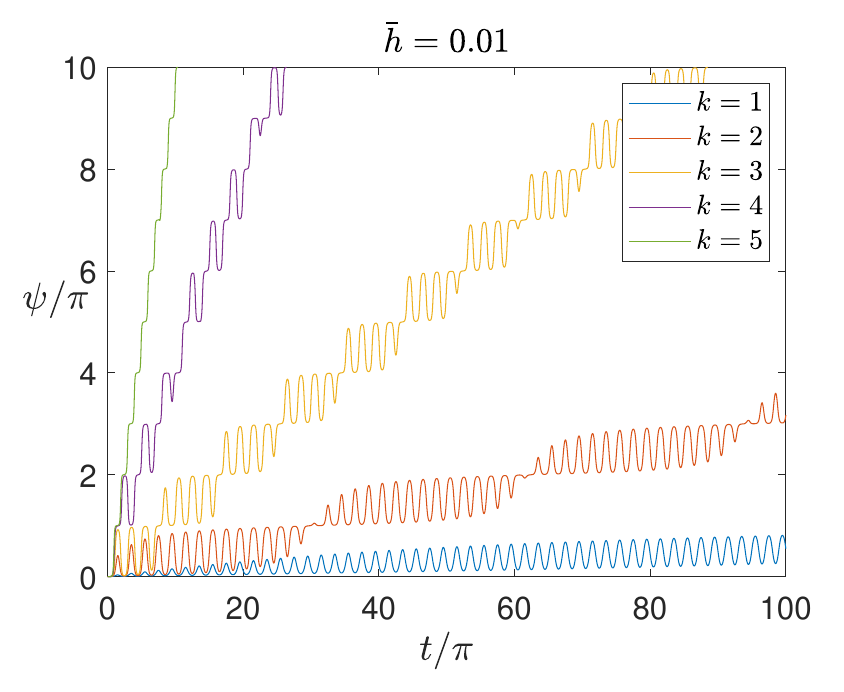}
\end{minipage}
}
\subfigure[ ]{
\label{fig.solver_phases_3}
\begin{minipage}{0.3\textwidth}
\centering
\includegraphics[width=\textwidth]{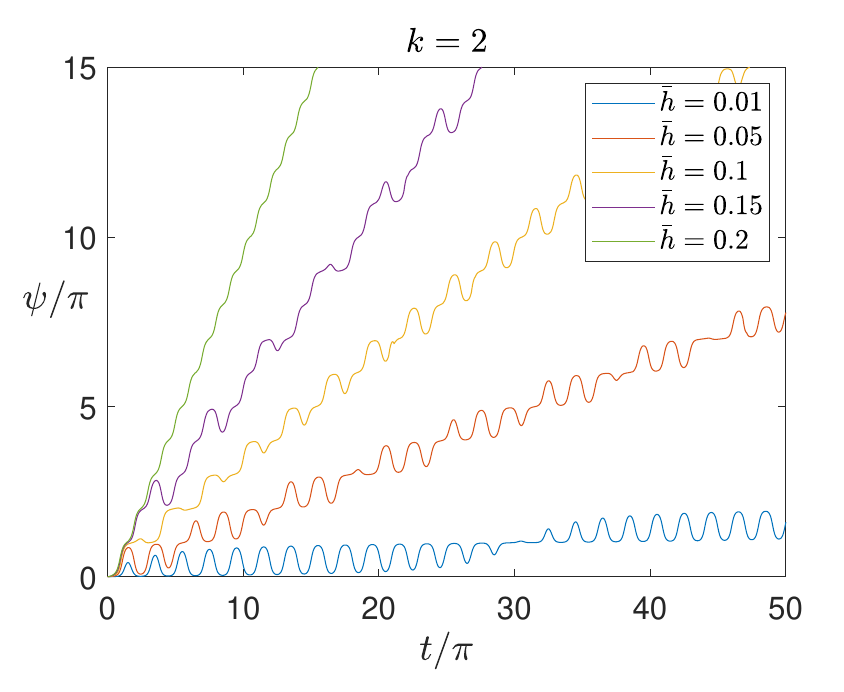}
\end{minipage}
}
\caption{\label{fig.solver_phases} 
\textbf{(a)} Schematic crossover diagram of a solution of Eq.~(\ref{eqn6-0a}) in favor 
for $\psi(t)$. In a `oscillating region' ($k\bar{h}\ll 1$),  $\psi(t)$ comprises of two 
oscillations with short periodicity $T_1=2\pi$ and a longer periodicity $T_2={\cal O}(\pi/(k\bar{h}))$. 
In a `stepping region' ($k\bar{h}\ge 1$), $\psi(t)$ takes a constant value 
of $n\pi$ around $t=n\pi$ and $\psi(t)$ increases 
abruptly from $n\pi$ to $(n+1)\pi$ around $t=(2n+1)\pi/2$. 
\textbf{(b, c)} A crossover from the oscillating region to the 
stepping region. 
}
\end{figure}

In this section, we solve the spin-charge coupled Josephson equations under a particular physical circumstance depicted in Fig.~2(a) of the 
main text. Thereby, the electron layer is externally connected 
to a closed electric circuit with an electric 
resistance $R_a$, and the hole layer is connected to 
another external circuit with an electric 
resistance $R_b$. 
An exchange field is induced in the hole layer through a magnetic proximity effect. 
By using a spatial variation of the exchange field, we apply
a spin voltage across the junction between two domains; 
$V_{Sb}\ne 0$, $V_{Sa}=0$. According to the Josephson 
equations, the spin voltage results in a linear increase 
of $\tilde{\varphi}_{0}$ in time, which leads 
to both a.c. charge supercurent $I_C$ and a.c. spin supercurrent $I_S$. 
Leads in the external circuits do not conserve spin angular momenta 
in general. Thus, the spin component of 
the supercurrents injected into the external circuits 
shall decay quickly and it has no significant impact on the spin 
voltage. In this respect, we can assume that the spin voltage is determined 
only by the static exchange field by the proximity effect. 
$V_{S}=V_{Sb}\pm V_{Sa}$ thus given is constant in time. 
On the one hand, the charge component of the supercurrents 
induce an a.c. charge voltages 
in both electron and hole layers; 
$V_{Ca}=I_{Ca}R_a$ and $V_{Cb}=I_{Cb}R_b$. From 
$V_{C}=V_{Cb}-V_{Ca}$ and $I_{Ca}=-I_{Cb}=-I_C$, 
Eqs.~(\ref{eqn5-37}, \ref{eqn5-39}) lead to a closed 
equation of motion for $\tilde{\psi}(t)$, 
\begin{align}
\label{eqn6-0}
 \frac{d\tilde{\psi}}{dt} = - I_{0} R \!\ \bigg(\sin \big(\tilde{\psi}\big) \cos\big(\mp V_St\big) 
 - \bar{h} \!\ \cos \big(\tilde{\psi}\big) \sin \big(\mp V_St\big) \bigg), 
\end{align}
with $R \equiv R_a+R_b$. With rescaling of the relevant variables, 
\begin{align}
    V_S t \equiv s, \quad k \equiv \frac{I_0R}{V_S}, 
\end{align}
we have 
\begin{align}
    \frac{d\tilde{\psi}}{ds} = - k \!\ \bigg(\sin \big(\tilde{\psi}\big) \cos\big(s\big) 
 \pm \bar{h} \!\ \cos \big(\tilde{\psi}\big) \sin \big(s\big) \bigg). \label{eqn6-0a}
\end{align}
In this section, we will describe (numerical) solution of this non-linear differential equation 
in favor for $\tilde{\psi}(s)$. Without loss of generality, we take the minus sign, i.e. the longitudinal phase (Eq.~(\ref{eqn2-17}))
in Eq.~(\ref{eqn6-0a}), and $k$ and $\bar{h}$ can be assumed to be positive. $\bar{h} 
\equiv {\sf{h}} (1+A_x/A_0)$ is supposed to be much small than 1; 
$|{\sf{h}}|\equiv |h'/h_c|\ll 1$ and $A_x/A_0 = {\cal O}(1)$. 
Thus, we discuss the solution only in a region of $0\le \bar{h}\ll 1$. For simplicity, 
we remove the tilde from $\tilde{\psi}(s)$ and call $s$ as $t$, 
$\tilde{\psi}(s) \rightarrow \psi(t)$.

\begin{figure}[t]
\centering
\subfigure[ ]{
\label{fig.solver_oscillating_long_1}
\begin{minipage}{0.3\textwidth}
\centering
\includegraphics[width=\textwidth]{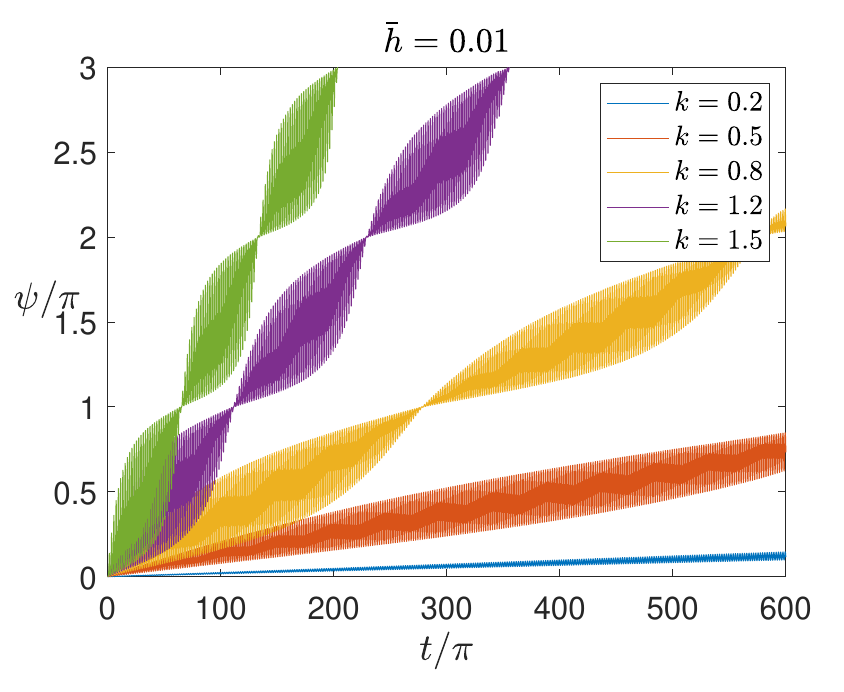}
\end{minipage}
}
\subfigure[ ]{
\label{fig.solver_oscillating_long_2}
\begin{minipage}{0.3\textwidth}
\centering
\includegraphics[width=\textwidth]{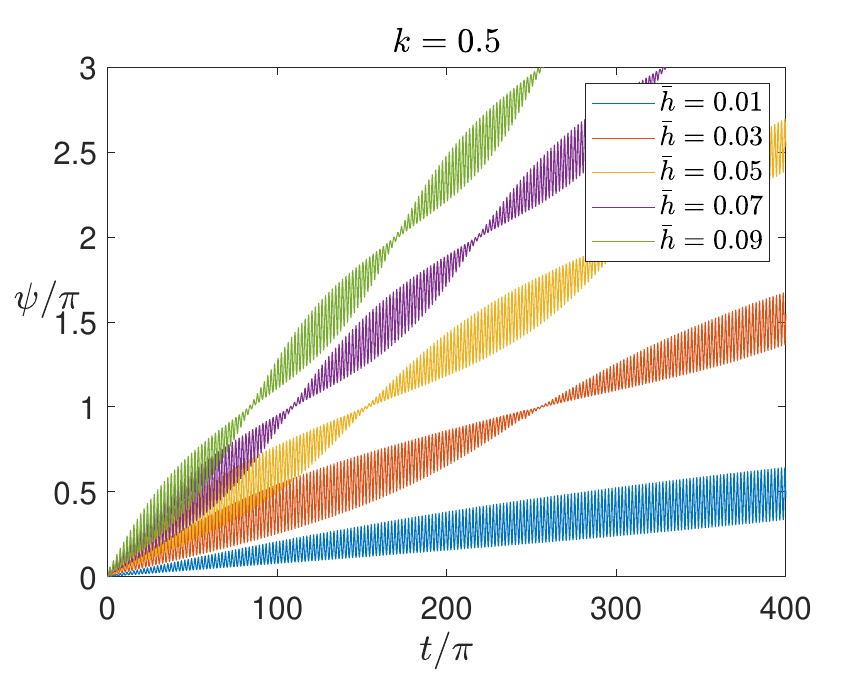}
\end{minipage}
}
\subfigure[ ]{
\label{fig.solver_oscillating_long_3}
\begin{minipage}{0.3\textwidth}
\centering
\includegraphics[width=\textwidth]{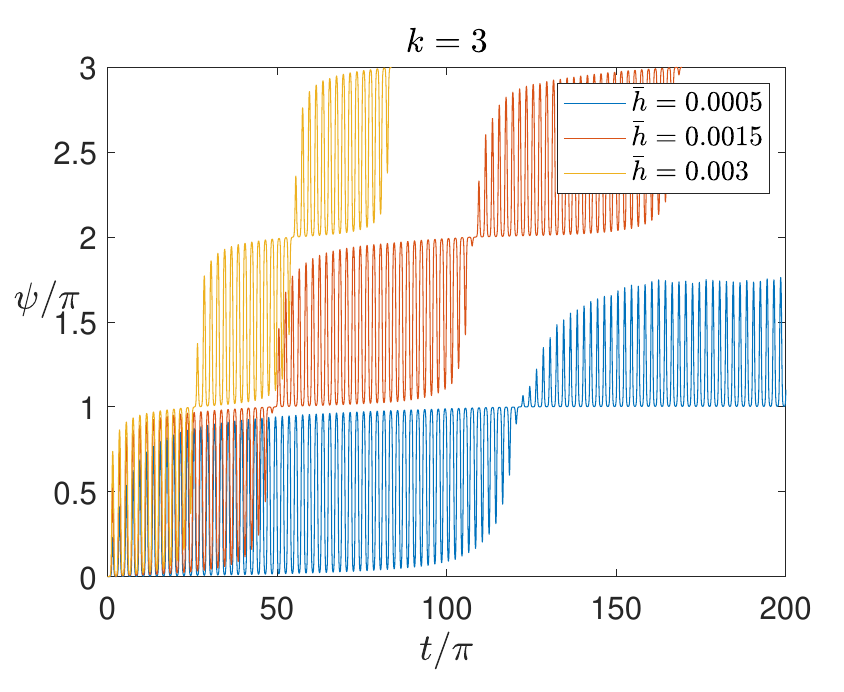}
\end{minipage}
}
\caption{\label{fig.solver_oscillating_long} 
$\psi(t)$ in the oscillating region in a longer time scale. 
$\psi(t)$ comprises of two oscillations with a short periodicity and a longer periodicity. 
The short periodicity is $T_1=2\pi$ (see Fig.~\ref{fig.solver_oscillating_short}), while the longer periodicity changes with $k\bar{h}$. 
}
\end{figure}

\begin{figure}[t]
\centering
\subfigure[ ]{
\label{fig.solver_oscillating_short_1}
\begin{minipage}{0.3\textwidth}
\centering
\includegraphics[width=\textwidth]{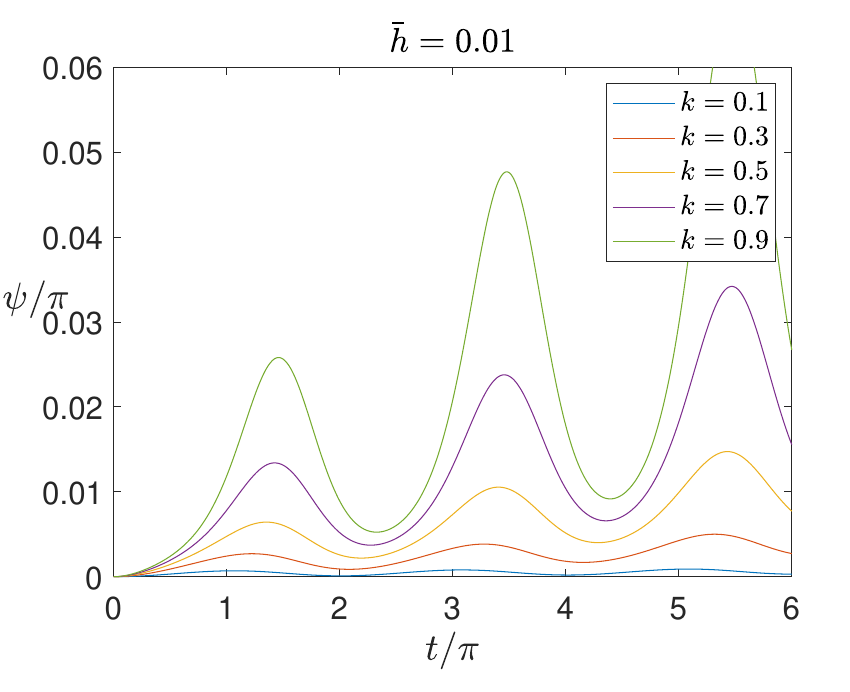}
\end{minipage}
}
\subfigure[ ]{
\label{fig.solver_oscillating_short_2}
\begin{minipage}{0.3\textwidth}
\centering
\includegraphics[width=\textwidth]{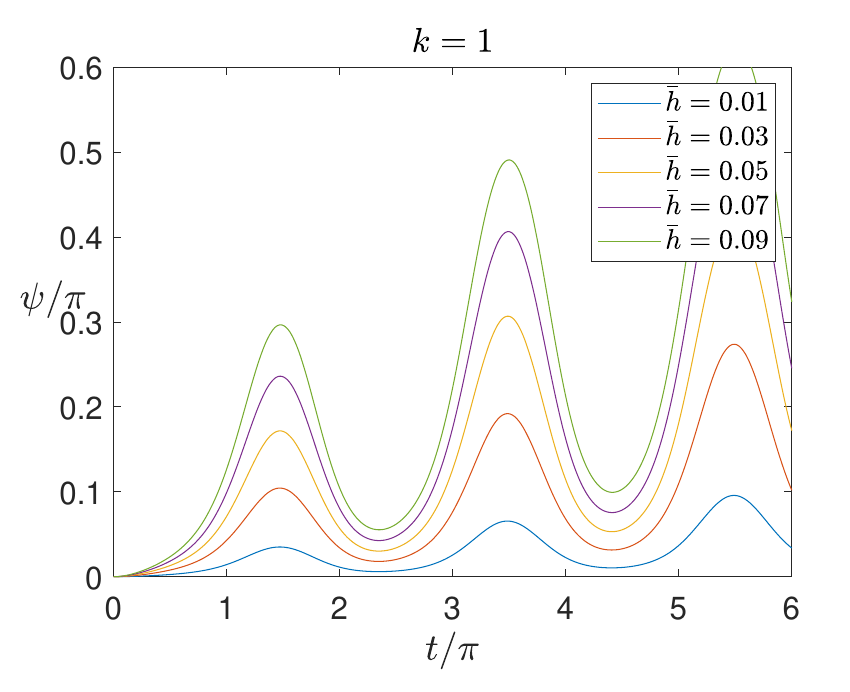}
\end{minipage}
}
\caption{\label{fig.solver_oscillating_short} 
$\psi(t)$ in the oscillating region in a shorter time scale. The short 
periodicity $T_1$ is around $2\pi$. 
}
\end{figure}

\begin{figure}[t]
\centering
\subfigure[ ]{
\label{fig.solver_stepping_1}
\begin{minipage}{0.3\textwidth}
\centering
\includegraphics[width=\textwidth]{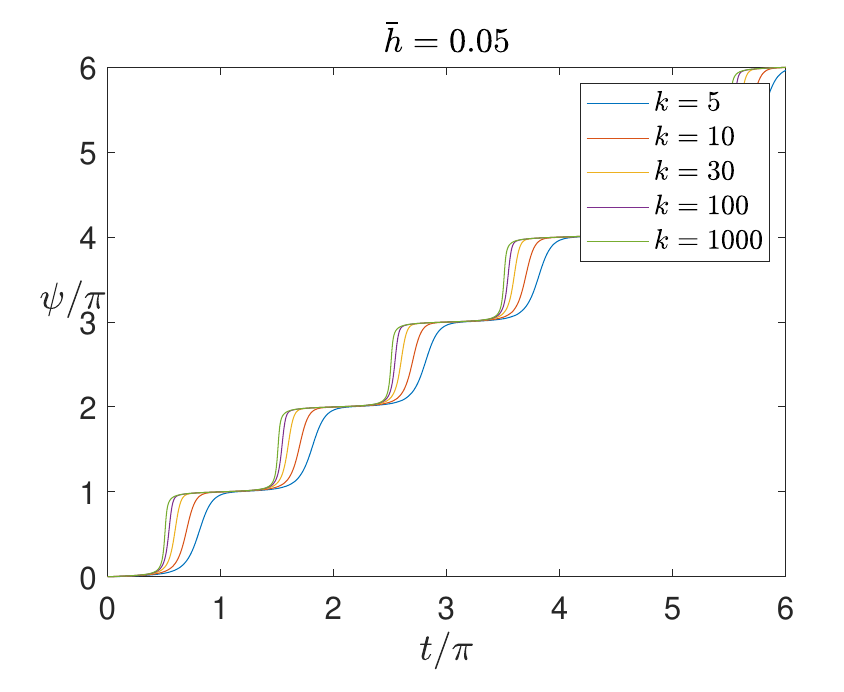}
\end{minipage}
}
\subfigure[ ]{
\label{fig.solver_stepping_2}
\begin{minipage}{0.3\textwidth}
\centering
\includegraphics[width=\textwidth]{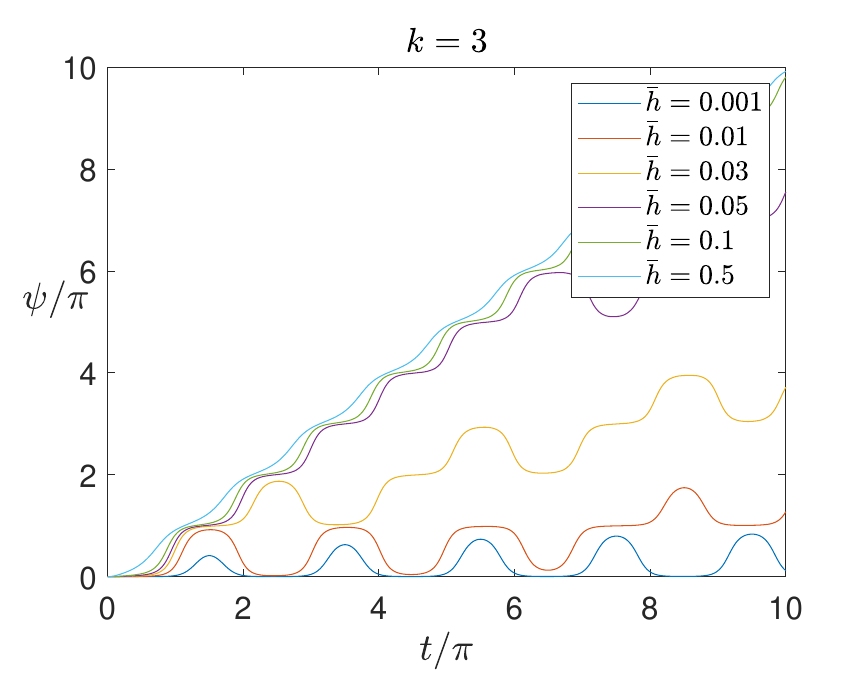}
\end{minipage}
}
\caption{\label{fig.solver_stepping} 
\textbf{(a)} $\psi(t)$ in the stepping region. $\psi(t)$ takes a constant 
value of $n\pi$ around $t=n\pi$, while it changes abruptly from 
$n\pi$ to $(n+1)\pi$ around $t=(2n+1)\pi/2$. \textbf{(b)} $\psi(t)$ 
in the stepping region ($\bar{h}=0.5, \!\ 0.1, \!\ 0.05$), in 
the crossover region $(\bar{h}=0.03)$ and in the oscillating 
region ($\bar{h}=0.01, \!\ 0.001$).}
\end{figure}

When $\bar{h}=0$, the solution is oscillatory in time with $2\pi$ 
periodicity ($T_1=2\pi$); 
\begin{equation}
\label{eqn6-1}
\frac{\psi(t)}{2}=\arctan(\tan(\frac{\psi(0)}{2})e^{-k \sin(t)}). 
\end{equation} 
The solution respects a time-reversal symmetry 
$\psi(\pi-t)=\psi(t)$. The amplitude of the oscillation gets larger for 
larger $k$, while it is always bounded by $\pi$; 
$\psi(t)$ is in the same branch of the arctan function of Eq.~(\ref{eqn6-1}).   
A finite $\bar{h}$ breaks the time-reversal symmetry, and the solution 
acquires a $t$-asymmetric component that increases linearly in time $t$. 
The form of the differential equation indicates that the $t$-asymmetric component 
must be scaled by $k\bar{h}t$ for $k\bar{h}\ll 1$; without loss of generality,
we take $\psi(0)=0$, such that
$\langle \psi(t) \rangle \sim k \bar{h} t$, where $\langle \psi(t) \rangle$ 
is an average of $\psi(t)$ over a time scale much longer than $T_1$ and 
much shorter than $1/(k\bar{h})$.
$\langle \psi(t) \rangle$ modifies the short-periodicity ($T_1$) ocillating
amplitute by the factor $\mathrm{sin}(\psi)$ in front of $\mathrm{cos}(t)$.
Overall behaviours of numerical solutions are consistent with 
this indication (see Fig.~\ref{fig.solver_oscillating_long}). 
Due to the $t$-asymmetric component, the solution with finite $\bar{h}$ with 
$k\bar{h}\ll 1$ comprises of two oscillations: one with the shorter periodicity 
$T_1=2\pi$, and the other with a longer peridocity, 
$T_2 = {\cal O}(\pi/(k\bar{h}))$ (Figs.~\ref{fig.solver_oscillating_long}, \ref{fig.solver_oscillating_short}). 

When $k\bar{h} \ll 1$, the two oscillations 
are clearly distinguishable (`oscillating region'). When $k\bar{h} = {\cal O}(1)$, 
the two periodicities become on the same order and the solution shows a crossover 
from the oscillating region to a `stepping region' 
(Figs.~\ref{fig.solver_phases_2}, \ref{fig.solver_phases_3}). When $k\gg k\bar{h} > 1$, the solution converges 
into the stepping behavior, where $\psi(t)$ shows a plateau 
($\psi(t)=n\pi$) around $t=n\pi$, and $\psi(t)$ 
increases abruptly from $n\pi$ to $(n+1)\pi$ around 
$t=(2n+1)\pi/2$ (Fig.~\ref{fig.solver_stepping}).


\subsection{\label{sec7} Derivation of classical ground-state phase diagram with Rashba coupling}

In this section, we describe the minimization of 
the Lagrangian in the presence of the antisymmetric vector-product (AVP) type 
 interaction ($D \ne 0$), Eq.~(\ref{eqn1-12}). 
In the absence of the AVP type interaction ($D=0$), the classical Lagrangian  
is minimized by spatial uniform configurations of 
$\vec{\Phi}^{\prime}(\vec{r})$ and 
$\vec{\Phi}^{\prime\prime}(\vec{r})$. To discuss the minimization 
in the presence of the AVP type interaction, let us decompose these 
four-component vectors into their amplitude parts ($\Phi^{\prime}(\vec{r})$ 
and $\Phi^{\prime\prime}(\vec{r})$) and four-component 
unit vector parts ($\vec{\psi}^{\prime}(\vec{r})$ and 
$\vec{\psi}^{\prime\prime}(\vec{r})$); $\vec{\Phi}^{\prime}(\vec{r}) 
\equiv \Phi^{\prime}(\vec{r}) \vec{\psi}^{\prime}(\vec{r})$ and 
$\vec{\Phi}^{\prime\prime}(\vec{r}) \equiv \Phi^{\prime\prime}(\vec{r})
\vec{\psi}^{\prime\prime}(\vec{r})$.  
Spatial gradients of the amplitude parts do not lower the AVP type interaction 
energy because of its anti-symmetric form, e.g. 
\begin{align}
    D [(\Phi^{\prime}\psi^{\prime}_z) \partial_x 
    (\Phi^{\prime}\psi^{\prime}_x) - 
    (\Phi^{\prime}\psi^{\prime}_x) \partial_x 
    (\Phi^{\prime}\psi^{\prime}_z)] = D (\Phi^{\prime})^2 
    [\psi^{\prime}_z \partial_x (\psi^{\prime}_x) - 
    \psi^{\prime}_x \partial_x (\psi^{\prime}_z)].
\end{align}
Thus, we take the amplitude parts to be spatially uniform, 
$\vec{\Phi}^{\prime}(\vec{r}) 
\equiv \Phi^{\prime} \vec{\psi}^{\prime}(\vec{r})$ and 
$\vec{\Phi}^{\prime\prime}(\vec{r}) \equiv \Phi^{\prime\prime} 
\vec{\psi}^{\prime\prime}(\vec{r})$. 

The classical Lagrangian of Eq.~(\ref{eqn1-12}) 
consists of three parts:
\begin{equation}
\label{eqn7-1}
\frac{S}{\beta} = S_0[\vec{\Phi}^{\prime},\vec{\Phi}^{\prime}]
+S_1[\vec{\Phi}^{\prime}]+S_1[\vec{\Phi}^{\prime\prime}]
\equiv 
\int\mathrm{d}^2\vec{r}[\mathcal{L}_0(\vec{\Phi}',\vec{\Phi}'')+\mathcal{L}_1(\vec{\Phi}',\partial_i\vec{\Phi}')+\mathcal{L}_1(\vec{\Phi}'',\partial_i\vec{\Phi}'')]. \quad (i=x,y)
\end{equation}
The first part describes a coupling between $\vec{\Phi}^{\prime}(\vec{r})$ 
and $\vec{\Phi}^{\prime\prime}(\vec{r})$. It is free from the spatial 
gradients of the fields,
\begin{align}
S_0[\vec{\Phi}^{\prime},\vec{\Phi}^{\prime}] & \equiv\int\mathrm{d}^2\vec{r}\mathcal{L}_0(\vec{\Phi}',\vec{\Phi}''), \nonumber \\
{\cal L}_{0}(\vec{\Phi}^{\prime},\vec{\Phi}^{\prime\prime}) 
&= A ({\Phi^{\prime}}^2+{\Phi^{\prime\prime}}^2) 
+ B [{\Phi^{\prime}}^4+{\Phi^{\prime\prime}}^4+6
{\Phi^{\prime}}^2{\Phi^{\prime\prime}}^2]  -2 \Phi^{\prime}\Phi^{\prime\prime} g (\Phi^{\prime},\Phi^{\prime\prime},
\vec{\psi}^{\prime},\vec{\psi}^{\prime\prime}), \nonumber \\
 g(\Phi^{\prime},\Phi^{\prime\prime},
\vec{\psi}^{\prime},\vec{\psi}^{\prime\prime}) 
&=  2 B \Phi^{\prime}\Phi^{\prime\prime} 
\big(\vec{\psi}^{\prime}({\vec{r}})\cdot \vec{\psi}^{\prime\prime}(\vec{r})\big)^2 
+ h \big(\psi^{\prime}_y(\vec{r}) \psi^{\prime\prime}_z(\vec{r}) 
- \psi^{\prime}_z(\vec{r}) \psi^{\prime\prime}_y(\vec{r})\big) 
- h^{\prime} \big(\psi^{\prime}_0(\vec{r}) \psi^{\prime\prime}_x(\vec{r}) 
- \psi^{\prime}_x(\vec{r}) \psi^{\prime\prime}_0(\vec{r})\big). \label{eqn7-1}
\end{align}
The other two parts depend on $\vec{\Phi}^{\prime}(\vec{r})$ and 
$\vec{\Phi}^{\prime\prime}(\vec{r})$ separately and they 
depend on their spatial gradients, e.g. 
\begin{equation}
\label{eqn7-2}
S_1[\Phi^{\prime}\vec{\psi}^{\prime}]=\Phi'^2\int \mathrm{d}^2\vec{r}\{\lambda(\nabla\psi^{\prime}_\mu)\cdot(\nabla\psi^{\prime}_\mu)-D(\psi^{\prime}_z\partial_x\psi^{\prime}_x-\psi^{\prime}_x\partial_x\psi^{\prime}_z-\psi^{\prime}_y\partial_y\psi^{\prime}_z+\psi^{\prime}_z\partial_y\psi^{\prime}_y+\psi^{\prime}_x\partial_y\psi^{\prime}_0-\psi^{\prime}_y\partial_x\psi^{\prime}_0+\psi^{\prime}_0\partial_x\psi^{\prime}_y-\psi^{\prime}_0\partial_y\psi^{\prime}_x)\}.
\end{equation}
We first minimize $S_1[\Phi^{\prime}\vec{\psi}^{\prime}]$ 
and $S_1[\Phi^{\prime\prime}\vec{\psi}^{\prime\prime}]$ 
in terms of the four-component vectors $\vec{\psi}^{\prime}$ and $\vec{\psi}^{\prime\prime}$ respectively.   
We then show that $\vec{\Phi}^{\prime}$ 
and $\vec{\Phi}^{\prime\prime}$ thus determined 
also maximally minimize $S_0[\vec{\Phi}^{\prime},\vec{\Phi}^{\prime}]$ 
by optimizing the amplitude parts, $\Phi^{\prime}$ and 
$\Phi^{\prime\prime}$.

To minimize $S_1[\Phi^{\prime}\vec{\psi}^{\prime}]$ 
in terms of $\vec{\psi}^{\prime}(\vec{r}) \equiv 
(\psi^{\prime}_0(\vec{r}),\psi^{\prime}_x(\vec{r}),
\psi^{\prime}_y(\vec{r}),\psi^{\prime}_z(\vec{r}))$, 
take the Fourier transformation of $\psi^{\prime}_{\mu}(\vec{r})$, 
\begin{align}
\psi^{\prime}_{\mu,\vec{k}}=\int\mathrm{d}^2\vec{r}\mathrm{e}^{-\mathrm{i}\vec{k}\cdot\vec{r}}\psi^{\prime}_\mu(\vec{r}), \quad \psi^{\prime}_\mu(\vec{r})=\frac{1}{V}\sum_{\vec{k}}\mathrm{e}^{\mathrm{i}\vec{k}\cdot\vec{r}}\psi^{\prime}_{\mu,\vec{k}},
\end{align}
for $\mu=0,x,y,z$ with  ${\psi^{\prime}}^*_{\mu,\vec{k}}=\psi^{\prime}_{\mu,-\vec{k}}$. 
The 
Fourier component $\psi^{\prime}_{\mu,\vec{k}}$ is given by two real-valued 
functions $\alpha_{\mu,\vec{k}}$ and $\beta_{\mu,\vec{k}}$ as  
$\psi^{\prime}_{\mu,\vec{k}}\equiv  \alpha_{\mu,\vec{k}}+{\rm i}
\beta_{\mu,\vec{k}}$. They are even and odd functions in $\vec{k}$ 
respectively. $S_1 [\Phi^{\prime}\vec{\psi}^{\prime}]$ is given by these two 
functions:
\begin{align}
S_1[\Phi^{\prime}\vec{\psi}^{\prime}]&=\frac{2\Phi'^2\lambda}{V}\sum_{k_x>0}k^2 
w^2_{\vec{k}}
+ \frac{4\Phi'^2D}{V}\sum_{k_x>0} f(\alpha_{\mu,\vec{k}},\beta_{\mu,\vec{k}}),\label{eqn7-3a}\\
f(\alpha_{\mu,\vec{k}},\beta_{\mu,\vec{k}})&\equiv k_x(\alpha_{z,\vec{k}}\beta_{x,\vec{k}}-\alpha_{x,\vec{k}}\beta_{z,\vec{k}}+\alpha_{0,\vec{k}}\beta_{y,\vec{k}}-\alpha_{y,\vec{k}}\beta_{0,\vec{k}})+k_y(\alpha_{z,\vec{k}}\beta_{y,\vec{k}}-\alpha_{y,\vec{k}}\beta_{z,\vec{k}} +\alpha_{x,\vec{k}}\beta_{0,\vec{k}}
-\alpha_{0,\vec{k}}\beta_{x,\vec{k}}),  \label{eqn7-3}
\end{align}
with $w^2_{\vec{k}}\equiv \sum_{\mu} ( \alpha_{\mu,\vec{k}}\alpha_{\mu,\vec{k}}+  \beta_{\mu,\vec{k}}\beta_{\mu,\vec{k}})$.
For given $w_{\vec{k}}$, 
$f(\alpha_{\mu,\vec{k}},\beta_{\mu,\vec{k}})$ in Eq.~(\ref{eqn7-3a}) shall be minimized 
for each $\hat{k}$ (the subscribe $\vec{k}$ will be omitted for convenience):
\begin{equation}
\label{eqn7-5}
f(\alpha_\mu,\beta_\mu)=k_x(\alpha_z\beta_x-\alpha_x\beta_z+\alpha_0\beta_y-\alpha_y\beta_0)+k_y(\alpha_z\beta_y-\alpha_y\beta_z-\alpha_0\beta_x+\alpha_x\beta_0)=-\hat{k}\cdot(\hat{\alpha}'\times\hat{\beta}'+\alpha'_0\hat{\beta}'-\beta'_0\hat{\alpha}').
\end{equation}
In the right hand side of Eq.~(\ref{eqn7-5}), the three-component vectors 
$\hat{\alpha}^{\prime}$, $\hat{\beta}^{\prime}$, $\hat{k}$ are introduced as, 
\begin{align}
\label{eqn7-6}
&\hat{\alpha}^{\prime} \equiv (\alpha^{\prime}_x,\alpha^{\prime}_y,\alpha^{\prime}_z), \quad 
\hat{\beta}^{\prime} \equiv (\beta^{\prime}_x,\beta^{\prime}_y,\beta^{\prime}_z), \quad 
\hat{k} \equiv (k_x,k_y,0) \equiv (\vec{k},0), \nonumber \\
&\alpha_x=\alpha'_y,\quad\alpha_y=-\alpha'_x,\quad\alpha_z=\alpha'_z,\quad\alpha_0=\alpha'_0;\nonumber\\
&\beta_x=\beta'_y,\quad\beta_y=-\beta'_x,\quad\beta_z=\beta'_z,\quad\beta_0=\beta'_0.
\end{align}
The function $f(\alpha_\mu,\beta_\mu)$ can be easily minimized for the special $\hat{k}$. For $\vec{k}=k\vec{e}_x$, it is minimized 
by 
\begin{align}
\label{eqn7-7}
\alpha'_y=\frac{w}{\sqrt{2}}\mathrm{cos}\zeta\mathrm{cos}\delta_1,\quad \beta'_y=-\frac{w}{\sqrt{2}}\mathrm{cos}\zeta\mathrm{sin}\delta_1,&\quad \alpha'_z=\frac{w}{\sqrt{2}}\mathrm{cos}\zeta\mathrm{sin}\delta_1,\quad \beta'_z=\frac{w}{\sqrt{2}}\mathrm{cos}\zeta\mathrm{cos}\delta_1,\nonumber\\
\alpha'_0=\frac{w}{\sqrt{2}}\mathrm{sin}\zeta\mathrm{cos}\delta_2,\quad \beta'_0=-\frac{w}{\sqrt{2}}\mathrm{sin}\zeta\mathrm{sin}\delta_2,&\quad \alpha'_x=\frac{w}{\sqrt{2}}\mathrm{sin}\zeta\mathrm{sin}\delta_2,\quad \beta'_x=\frac{w}{\sqrt{2}}\mathrm{sin}\zeta\mathrm{cos}\delta_2,
\end{align}
with arbitrary U(1) variables $\zeta$, $\delta_1$ and $\delta_2$. 
For $\vec{k}=k\vec{e}_y$, we take a substitution of  $\alpha'_1\rightarrow\alpha'_2$, $\alpha'_2\rightarrow -\alpha'_1$, $\alpha'_3\rightarrow\alpha'_3$, $\alpha'_0\rightarrow\alpha'_0$ in  Eq.~(\ref{eqn7-7}), and change 
$\vec{\beta'}$ similarly. For general  $\vec{k}=k(\mathrm{cos}\omega\vec{e}_x+\mathrm{sin}\omega\vec{e}_y)$, 
the function $f(\alpha_\mu,\beta_\mu)$ is minimized by a combination of these two, 
\begin{equation}
\label{eqn7-8}
\vec{\alpha}'=\frac{w}{\sqrt{2}}[(-\mathrm{sin}\omega\mathrm{cos}\zeta\mathrm{cos}\delta_1+\mathrm{cos}\omega\mathrm{sin}\zeta\mathrm{sin}\delta_2)\vec{e}_x+(\mathrm{sin}\omega\mathrm{sin}\zeta\mathrm{sin}\delta_2+\mathrm{cos}\omega\mathrm{cos}\zeta\mathrm{cos}\delta_1)\vec{e}_y+\mathrm{cos}\zeta\mathrm{sin}\delta_1\vec{e}_z+\mathrm{sin}\zeta\mathrm{cos}\delta_2\vec{e}_0],
\end{equation}
\begin{equation}
\label{eqn7-9}
\vec{\beta}'=\frac{w}{\sqrt{2}}[(\mathrm{sin}\omega\mathrm{cos}\zeta\mathrm{sin}\delta_1+\mathrm{cos}\omega\mathrm{sin}\zeta\mathrm{cos}\delta_2)\vec{e}_x+(\mathrm{sin}\omega\mathrm{sin}\zeta\mathrm{cos}\delta_2-\mathrm{cos}\omega\mathrm{cos}\zeta\mathrm{sin}\delta_1)\vec{e}_y+\mathrm{cos}\zeta\mathrm{cos}\delta_1\vec{e}_z-\mathrm{sin}\zeta\mathrm{sin}\delta_2\vec{e}_0].
\end{equation}
With the solution Eqs.~(\ref{eqn7-8}--\ref{eqn7-9}), $S_1[\Phi^{\prime}\vec{\psi}^{\prime}(\vec{r})]$ 
is minimized by the following 
$\vec{\psi}^{\prime}(\vec{r})$ for a given $\Phi^{\prime}$ 
and $w_{\vec{k}}$:
\begin{align}
\label{eqn7-10}
\vec{\psi}^{\prime}(\vec{r})&=\frac{1}{V}\sum_{k_x>0}\sqrt{2}w_{\vec{k}}\{[-\mathrm{sin}\omega\mathrm{sin}\zeta\mathrm{sin}(\vec{k}\cdot\vec{r}-\delta_2)+\mathrm{cos}\omega\mathrm{cos}\zeta\mathrm{cos}(\vec{k}\cdot\vec{r}-\delta_1)]\vec{e}_x\nonumber\\
&+[\mathrm{sin}\omega\mathrm{cos}\zeta\mathrm{cos}(\vec{k}\cdot\vec{r}-\delta_1)+\mathrm{cos}\omega\mathrm{sin}\zeta\mathrm{sin}(\vec{k}\cdot\vec{r}-\delta_2)]\vec{e}_y 
-\mathrm{cos}\zeta\mathrm{sin}(\vec{k}\cdot\vec{r}-\delta_1)\vec{e}_z 
+\mathrm{sin}\zeta\mathrm{cos}(\vec{k}\cdot\vec{r}-\delta_2)\vec{e}_0\},
\end{align}
\begin{equation}
\label{eqn7-11}
S_1[\Phi^{\prime}\vec{\psi}^{\prime}(\vec{r})]=\frac{2\Phi'^2}{V}
\sum_{k_x>0}\big(\lambda k^2-Dk\big) \!\ w_{\vec{k}}^2.
\end{equation}
Here $\omega$ in Eq.~(\ref{eqn7-10}) is a function of $\vec{k}/k$; $\vec{k} \equiv k(\mathrm{cos}\omega\vec{e}_x+\mathrm{sin}\omega\vec{e}_y)$, while 
$\zeta$, $\delta_1$ and $\delta_2$ are arbitrary functions of $\vec{k}$. 
From Eq.~(\ref{eqn7-11}), $S_1[\Phi^{\prime}\vec{\psi}^{\prime}(\vec{r})]$ can 
be maximally minimized by those $w_{\vec{k}}$ that is non-zero only when 
$k=D/(2\lambda)$;  
\begin{equation}
\label{eqn7-12}
\vec{k}=k(\mathrm{cos}\omega\vec{e}_x+\mathrm{sin}\omega\vec{e}_y)=\frac{D}{2\lambda}(\mathrm{cos}\omega\vec{e}_x+\mathrm{sin}\omega\vec{e}_y).
\end{equation}
The action thus minimized depends only on $\Phi^{\prime}$,  
\begin{equation}
\label{eqn7-13}
S_1[\Phi^{\prime}\vec{\psi}^{\prime}(\vec{r})]=-\frac{\Phi'^2D^2}{2\lambda V}\sum_{k_x>0}w_{\vec{k}}^2=-\frac{\Phi'^2D^2 V}{4\lambda}.
\end{equation}
In the right hand side, we use a global constraint,  
$\sum_{\vec{k}} w^2_{\vec{k}} = V^2$, 
that comes from  the local constraint  $\sum_{\mu}\psi^{\prime}_\mu(\vec{r})\psi^{\prime}_\mu(\vec{r})=1$, 
\begin{equation}
\label{eqn7-4}
\sum_{\mu} \int \mathrm{d}^2\vec{r}\psi^{\prime}_\mu(\vec{r})
\psi^{\prime}_\mu(\vec{r})=\frac{1}{V}\sum_{\mu,\vec k}{\psi^{\prime}}^*_{\mu,\vec{k}}
\psi^{\prime}_{\mu,\vec{k}}=\frac{1}{V}\sum_{\vec k}w^2_{\vec{k}}=V. 
\end{equation} 
In Sec.~\ref{sec11}, we show that it is impossible that  
$\psi_\mu(\vec{r})$ given by Eq.~(\ref{eqn7-10}) 
consists of two wavevector Fourier components. Specifically, 
we prove that if $\vec{\psi}(\vec{r})$ in Eq.~(\ref{eqn7-10}) consists 
of the two wavevector Fourier components, $\vec{k}_1$ and $\vec{k}_2$; 
\begin{align}
w_{\vec{k}} = \left\{\begin{array}{cl} 
0 & {\rm for} \!\ \!\ \vec{k} \ne \vec{k}_1 \!\ 
\!\ {\rm  and}  \!\ 
\!\ \vec{k} \ne \vec{k}_2, \\
w_1 \ne 0 & {\rm for} \!\ \!\ \vec{k} = \vec{k}_1, \\
w_2 \ne 0 & {\rm for} \!\ \!\ \vec{k} = \vec{k}_2 \ne \vec{k}_1,-\vec{k}_1, \\
\end{array}\right. 
\end{align}
$\vec{\psi}(\vec{r})$ thus given cannot satisfy 
the local constraint 
$(\vec{\psi}(\vec{r})\cdot \vec{\psi}(\vec{r}) =1)$ in any way.
The conclusion can be generalized into a case with more than the two momenta. 
Thus, we regard $\vec{\psi}(\vec{r})$ is composed 
only of one component of momentum $\vec{k}$ on $|\vec{k}|=D/(2\lambda)$ and take 
$w=\frac{V}{\sqrt{2}}$ from the global constraint. In conclusion, 
$S_{1}[\Phi^{\prime}\vec{\psi}^{\prime}(\vec{r})]$ is maximally minimized by, 
\begin{align}
\label{eqn7-13a}
\vec{\psi}^{\prime}(\vec{r})&=\{[-\mathrm{sin}\omega^{\prime}\mathrm{sin}\zeta\mathrm{sin}
(\vec{k}^{\prime}\cdot\vec{r}-\delta^{\prime}_2)+\mathrm{cos}\omega^{\prime}\mathrm{cos}\zeta^{\prime}
\mathrm{cos}(\vec{k}^{\prime}\cdot\vec{r}-\delta^{\prime}_1)]\vec{e}_x\nonumber\\
&+[\mathrm{sin}\omega^{\prime}\mathrm{cos}\zeta^{\prime}\mathrm{cos}(\vec{k}^{\prime}\cdot\vec{r}-\delta^{\prime}_1)
+\mathrm{cos}\omega^{\prime}\mathrm{sin}\zeta^{\prime}\mathrm{sin}(\vec{k}^{\prime}\cdot\vec{r}-\delta^{\prime}_2)]
\vec{e}_y -\mathrm{cos}\zeta^{\prime}\mathrm{sin}(\vec{k}^{\prime}\cdot\vec{r}-\delta^{\prime}_1)\vec{e}_z 
+\mathrm{sin}\zeta^{\prime}\mathrm{cos}(\vec{k}^{\prime}\cdot\vec{r}-\delta^{\prime}_2)\vec{e}_0\},
\end{align}
with $\vec{k}^{\prime}=\frac{D}{2\lambda}(\cos\omega^{\prime} \vec{e}_x+\sin\omega^{\prime}\vec{e}_y)$ 
together with arbitrary U(1) phase variables, $\zeta^{\prime}$, $\delta^{\prime}_1$ and 
$\delta^{\prime}_2$. Note that this satisfies the local constraint, $\vec{\psi}(\vec{r}) 
\cdot \vec{\psi}(\vec{r})=1$. Likewise, $S_{1}[\Phi^{\prime\prime}\vec{\psi}^{\prime\prime}(\vec{r})]$ 
is maximally minimized by $\psi^{\prime\prime}(\vec{r})$ 
given by the same type of the unit vector 
as Eq.~(\ref{eqn7-13a}) with another set of 
the U(1) variables of 
$\omega^{\prime\prime}$ ($\vec{k}^{\prime\prime}$), $\zeta^{\prime\prime}$, $\delta^{\prime\prime}_1$ and $\delta^{\prime\prime}_2$. 

Finally, we minimize $S_0[\vec{\Phi}^{\prime},\vec{\Phi^{\prime\prime}}]$ within a `manifold' 
of Eq.~(\ref{eqn7-13a}) for 
$\vec{\Phi}^{\prime}(\vec{r})\equiv \Phi^{\prime}\vec{\psi}^{\prime}(\vec{r})$ and that for  $\vec{\Phi}^{\prime\prime}(\vec{r}) \equiv \Phi^{\prime\prime}\vec{\psi}^{\prime\prime}(\vec{r})$. 
The minimization is carried out in a parameter space subtended by 
$\omega^{\prime}$ ($\vec{k}^{\prime}$), $\zeta^{\prime}$, $\delta^{\prime}_1$, 
$\delta^{\prime}_{2}$, $\omega^{\prime\prime}$ ($\vec{k}^{\prime\prime}$), 
$\zeta^{\prime\prime}$, $\delta^{\prime\prime}_1$, 
$\delta^{\prime\prime}_{2}$, $\Phi^{\prime}$ and $\Phi^{\prime\prime}$. 
To begin with, we consider to maximize $g$ in Eq.~(\ref{eqn7-1}) at a given spatial point $\vec{r}$. 
The maximization leads to $\omega^{\prime}=\omega^{\prime\prime}=\pi/2$ ($\vec{k}^{\prime}=\vec{k}^{\prime\prime}
= \frac{D}{2\lambda} \vec{e}_y$), while $\zeta^{\prime}=\zeta^{\prime\prime}=\pi/2$ for 
$|h|\le |h^{\prime}|$, and $\zeta^{\prime}=\zeta^{\prime\prime}=0$ for $|h^{\prime}|\le |h|$. 
For each of these two cases, we then maximize $g$ in Eq.~(\ref{eqn7-1}) 
in terms of $\delta^{\prime}_2-\delta^{\prime\prime}_2$ 
and $\delta^{\prime}_1-\delta^{\prime\prime}_1$ respectively, and finally minimize the 
whole action  in terms of $\Phi^{\prime}$ and $\Phi^{\prime\prime}$;  
\begin{align}
\label{eqn7-14}
s&\equiv\frac{S}{V}=A'(\Phi'^2+\Phi''^2)+B(\Phi'^4+\Phi''^4+6\Phi'^2\Phi''^2)-2\Phi'\Phi''g, 
\nonumber \\
g &\equiv \left\{\begin{array}{cc}
2B \Phi^{\prime}\Phi^{\prime\prime}
\cos(\delta^{\prime}_1-\delta^{\prime\prime}_1) 
+ h \sin(\delta^{\prime}_1-\delta^{\prime\prime}_1) & 
{\rm when} \!\ \!\ \!\ |h^{\prime}|<|h|, \\
2B \Phi^{\prime}\Phi^{\prime\prime} 
\cos(\delta^{\prime}_2-\delta^{\prime\prime}_2) 
- h^{\prime} \sin(\delta^{\prime}_2-\delta^{\prime\prime}_2) & 
{\rm when} \!\ \!\ \!\ |h|<|h^{\prime}|, \\
\end{array}\right.  
\end{align}
with $A'=-(\alpha-\frac{2}{g}+\frac{D^2}{4\lambda})$. Here the $D^2$ term in $A^{\prime}$ 
comes from Eq.~(\ref{eqn7-13}). 
The maximization of $g$ in terms of $\delta^{\prime}_2-\delta^{\prime\prime}_2=\alpha_1$ 
and $\delta^{\prime}_1-\delta^{\prime\prime}_1=\alpha_2$ 
and the minimization of the classical action $s$ in 
terms of $\Phi^{\prime}$ and $\Phi^{\prime\prime}$ 
are essentially 
same as in Eqs.~(\ref{eqn2-4-7}--\ref{eqn2-15d}). Thereby, Eqs.~(\ref{eqn2-16}--\ref{eqn2-19}) are still 
valid classical solutions, except for the following substitutions,  
\begin{equation}
\label{eqn7-15}
\varphi_0\rightarrow\varphi_0-Ky,\quad K\equiv\frac{D}{2\lambda},
\end{equation}
\begin{equation}
\label{eqn7-15d}
h_c=\alpha-\frac{2}{g}\rightarrow h_c=\alpha-\frac{2}{g}+\frac{D^2}{4\lambda}.
\end{equation}
To summarize, we have the following four phases.\\  
{\bf For $|h'|<|h|<h_c$ (regular helicoidal phase)}:
\begin{align}
\label{eqn7-16}
&\vec{\phi}=\rho\mathrm{cos}\theta(\mathrm{cos}(\varphi_0-Ky)\vec{e}_y+
\mathrm{sin}(\varphi_0-Ky)\vec{e}_z)+\mathrm{i}\rho\mathrm{sin}\theta[
\mathrm{cos}(\varphi+\varphi_0-Ky)\vec{e}_y+\mathrm{sin}(\varphi+\varphi_0-Ky)\vec{e}_z],\nonumber\\
&\rho=\sqrt{\frac{h_c}{2|\gamma|}},\quad\mathrm{sin}\varphi\mathrm{sin}2\theta=\frac{h}{h_c}.
\end{align}
{\bf For $|h|<|h'|<h_c$ (regular helical phase)}:
\begin{align}
\label{eqn7-17}
&\vec{\phi}=\rho[-\mathrm{sin}\theta\mathrm{cos}(\varphi+\varphi_0-Ky)\vec{e}_0+\mathrm{cos}\theta\mathrm{sin}(\varphi_0-Ky)\vec{e}_x]+\mathrm{i}\rho[\mathrm{cos}\theta\mathrm{cos}(\varphi_0-Ky)\vec{e}_0
+\mathrm{sin}\theta\mathrm{sin}(\varphi+\varphi_0-Ky)\vec{e}_x],\nonumber\\
&\rho=\sqrt{\frac{h_c}{2|\gamma|}},\quad\mathrm{sin}\varphi\mathrm{sin}2\theta=-\frac{h'}{h_c}.
\end{align}
{\bf For $|h'|<|h|$, $h_c<|h|$ (saturated helicoidal phase)}:
\begin{equation}
\label{eqn7-18}
\vec{\phi}=\rho(\mathrm{cos}(\varphi_0-Ky)\vec{e}_y+\mathrm{sin}(\varphi_0-Ky)\vec{e}_z)
-\mathrm{i}\rho\mathrm{sgn}(h)[\mathrm{sin}(\varphi_0-Ky)\vec{e}_y
-\mathrm{cos}(\varphi_0-Ky)\vec{e}_z],\quad \rho=\sqrt{\frac{h_c+|h|}{8|\gamma|}}.
\end{equation}
{\bf For $|h|<|h'|$, $h_c<|h'|$ (saturated helical phase)}:
\begin{equation}
\label{eqn7-19}
\vec{\phi}=\rho[\mathrm{sgn}(-h')\mathrm{sin}(\varphi_0-Ky)\vec{e}_0 
+\mathrm{sin}(\varphi_0-Ky)\vec{e}_x]+\mathrm{i}\rho[
\mathrm{cos}(\varphi_0-Ky)\vec{e}_0+\mathrm{sgn}(-h')\mathrm{cos}(\varphi_0-Ky)\vec{e}_x],
\quad\rho=\sqrt{\frac{h_c+|h'|}{8|\gamma|}}.
\end{equation}
Here we call ground-state configurations of 
Eqs.~(\ref{eqn7-16}, \ref{eqn7-18}) with the substitutions as regular and saturated 
helicoidal phase and those of 
Eqs.~(\ref{eqn7-17}, \ref{eqn7-19}) as regular and saturated 
helical phase. Both helical and helicoidal phases have a nonzero momentum $K$, 
breaking the translational symmetry along $y$-axis. The phase diagram with $D\ne 0$ 
is given by Fig.~\ref{fig.phases_1} where `transverse' and `longitudinal' are 
replaced by `helicoidal' and `helical' respectively. The helicoidal phases were 
introduced and studied in Ref.~\cite{chen2019}. A schematic picture of 
the helical phase is shown in Fig.~\ref{fig.phases_2}.

\subsection{\label{sec8} Derivation of spin-charge coupled Josephson equation with Rashba coupling}
In this section, we derive spin-charge coupled Josephson equations 
in the helicoidal and helical phases. We 
first apply Noether's theorem to the $\phi^4$-type effective Lagrangian, Eq.~(\ref{eqn1-11}), to express (conserved) charge and spin currents in 
terms of the four-components excitonic fields and their spatial gradient 
terms. We then substitute the classical solutions,  Eqs.~(\ref{eqn7-16}--\ref{eqn7-19}), into the expressions. This leads to the spin-charge coupled Josephson equations in the helical and helicoidal phases.    

In practice, the magnetic exchange fields within the $xy$ plane can be experimentally 
controlled by an external magnetic field in the plane. The magnetic 
field causes magnetic vector potentials in the 
electron and hole layers. In the presence of the vector potentials, 
$\partial_i$ in Eqs.~(1, \ref{eqn0-0}) ($i=x,y$) are replaced by 
$\partial_i +{\mathrm{i}} A_{a,i}$ in the electron layer 
and by $\partial_i +{\mathrm{i}} A_{b,i}$ 
in the hole layer. The two vector potentials, $A_{a,i}$ and $A_{b,i}$, are 
generally different from each other, as the electron and hole layers are 
spatially separated along the $z$ direction. An integral 
of the difference between them, $\tilde{A}_i \equiv A_{b,i}-A_{a,i}$, 
is the magnetic flux penetrating through the 
separation layer. The difference appears in the $\phi^4$-type effective 
Lagrangian in a covariant way; 
$\partial_{i}$ in Eq.~(\ref{eqn1-11}) is replaced by $\partial_{i} 
- {\mathrm{i}} \tilde{A}_{i}$.

Noether's theorem associates a global continuous symmetry in Lagrangian 
${\cal L}(\vec{\phi},\partial_i \vec{\phi})$ with a conserved current.~\cite{peskin1995} 
Suppose that a Lagrangian is invariant under the following transformation, 
\begin{equation}
\label{eqn8-1}
\phi_\nu\rightarrow\phi_\nu+\epsilon\Delta\phi_\nu,\quad \phi_\nu^{*}\rightarrow\phi_\nu^{*}+\epsilon\Delta\phi_\nu^{*},
\end{equation}
with a small $\epsilon$. 
Noether's theorem dictates that a system must have a 
conserved current defined as follows:
\begin{equation}
\label{eqn8-2}
J_\mu=\frac{\partial\mathcal{L}}{\partial(\partial_\mu \phi_\nu)}\Delta \phi_\nu+\frac{\partial\mathcal{L}}{\partial(\partial_\mu \phi_\nu^*)}\Delta \phi_\nu^*.
\end{equation}
The effective $\phi^4$-type lagrangian has the pseudospin rotational symmetry around 
$x$ axis;
\begin{align}
    \left\{\begin{array}{l} 
    \left(\begin{array}{c} 
    -{\rm i}\phi_0 \\
    \phi_x \\
    \end{array}\right) \rightarrow \left(\begin{array}{c} 
    -{\rm i}\tilde{\phi}_0 \\
    \tilde{\phi}_x \\
    \end{array}\right) = \left(\begin{array}{cc}
    \cos (\delta \varphi_0) & -\sin (\delta \varphi_0) \\
    \sin (\delta \varphi_0) & \cos (\delta \varphi_0) \\
    \end{array}\right) \left(\begin{array}{c} 
    -{\rm i}\phi_0 \\
    \phi_x \\
    \end{array}\right), \\
    \left(\begin{array}{c} 
    \phi_y \\
    \phi_z \\
    \end{array}\right) \rightarrow \left(\begin{array}{c} 
    \tilde{\phi}_y \\
    \tilde{\phi}_z \\
    \end{array}\right) = \left(\begin{array}{cc}
    \cos (\delta \varphi_0) & -\sin (\delta \varphi_0) \\
    \sin (\delta \varphi_0) & \cos (\delta \varphi_0) \\
    \end{array}\right) \left(\begin{array}{c} 
    \phi_y \\
    \phi_z \\
    \end{array}\right), \\ 
    \end{array}\right. 
    \label{eqn8-2a}
\end{align}
and the relative U(1) gauge symmetry;
\begin{align}
    \vec{\phi} \rightarrow  e^{i\psi} \vec{\phi}.  \label{eqn8-2b}
\end{align}
Note that in the presence of the Rashba coupling $\xi_{e} \ne 0$, 
the pseudospin rotational symmetry around the $x$ axis is nothing but 
the spin rotational symmetry in the hole layer in Eqs.~(1, \ref{eqn0-0});  
taking $\varphi_a=0$ and $\varphi_b=\mp\delta \varphi_0$ in 
Eq.~(6) of the main text leads to Eq.~(\ref{eqn8-2a}).
For the relative $U(1)$ gauge symmetry, taking $\psi_b-\psi_a=-\psi$
in Eq.~(7) of the main text leads to Eq.~(\ref{eqn8-2b}).

To calculate charge and spin Noether current in the hole layer, we choose $\varphi_b$ and $\psi_b-\psi_a$ 
to be positive. Namely, taking $\delta \varphi_0$ and $\psi$ in Eqs.~(\ref{eqn8-2a}, \ref{eqn8-2b}) 
to be $\mp \epsilon$ and $-\epsilon$ with infinitesimally small positive $\epsilon$, we obtain  
\begin{equation}
\label{eqn8-5}
\phi_y\rightarrow\phi_y\pm\epsilon\phi_z,\quad\phi_z\rightarrow\phi_z\mp\epsilon\phi_y,
\end{equation}
\begin{equation}
\label{eqn8-6}
-\mathrm{i}\phi_0\rightarrow-\mathrm{i}\phi_0\pm\epsilon\phi_x,\quad\phi_x\rightarrow\phi_x\mp\epsilon(-\mathrm{i}\phi_0),
\end{equation}
(and $\phi^*_\nu$ changes accordingly) for the spin rotational symmetry 
and
\begin{equation}
\label{eqn8-3}
\phi_\nu\rightarrow\phi_\nu-\mathrm{i}\epsilon\phi_\nu,\quad\phi_\nu^*\rightarrow\phi_\nu^*+\mathrm{i}\epsilon\phi_\nu^*,
\end{equation}
for the relative gauge symmetry respectively. 
Accordingly, the corresponding conserved currents are calculated from 
Noether's theorem:
\begin{equation}
\label{eqn8-4}
J^C_\mu=-\frac{\partial\mathcal{L}}{\partial(\partial_\mu\phi_\nu)}\mathrm{i}\phi_\nu+\frac{\partial\mathcal{L}}{\partial(\partial_\mu\phi_\nu^*)}\mathrm{i}\phi_\nu^*,
\end{equation}
and 
\begin{align}
\label{eqn8-7}
J^S_\mu&=\pm\big{\{}\frac{\partial\mathcal{L}}{\partial(\partial_\mu \phi_y)}\phi_z-\frac{\partial\mathcal{L}}{\partial(\partial_\mu \phi_z)}\phi_y+\frac{\partial\mathcal{L}}{\partial(\partial_\mu \phi_y^*)}\phi_z^*-\frac{\partial\mathcal{L}}{\partial(\partial_\mu \phi_z^*)}\phi_y^*\nonumber\\
&+\frac{\partial\mathcal{L}}{\partial[\partial_\mu (-\mathrm{i}\phi_0)]}\phi_x-\frac{\partial\mathcal{L}}{\partial(\partial_\mu \phi_x)}(-\mathrm{i}\phi_0)+\frac{\partial\mathcal{L}}{\partial[\partial_\mu (\mathrm{i}\phi_0^*)]}\phi_x^*-\frac{\partial\mathcal{L}}{\partial(\partial_\mu \phi_x^*)}(\mathrm{i}\phi_0^*)\big{\}}, 
\end{align}
respectively. Here $\mathcal{L}$ is the $\phi^4$-type Lagrangian density 
(see Eq.~(\ref{eqn1-11})).  

A substitution of the magnetic vector potentials into Eq.~(\ref{eqn1-11}) 
leads to the following Lagrangian density:
\begin{align}
\label{eqn8-8}
\mathcal{L}&=-\eta\phi^*_\nu(\partial_\tau-\mathrm{i}\tilde{A}_\tau)\phi_\nu+\lambda[(\partial_i+\mathrm{i}\tilde{A}_i)\phi_\nu^*][(\partial_i-\mathrm{i}\tilde{A}_i)\phi_\nu]\nonumber\\
&-D[\phi_z^*(\partial_x\phi_x)-\phi_x^*(\partial_x\phi_z)-\phi_y^*(\partial_y\phi_z)+\phi_z^*(\partial_y\phi_y)-\mathrm{i}\phi^*_x(\partial_y\phi_0)+\mathrm{i}\phi^*_y(\partial_x\phi_0)+\mathrm{i}\phi^*_0(\partial_x\phi_y)-\mathrm{i}\phi^*_0(\partial_y\phi_x)]+..., 
\end{align}
where $D=2K\lambda$ from Eq.~(\ref{eqn1-18}), $i=x,y$ and $\nu=0,x,y,z$. Here 
``..." denotes other terms without the spatial gradients of 
$\phi_\nu$ and $\phi^*_\nu$; they do not contribute to the Noether currents. 
Putting Eq.~(\ref{eqn8-8}) into Eq.~(\ref{eqn8-4}), 
we get the (hole-layer) charge Noether current: 
\begin{equation}
\label{eqn8-9}
J_i^C=2\mathrm{i}\lambda\phi_\nu^*(\partial_i-\mathrm{i}\tilde{A}_i)\phi_\nu+\mathrm{i}D[\delta_{i,x}(\phi_z^*\phi_x-\phi^*_x\phi_z)-\delta_{i,y}(\phi^*_y\phi_z-\phi^*_z\phi_y)+\mathrm{i}\delta_{i,x}(\phi_0^*\phi_y+\phi_y^*\phi_0)-\mathrm{i}\delta_{i,y}(\phi^*_0\phi_x+\phi^*_x\phi_0)],
\end{equation}
where we have used $\phi_\nu\partial_i\phi^*_\nu=-\phi^*_\nu\partial_i\phi_\nu$. Putting Eq.~(\ref{eqn8-8}) into Eq.~(\ref{eqn8-7}), we get the (hole-layer) spin Noether current:
\begin{align}
\label{eqn8-10}
J_i^S&=\pm 2\lambda[\phi_z^*(\partial_i-\mathrm{i}\tilde{A}_i)\phi_y-\phi_y^*(\partial_i-\mathrm{i}\tilde{A}_i)\phi_z-\mathrm{i}\phi^*_x(\partial_i-\mathrm{i}\tilde{A}_i)\phi_0-\mathrm{i}\phi_0^*(\partial_i-\mathrm{i}\tilde{A}_i)\phi_x]\nonumber\\
&\mp D[\delta_{i,x}(\mathrm{i}\phi_0^*\phi_z+\phi_x^*\phi_y)+\delta_{i,y}(\phi^*_z\phi_z+\phi^*_y\phi_y)+\delta_{i,x}(\mathrm{i}\phi_z^*\phi_0-\phi_y^*\phi_x)+\delta_{i,y}(\phi_z^*\phi_z+\phi_0^*\phi_0)].
\end{align}

Let us next substitute the classical solutions of  Eqs.~(\ref{eqn7-16}--\ref{eqn7-19}) into these expressions 
for the Noether currents. Thereby, the spatial gradients in the 
expressions apply not only to an explicit spatial-coordinate dependence of 
the classical solutions but also to the slowly-varying gapless 
modes, $\varphi_0$ and $\psi$. Thus, the spatial derivatives 
in Eqs.~(\ref{eqn8-9}, \ref{eqn8-10}) can be decomposed into:
\begin{equation}
\label{eqn8-11}
\partial_i=(\partial_i\psi)\frac{\partial}{\partial\psi}+(\partial_i\varphi_0)\frac{\partial}{\partial\varphi_0}+\partial'_i,
\end{equation}
where $\partial'_i$ applies only to the explicit spatial-coordinate 
dependence. From Eqs.~(\ref{eqn7-16}, \ref{eqn7-18}), these partial derivatives 
take the following forms in the helicoidal phases,
\begin{equation}
\label{eqn8-12}
\frac{\partial\phi_y}{\partial\psi}=\mathrm{i}\phi_y,\quad \frac{\partial\phi_z}{\partial\psi}=\mathrm{i}\phi_z,\quad \frac{\partial\phi_y}{\partial\varphi_0}=-\phi_z,\quad \frac{\partial\phi_z}{\partial\varphi_0}=\phi_y,\quad \partial'_i\phi_y=K\phi_z\delta_{i,y},\quad \partial'_i\phi_z=-K\phi_y\delta_{i,y}.
\end{equation}
From Eqs.~(\ref{eqn7-17}, \ref{eqn7-19}), we 
obtain a set of similar relations for the helical phases. Using them 
in Eqs.~(\ref{eqn8-9}, \ref{eqn8-10})
together with $D=2K\lambda$, and recover units by substitutions $\tilde{A}_i\rightarrow e\tilde{A}_i/\hbar c$, we finally obtain 
results for the charge current and the spin current:
\begin{equation}
\label{eqn8-13}
J^C_i=-\frac{\lambda h_c}{|\gamma|}[(\partial_i\psi-
\frac{e}{\hbar c}\tilde{A}_i)-{\sf{h}}\partial_i\varphi_0],
\end{equation}
\begin{equation}
\label{eqn8-14}
J^S_i=\mp\frac{\lambda h_c}{|\gamma|}[\partial_i\varphi_0-
{\sf{h}}(\partial_i\psi-\frac{e}{\hbar c}\tilde{A}_i)]. 
\end{equation}
Eqs.~(\ref{eqn8-13}, \ref{eqn8-14}) 
are consistent with 
Eqs.~(15, 16) in the main text
in a small-phase-difference limit
up to the coefficients ($I_0$ and $\bar{h}_{\pm}$) 
respectively. The consistency concludes that the  
qualitative properties of the spin-charge coupled Josephson 
effect holds true also in the presence of the SOC in the electron layer.

\subsection{\label{sec10} Delta functions between real numbers and bilinear Grassmann numbers in Eq.~(\ref{eqn5-7})}

In this section, we define a delta function between a real number and a bilinear Grassmann number and discuss its properties. The discussion 
can be regarded as a proof for Eq.~(\ref{eqn5-7}). For simplicity, we discuss normal integrals (i.e. zero-dimensional field theory), and the results are easy to be generalized for path integrals (i.e. finite-dimensional field theory). Below $(\psi,\psi^*)$ is a pair of conjugate Grassmann numbers, while $N$ and $\mu$ are real numbers. We define a delta function:
\begin{equation}
\label{eqn10-1}
\delta(N-\psi^*\psi)\equiv\delta(N)-\delta'(N)\psi^*\psi,
\end{equation}
where $\delta(N)$ is a delta function for real numbers, $\delta'(N)\equiv\mathrm{d}\delta(N)/\mathrm{d}N$. It leads to following two equations:
\begin{equation}
\label{eqn10-2}
\int\mathrm{d}N\delta(N-\psi^*\psi)f(N)=\int\mathrm{d}N[\delta(N)f(N)+\psi^*\psi\delta(N)f'(N)]=f(0)+\psi^*\psi f'(0)=f(\psi^*\psi).
\end{equation}
\begin{equation}
\label{eqn10-3}
\int\frac{\mathrm{d}\mu}{2\pi}\mathrm{e}^{\mathrm{i}\mu(N-\psi^*\psi)}=\int\frac{\mathrm{d}\mu}{2\pi}\mathrm{e}^{\mathrm{i}\mu N}(1-\mathrm{i}\mu\psi^*\psi)
=\int\frac{\mathrm{d}\mu}{2\pi}(1-\psi^*\psi\frac{\partial}{\partial N}) 
\mathrm{e}^{\mathrm{i}\mu N}=\delta(N)-\psi^*\psi\delta'(N)=\delta(N-\psi^*\psi).
\end{equation}
Eq.~(\ref{eqn10-2}) and Eq.~(\ref{eqn10-3}) lead to the second line and the third line of Eq.~(\ref{eqn5-7}) respectively. Note that the delta function thus defined does not satisfy a reciprocal equation of Eq.~(\ref{eqn10-2}):
\begin{equation}
\label{eqn10-4}
\int\mathrm{d}\psi^*\mathrm{d}\psi\delta(N-\psi^*\psi)g(\psi^*\psi)\neq \int\mathrm{d}\psi^*\mathrm{d}\psi\bar{\delta}(N-\psi^*\psi)g(\psi^*\psi)=g(N),
\end{equation}
where we define a reciprocal delta function  as:
\begin{equation}
\label{eqn10-5}
\bar{\delta}(N-\psi^*\psi)\equiv -N-\psi^*\psi.
\end{equation}
The right hand side of Eq.~(\ref{eqn10-4}) holds true for an arbitrary linear function $g(x)\equiv a + b x$, 
\begin{equation}
\label{eqn10-6}
\int\mathrm{d}\psi^*\mathrm{d}\psi\bar{\delta}(N-\psi^*\psi)g(\psi^*\psi)=\int\mathrm{d}\psi\mathrm{d}\psi^*(N+\psi^*\psi)(a+b\psi^*\psi)=a+bN=g(N),
\end{equation}
where we used $\mathrm{d}\psi\mathrm{d}\psi^*=-\mathrm{d}\psi^*\mathrm{d}\psi$.

Although Eq.~(\ref{eqn5-7}) is verified, the physical meanings of $N_C$ and $N_S$ are still not apparent due to the uncommon properties of the delta function (Eq.~(\ref{eqn10-4})). Below we further verify Eq.~(\ref{eqn5-6-1}) in the sense of expectation values, i.e.
\begin{equation}
\label{eqn10-7}
\big\langle N_C-\frac{1}{2}\sum_{\alpha} \Big[{\bm b}^{\dagger}_{1\alpha}{\bm b}_{1\alpha}-{\bm b}^{\dagger}_{2\alpha}{\bm b}_{2\alpha}\Big]\big\rangle=\big\langle N_S-\frac{1}{2}\sum_{\alpha} \Big[{\bm b}^{\dagger}_{1\alpha}{\bm \sigma}_x {\bm b}_{1\alpha}-{\bm b}^{\dagger}_{2\alpha}{\bm \sigma}_x{\bm b}_{2\alpha}\Big]\big\rangle=0,
\end{equation}
where the expectations values are taken with respect to the partition function in the last line of Eq.~(\ref{eqn5-7}). We can verify it by a proof of a equation:
\begin{equation}
\label{eqn10-8}
\langle N-\psi^*\psi\rangle=-\mathrm{i}\frac{1}{\mathcal{Z}[0]}\frac{\delta\mathcal{Z}[\mu']}{\delta\mu'}|_{\mu'=0}=0,
\end{equation}
where
\begin{equation}
\label{eqn10-9}
\mathcal{Z}[\mu']\equiv\int\mathcal{D}\mu\mathcal{D}N\mathcal{D}\psi^*\mathcal{D}\psi\mathrm{e}^{\mathrm{i}\int\mathrm{d}\tau(\mu+\mu')(N-\psi^*\psi)-\mathcal{S}[\psi,\psi^*]}.
\end{equation}
Eq.~(\ref{eqn10-8}) can be proved by identities of variations and path integrals;
\begin{align}
\label{eqn10-10}
\langle N-\psi^*\psi\rangle&=\frac{-\mathrm{i}}{\mathcal{Z}[0]}\int\mathcal{D}\mu\frac{\delta}{\delta\mu'}|_{\mu'=0}\int\mathrm{D}N\mathcal{D}\psi^*\mathcal{D}\psi\mathrm{e}^{\mathrm{i}\int\mathrm{d}\tau(\mu+\mu')(N-\psi^*\psi)-\mathcal{S}[\psi,\psi^*]}\nonumber\\
&=\frac{-\mathrm{i}}{\mathcal{Z}[0]}\int\mathcal{D}\mu\frac{\delta}{\delta\mu}\int\mathrm{D}N\mathcal{D}\psi^*\mathcal{D}\psi\mathrm{e}^{\mathrm{i}\int\mathrm{d}\tau\mu(N-\psi^*\psi)-\mathcal{S}[\psi,\psi^*]}=0,
\end{align}
where the last equation holds because it is a surface term. Eq.~(\ref{eqn10-8}) can be easily generalized to Eq.~(\ref{eqn10-7}), so $N_{C/S}$ indeed has the physical meaning of the charge/spin density difference in the hole layer.

\subsection{\label{sec11} Impossibility that Eq.~(\ref{eqn7-10}) consists of two 
wavevector Fourier components}

In this section, we show it impossible that $\psi^{\prime}_{\mu}(\vec{r})$ in Eq.~(\ref{eqn7-10}) 
is given by two wavevectors Fourier components. To be specific, we show that $\psi^{\prime}_{\mu}(\vec{r})$ 
thus given cannot satisfy the local constraint, 
$\vec{\psi}^{\prime}(\vec{r})\cdot \vec{\psi}^{\prime}(\vec{r}) =1$. For simplicity of the notation, 
we omit the prime in $\vec{\psi}^{\prime}(\vec{r})$ in this section: 
$\vec{\psi}^{\prime}(\vec{r}) \rightarrow \vec{\psi}(\vec{r})$. Suppose that $\vec{\psi}(\vec{r})$
is given by two wavevector Fourier components, $\vec{k}$ and $\vec{k}^{\prime}$ 
with $k_x>0$, $k^{\prime}_x>0$;  
\begin{equation}
\label{eqn11-1}
\vec{\psi}(\vec{r})=\frac{2}{V}\mathrm{Re}(\vec{\psi}_{\vec{k}} 
\mathrm{e}^{\mathrm{i}\vec{k}\cdot\vec{r}}+\vec{\psi}_{\vec{k}'}\mathrm{e}^{\mathrm{i}\vec{k}'\cdot\vec{r}})
\equiv\frac{\sqrt{2}}{V}(w\vec{\tilde{\psi}}_{\vec{k}}(\vec{r})+w'\vec{\tilde{\psi}}_{\vec{k}^{\prime}} (\vec{r})),
\end{equation}
Here $\vec{\tilde{\psi}}_{\vec{k}}(\vec{r})$ is a unit vector defined by Eq.~(\ref{eqn7-10}), 
\begin{align}
   \vec{\tilde{\psi}}_{\vec{k}}(\vec{r}) =&  [-\mathrm{sin}\omega\mathrm{sin}\zeta\mathrm{sin}(\vec{k}\cdot\vec{r}-\delta_2)+\mathrm{cos}\omega\mathrm{cos}\zeta\mathrm{cos}(\vec{k}\cdot\vec{r}-\delta_1)]\vec{e}_x\nonumber\\
&+[\mathrm{sin}\omega\mathrm{cos}\zeta\mathrm{cos}(\vec{k}\cdot\vec{r}-\delta_1)+\mathrm{cos}\omega\mathrm{sin}\zeta\mathrm{sin}(\vec{k}\cdot\vec{r}-\delta_2)]\vec{e}_y 
-\mathrm{cos}\zeta\mathrm{sin}(\vec{k}\cdot\vec{r}-\delta_1)\vec{e}_z 
+\mathrm{sin}\zeta\mathrm{cos}(\vec{k}\cdot\vec{r}-\delta_2)\vec{e}_0, 
\end{align}
with $\vec{k} \equiv \frac{D}{2\lambda}(\cos\omega \vec{e}_x + \sin\omega \vec{e}_y)$ and 
arbitrary U(1) phase variables $\zeta$, $\delta_1$, $\delta_2$. The other unit vector 
$\vec{\tilde{\psi}}_{\vec{k}^{\prime}}(\vec{r})$ is defined in the same way with 
$\vec{k}^{\prime} \equiv \frac{D}{2\lambda}(\cos\omega^{\prime} \vec{e}_x + \sin\omega^{\prime} \vec{e}_y)$, 
$\zeta^{\prime}$, $\delta^{\prime}_1$, $\delta^{\prime}_2$. 
Since $k_x>0$, $k^{\prime}_x>0$ and $\vec{k}\ne \vec{k}^{\prime}$, 
$2\tau \equiv \omega-\omega^{\prime}$ must satisfy 
$\sin \tau \ne 0$ and $\cos\tau \ne 0$. 

In the following, we will argue that  
it is impossible that $\psi_{\mu}(\vec{r})$ thus given 
satisfies the local constraint. Firstly, the norm of $\psi_{\mu}(\vec{r})$ 
is given by 
\begin{equation}
\label{eqn11-2}
\vec{\psi}(\vec{r})\cdot \vec{\psi}(\vec{r})=\frac{2}{V^2}(w^2+w'^2) 
+\frac{4}{V^2}ww' \vec{\tilde{\psi}}_{\vec{k}}(\vec{r})
\cdot \vec{\tilde{\psi}}_{\vec{k}'}(\vec{r}), 
\end{equation}
where the second term in the right hand side generally 
depends on spatial coordinate $\vec{r}$,  
\begin{align}
\label{eqn11-3}
2\vec{\tilde{\psi}}_{\vec{k}}(\vec{r})\cdot \vec{\tilde{\psi}}_{\vec{k}'}(\vec{r})=& [(1-\mathrm{cos}(\omega-\omega'))(\mathrm{sin}\zeta\mathrm{sin}\zeta'\mathrm{cos}(\delta_2+\delta_2')-\mathrm{cos}\zeta\mathrm{cos}\zeta'\mathrm{cos}(\delta_1+\delta_1'))\nonumber\\
& \!\ \!\ \!\ +\mathrm{sin}(\omega-\omega')(\mathrm{sin}\zeta\mathrm{cos}\zeta'\mathrm{sin}(\delta_2+\delta_1')-\mathrm{cos}\zeta\mathrm{sin}\zeta'\mathrm{sin}(\delta_2'+\delta_1))]\mathrm{cos}((\vec{k}+\vec{k}')\cdot\vec{r})\nonumber\\
 & +[(1-\mathrm{cos}(\omega-\omega'))(\mathrm{sin}\zeta\mathrm{sin}\zeta'\mathrm{sin}(\delta_2+\delta_2')-\mathrm{cos}\zeta\mathrm{cos}\zeta'\mathrm{sin}(\delta_1+\delta_1'))\nonumber\\
&\!\ \!\ \!\ +\mathrm{sin}(\omega-\omega')(-\mathrm{sin}\zeta\mathrm{cos}\zeta'\mathrm{cos}(\delta_2+\delta_1')+\mathrm{cos}\zeta\mathrm{sin}\zeta'\mathrm{cos}(\delta_2'+\delta_1))]\mathrm{sin}((\vec{k}+\vec{k}')\cdot\vec{r})\nonumber\\
& +[(1+\mathrm{cos}(\omega-\omega'))(\mathrm{sin}\zeta\mathrm{sin}\zeta'\mathrm{cos}(\delta_2-\delta_2')+\mathrm{cos}\zeta\mathrm{cos}\zeta'\mathrm{cos}(\delta_1-\delta_1'))\nonumber\\
& \!\ \!\ \!\ +\mathrm{sin}(\omega-\omega')(\mathrm{sin}\zeta\mathrm{cos}\zeta'\mathrm{sin}(\delta_2-\delta_1')-\mathrm{cos}\zeta\mathrm{sin}\zeta'\mathrm{sin}(\delta_2'-\delta_1))]\mathrm{cos}((\vec{k}-\vec{k}')\cdot\vec{r})\nonumber\\
 & +[(1+\mathrm{cos}(\omega-\omega'))(\mathrm{sin}\zeta\mathrm{sin}\zeta'\mathrm{sin}(\delta_2-\delta_2')+\mathrm{cos}\zeta\mathrm{cos}\zeta'\mathrm{sin}(\delta_1-\delta_1'))\nonumber\\
&\!\ \!\ \!\ +\mathrm{sin}(\omega-\omega')(-\mathrm{sin}\zeta\mathrm{cos}\zeta'\mathrm{cos}(\delta_2-\delta_1')-\mathrm{cos}\zeta\mathrm{sin}\zeta'\mathrm{cos}(\delta_2'-\delta_1))]\mathrm{cos}((\vec{k}-\vec{k}')\cdot\vec{r}).
\end{align} 
Thus, in order for $\vec{\psi}(\vec{r})$ to satisfy the 
local constraint, the coefficients in front of $\mathrm{cos}((\vec{k}+\vec{k}')\cdot\vec{r})$, $\mathrm{sin}((\vec{k}+\vec{k}')\cdot\vec{r})$, $\mathrm{cos}((\vec{k}-\vec{k}')\cdot\vec{r})$ and $\mathrm{cos}((\vec{k}-\vec{k}')\cdot\vec{r})$ in 
Eq.~(\ref{eqn11-3}) should be zero. 
Adding the coefficients in front of the first two terms together, we get:
\begin{equation}
\label{eqn11-4}
2\mathrm{sin}^2\tau[\mathrm{sin}\zeta\mathrm{sin}\zeta'\mathrm{e}^{\mathrm{i}(\delta_2+\delta_2')}-\mathrm{cos}\zeta\mathrm{cos}\zeta'\mathrm{e}^{\mathrm{i}(\delta_1+\delta_1')}]-2\mathrm{i}\mathrm{cos}\tau\mathrm{sin}\tau[\mathrm{sin}\zeta\mathrm{cos}\zeta'\mathrm{e}^{\mathrm{i}(\delta_2+\delta_1')}-\mathrm{cos}\zeta\mathrm{sin}\zeta'\mathrm{e}^{\mathrm{i}(\delta_1+\delta_2')}]=0.
\end{equation}
Adding the coefficients in front of the last two terms together, we get:
\begin{equation}
\label{eqn11-5}
2\mathrm{cos}^2\tau[\mathrm{sin}\zeta\mathrm{sin}\zeta'\mathrm{e}^{\mathrm{i}(\delta_2-\delta_2')}+\mathrm{cos}\zeta\mathrm{cos}\zeta'\mathrm{e}^{\mathrm{i}(\delta_1-\delta_1')}]-2\mathrm{i}\mathrm{cos}\tau\mathrm{sin}\tau[\mathrm{sin}\zeta\mathrm{cos}\zeta'\mathrm{e}^{\mathrm{i}(\delta_2-\delta_1')}+\mathrm{cos}\zeta\mathrm{sin}\zeta'\mathrm{e}^{\mathrm{i}(\delta_1-\delta_2')}]=0.
\end{equation}
As $\mathrm{sin}\tau\neq 0$ and $\mathrm{cos}\tau\neq 0$, these two lead to 
either one of the following two constraints on $\zeta$, $\zeta^{\prime}$, $\delta_1$, 
$\delta^{\prime}_1$, $\delta_2$, $\delta^{\prime}_2$ and 
$2\tau \equiv \omega-\omega^{\prime}$. One is 
\begin{align}
\label{eqn11-6}
&\mathrm{sin}\zeta\mathrm{sin}\zeta'\mathrm{e}^{\mathrm{i}(\delta_2+\delta_2')}-\mathrm{cos}\zeta\mathrm{cos}\zeta'\mathrm{e}^{\mathrm{i}(\delta_1+\delta_1')}=\mathrm{sin}\zeta\mathrm{cos}\zeta'\mathrm{e}^{\mathrm{i}(\delta_2+\delta_1')}-\mathrm{cos}\zeta\mathrm{sin}\zeta'\mathrm{e}^{\mathrm{i}(\delta_1+\delta_2')}\nonumber\\
&=\mathrm{sin}\zeta\mathrm{sin}\zeta'\mathrm{e}^{\mathrm{i}(\delta_2-\delta_2')}+\mathrm{cos}\zeta\mathrm{cos}\zeta'\mathrm{e}^{\mathrm{i}(\delta_1-\delta_1')}=\mathrm{sin}\zeta\mathrm{cos}\zeta'\mathrm{e}^{\mathrm{i}(\delta_2-\delta_1')}+\mathrm{cos}\zeta\mathrm{sin}\zeta'\mathrm{e}^{\mathrm{i}(\delta_1-\delta_2')}=0,
\end{align}
while the other is 
\begin{equation}
\label{eqn11-7}
\frac{\mathrm{sin}\zeta\mathrm{cos}\zeta'\mathrm{e}^{\mathrm{i}(\delta_2+\delta_1')}-\mathrm{cos}\zeta\mathrm{sin}\zeta'\mathrm{e}^{\mathrm{i}(\delta_1+\delta_2')}}{\mathrm{sin}\zeta\mathrm{sin}\zeta'\mathrm{e}^{\mathrm{i}(\delta_2+\delta_2')}-\mathrm{cos}\zeta\mathrm{cos}\zeta'\mathrm{e}^{\mathrm{i}(\delta_1+\delta_1')}}=-\frac{\mathrm{sin}\zeta\mathrm{sin}\zeta'\mathrm{e}^{\mathrm{i}(\delta_2-\delta_2')}+\mathrm{cos}\zeta\mathrm{cos}\zeta'\mathrm{e}^{\mathrm{i}(\delta_1-\delta_1')}}{\mathrm{sin}\zeta\mathrm{cos}\zeta'\mathrm{e}^{\mathrm{i}(\delta_2-\delta_1')}+\mathrm{cos}\zeta\mathrm{sin}\zeta'\mathrm{e}^{\mathrm{i}(\delta_1-\delta_2')}}=-\mathrm{i}(\mathrm{tan}\tau).
\end{equation}

We first explore a possibility of the second constraint, Eq.~(\ref{eqn11-7}). 
We simplify the first equation in Eq.~(\ref{eqn11-7}), to get $\mathrm{sin}^2\zeta\mathrm{e}^{2\mathrm{i}\delta_2}=\mathrm{cos}^2\zeta\mathrm{e}^{2\mathrm{i}\delta_1}$. Thus, the second 
constraint leads to 
\begin{equation}
\label{eqn11-8}
\delta_2=\delta_1+n\pi,\quad\zeta=\frac{\pi}{4},\frac{3\pi}{4},\frac{5\pi}{4},\frac{7\pi}{4},
\end{equation}
with an integer $n$. Putting Eq.~(\ref{eqn11-8}) back into Eq.~(\ref{eqn11-7}) again, 
we end up with 
$\pm 1=\mathrm{i}(\mathrm{tan}\tau)$. This is clearly impossible. 
We next explore a possibility of the first constraint, Eq.~(\ref{eqn11-6}). 
The first constraint is equivalent to the following eight equations
\begin{align}
\left\{\begin{array}{cc}
\mathrm{sin}\zeta\mathrm{sin}\zeta'\mathrm{cos}(\delta_2+\delta_2')-\mathrm{cos}\zeta\mathrm{cos}\zeta'\mathrm{cos}(\delta_1+\delta_1')=0,&\quad\mathrm{sin}\zeta\mathrm{cos}\zeta'\mathrm{sin}(\delta_2+\delta_1')-\mathrm{cos}\zeta\mathrm{sin}\zeta'\mathrm{sin}(\delta_1+\delta_2')=0,\\ 
\mathrm{sin}\zeta\mathrm{sin}\zeta'\mathrm{sin}(\delta_2+\delta_2')-\mathrm{cos}\zeta\mathrm{cos}\zeta'\mathrm{sin}(\delta_1+\delta_1')=0,&\quad\mathrm{sin}\zeta\mathrm{cos}\zeta'\mathrm{cos}(\delta_2+\delta_1')-\mathrm{cos}\zeta\mathrm{sin}\zeta'\mathrm{cos}(\delta_1+\delta_2')=0,\\
\mathrm{sin}\zeta\mathrm{sin}\zeta'\mathrm{cos}(\delta_2-\delta_2')+\mathrm{cos}\zeta\mathrm{cos}\zeta'\mathrm{cos}(\delta_1-\delta_1')=0,&\quad\mathrm{sin}\zeta\mathrm{cos}\zeta'\mathrm{sin}(\delta_2-\delta_1')+\mathrm{cos}\zeta\mathrm{sin}\zeta'\mathrm{sin}(\delta_1-\delta_2')=0,\\
\mathrm{sin}\zeta\mathrm{sin}\zeta'\mathrm{sin}(\delta_2-\delta_2')+\mathrm{cos}\zeta\mathrm{cos}\zeta'\mathrm{sin}(\delta_1-\delta_1')=0,&\quad\mathrm{sin}\zeta\mathrm{cos}\zeta'\mathrm{cos}(\delta_2-\delta_1')+\mathrm{cos}\zeta\mathrm{sin}\zeta'\mathrm{cos}(\delta_1-\delta_2')=0. \\ 
\end{array}\right. \label{eqn11-9}
\end{align}
Eq.~(\ref{eqn11-9}) leads to
\begin{align}
\label{eqn11-10}
\left\{\begin{array}{cc}
\mathrm{tan}(\delta_1+\delta'_1)=\mathrm{tan}(\delta_2+\delta'_2),&\quad\mathrm{tan}(\delta_2+\delta'_1)=\mathrm{tan}(\delta_1+\delta'_2),\\
\mathrm{tan}(\delta_1-\delta'_1)=\mathrm{tan}(\delta_2-\delta'_2),&\quad\mathrm{tan}(\delta_2-\delta'_1)=\mathrm{tan}(\delta_1-\delta'_2). \\
\end{array}\right. 
\end{align}
Eq.~(\ref{eqn11-10}) results in:
\begin{equation}
\label{eqn11-11}
2(\delta_2-\delta_1)=m\pi,\quad 2(\delta_2'-\delta_1')=n\pi,
\end{equation}
with integers $m$ and $n$. Putting Eq.~(\ref{eqn11-11}) into Eq.~(\ref{eqn11-10}), 
we can see that $m\pm n$ must be an even integer. Taking Eq.~(\ref{eqn11-11}) 
into Eq.~(\ref{eqn11-9}), we finally find the following four possibilities 
depending on the two integers $\frac{m+n}{2}$ and $\frac{m-n}{2}$, 
\begin{align}
\left\{\begin{array}{ll}
\mathrm{cos}(\zeta-\zeta')=\mathrm{cos}(\zeta+\zeta')=
\mathrm{sin}(\zeta - \zeta')=\mathrm{sin}(\zeta + \zeta') = 0, &  {\rm when} \!\ \!\ \frac{m+n}{2} 
\!\ \!\  {\rm is} \!\  \!\ {\rm odd} \!\ ({\rm even}) \!\ \!\ {\rm and} \!\ \!\  
\frac{m-n}{2} \!\ \!\ {\rm is} \!\  \!\ {\rm odd} \!\ ({\rm even}), \\
\mathrm{cos}(\zeta-\zeta')=\mathrm{sin}(\zeta-\zeta')=0, & 
{\rm when} \!\ \!\ \frac{m+n}{2} \!\  
\!\ {\rm is} \!\ \!\ {\rm odd} \!\ \!\ {\rm and} \!\ \!\   
\frac{m-n}{2} \!\ \!\ {\rm is} \!\ \!\ {\rm even}, \\
\mathrm{cos}(\zeta+\zeta')=\mathrm{sin}(\zeta+\zeta')=0, & 
{\rm when} \!\ \!\ \frac{m+n}{2} 
\!\ \!\ {\rm is} \!\  \!\ {\rm even} \!\ \!\ {\rm and} \!\ \!\   
\frac{m-n}{2} \!\ \!\  {\rm is} \!\ \!\ {\rm odd}. 
\end{array}\right. 
\end{align}
They are all impossible. This concludes that Eq.~(\ref{eqn11-3}) 
inevitably depends on the spatial coordinate, contradicting the local constraint. Thus, Eq.~(\ref{eqn11-1})
cannot be true.




%

\end{widetext}

\end{document}